\documentclass[letterpaper,preprintnumbers,amsmath,amssymb,nofootinbib,notitlepage,twocolumn]{revtex4-1}
\pdfoutput=1
\usepackage{amssymb}
\usepackage{xspace}
\usepackage{graphicx}
\usepackage{amsmath}
\usepackage{multirow}
\usepackage{verbatim}
\usepackage{diagbox}
\usepackage{eufrak}
\usepackage[usenames,dvipsnames]{color}
\usepackage[bookmarks=true, bookmarksnumbered=true, colorlinks=true, urlcolor=darkblue, citecolor=darkgreen, linkcolor=darkred, breaklinks=true]{hyperref}

\definecolor{darkblue}{rgb}{0,0,0.5}
\definecolor{darkgreen}{rgb}{0.1,0,0.3}
\definecolor{darkred}{rgb}{0.6,0,0}

\newcommand{\vect}[1]{\boldsymbol{#1}}
\newcommand{\nc}{\newcommand}
\nc{\ba}{\begin{eqnarray}}
\nc{\ea}{\end{eqnarray}}

\nc{\ga}{\gamma}
\nc{\om}{\omega}
\newcommand\ci{\chi}
\nc{\x}{{\bf x }}
\nc{\kk}{{\bf k }}
\nc{\f}{{\bf f }}
\nc{\e}{{\bf e }}
\nc{\gag}{g_{a \gamma}}
\nc{\ud}{\mathrm{d}}
\nc{\igev}{GeV$^{-1}$}
\nc{\ssi}{\sigma_{\mathrm{SI}}}
\nc{\ssd}{\sigma_{\mathrm{SD}}}
\nc{\tq}{\tilde \q}
\nc{\qmin}{q_{\mathrm{min}}}
\nc{\qmax}{q_{\mathrm{max}}}
\nc{\dmin}{\delta_{\mathrm{min}}}
\nc{\dmax}{\delta_{\mathrm{max}}}
\nc{\ie}{i.e.\xspace}
\nc{\del}{\partial}
\nc{\Cin}{C_{\mathrm{in}}}
\nc{\Cout}{C_{\mathrm{out}}}
\nc{\vrel}{v_{\mathrm{rel}}}
\nc{\shat}{\hat \sigma}
\nc{\ket}[1]{| #1 \rangle}
\nc{\bra}[1]{\langle #1 |}
\nc{\braket}[2]{\langle #1 | #2 \rangle}
\nc{\mn}{m_{\rm nuc}}
\nc{\mx}{m_\chi}
\nc{\sav}{\langle \sigma \rangle}
\nc{\qref}{q_{\rm ref}}
\begin{document}

%%%%%%%%%%%%%%%%%%%%%%%%%%%%%%%%%%%%%%%%%%%%%%%%%%%%%%%%%%%%%%%%%%%%%%%
\title{Thermal conduction by dark matter with velocity and momentum-dependent cross-sections}
\author{Aaron C. Vincent}
\email{vincent@ific.uv.es}
\affiliation{Instituto de F\'{\i}sica Corpuscular (IFIC)$,$
 CSIC-Universitat de Val\`encia$,$ \\  
 Apartado de Correos 22085$,$ E-46071 Valencia$,$ Spain}
\author{Pat Scott} 
\email{patscott@physics.mcgill.ca}
\affiliation{Department of Physics, McGill University 3600 Rue University, Montr\'eal, Qu\'ebec, Canada H3A 2T8}

\begin{abstract}
We use the formalism of Gould and Raffelt \cite{GouldRaffelt90a} to compute the dimensionless thermal conduction coefficients for scattering of dark matter particles with standard model nucleons via cross-sections that depend on the relative velocity or momentum exchanged between particles. Motivated by models invoked to reconcile various recent results in direct detection, we explicitly compute the conduction coefficients $\alpha$ and $\kappa$ for cross-sections that go as $\vrel^2$, $\vrel^4$, $\vrel^{-2}$, $q^2$, $q^4$ and $q^{-2}$, where $\vrel$ is the relative DM-nucleus velocity and $q$ is the momentum transferred in the collision. We find that a $\vrel^{-2}$ dependence can significantly enhance energy transport from the inner solar core to the outer core. The same can true for any $q$-dependent coupling, if the dark matter mass lies within some specific range for each coupling.  This effect can complement direct searches for dark matter; combining these results with state-of-the-art solar simulations should greatly increase sensitivity to certain DM models. It also seems possible that the so-called Solar Abundance Problem could be resolved by enhanced energy transport in the solar core due to such velocity- or momentum-dependent scatterings.
\end{abstract}

\maketitle

\section{Introduction}
If cosmological dark matter (DM) is similar in character to a WIMP (weakly-interacting massive particle), it may possess a weak-scale cross-section for interaction with ordinary nucleons.  In this case, collisions between DM particles in the halo of the Milky Way and nuclei in the Sun will slow some of the DM enough to gravitationally bind it to the Sun \cite{Press85,Griest87,Gould87b}. If enough is captured and remains in the Sun, DM may impact the structure and evolution of the Sun itself \cite{Steigman78,Spergel85,Faulkner85}. 

The effects of captured DM particles on the Sun and other stars have been a subject of investigation for many decades (see e.g.~\cite{Scott09,Turck12,Zurek13} for reviews).  Much work has concentrated on the detectable signals of annihilating DM in stellar cores, such as $\sim$GeV-scale neutrinos escaping from the Sun \cite{Krauss86, Gaisser86, Griest87, Gandhi:1993ce, Bottino:1994xp, Bergstrom98b, Barger02, Desai04, Desai08, IceCube09, IceCube09_KK, IC40DM, SuperK11, IC22Methods, Silverwood12, IC79} and the effects of core heating by annihilation products \cite{SalatiSilk89, BouquetSalati89a, Moskalenko07, Spolyar08, Bertone07, Fairbairn08, Scott08a, Iocco08a, Iocco08b, Scott09, Casanellas09, Ripamonti10, Zackrisson10a, Zackrisson10b, Scott11}, or the MeV-scale neutrinos from these decay products \cite{Rott13, Bernal:2012qh}.  The weakly-interacting nature of WIMP-like DM\footnote{Strictly, the WIMP paradigm implies certain weak-scale annihilation and nuclear scattering cross-sections, a GeV--TeV scale mass and a corresponding natural reproduction of the observed relic density via thermal production.  Here all that is important is the interaction with nucleons, so throughout this paper we use `WIMP' loosely to refer to any DM with the requisite nuclear interaction, regardless of its mass or annihilation cross-section.} can also make it a medium for energy transport in the Sun or other stars \cite{Steigman78,Spergel85,Faulkner85}, the implications of which have been studied extensively \cite{Gilliland86, Renzini87, Spergel88, Faulkner88, Bouquet89, BouquetSalati89b, Salati90, Dearborn90b, Dearborn90, GH90, CDalsgaard92, Faulkner93, Iocco12}.  Even when that energy transport is highly localised (\ie conductive), WIMPs may cool the core by absorbing heat, which can then be deposited at larger radii. These changes can be potentially constrained with the temperature-sensitive $^8$B neutrino flux, or complementarily, with precision helio/asteroseismological measurements of the sound speed, small frequency separations and gravity modes \cite{Lopes02a,Lopes02b,Bottino02,Frandsen10,Taoso10,Cumberbatch10,Lopes10,Lopes:2012,Lopes:2013,Casanellas:2013,Lopes:2014}.  

So far all studies of energy transport by DM particles have assumed that the cross-section for scattering between WIMPs and nuclei is independent of their relative velocity and the momentum exchanged in the collision.  This is also a common assumption in analyses of direct searches for dark matter.  Among these, DAMA \cite{Bernabei08}, CoGeNT \cite{CoGeNTAnnMod11}, CRESST-II \cite{CRESST11} and CDMS II \cite{Agnese:2013rvf} have reported excess events above their expected backgrounds, consistent with recoils caused by WIMP-like dark matter. On the other hand, XENON10 \cite{Angle:2011th}, XENON100 \cite{XENON2013}, COUPP \cite{COUPP12}, SIMPLE \cite{SIMPLE11}, Super-Kamiokande \cite{SuperK11} and IceCube \cite{IC79}, under the assumption of constant scattering cross-sections and a standard halo model, would together exclude such ``detections'' from being observations of WIMPs.  However, there are assumptions built into these exclusions. Some recent work (e.g.\ \cite{Bozorgnia:2013hsa, Mao:2013nda}) has focused on the astrophysical uncertainties.  A greater deal of interest has developed in scattering cross-sections with a non-trivial dependence upon the relative velocity or momentum transfer of the collision, as it appears that such velocity-- or momentum-dependent scattering could reconcile the various direct detection results \cite{Masso09, Feldstein:2009tr, Chang:2009yt, Feldstein10, Chang10, An10, Chang:2010en, Barger11, Fitzpatrick10, Frandsen:2013cna}.\footnote{This may be more difficult now that LUX \cite{LUX13} has released first results \cite{Gresham13}, however.}

The focus on constant cross-sections in stellar dark matter studies is therefore not because these are the only interesting DM models, but simply because the required theoretical background for including the effects of velocity or momentum-dependent couplings to nucleons in stellar structure simulations does not exist in the literature.  This is in contrast to direct detection, where such effects can be easily computed (e.g.\ \cite{Cirelli13}). Here we provide such calculations for a range of non-constant cross-sections.  This work will allow the full transport theory to be incorporated into a solar evolution code that also treats the capture and annihilation of DM, in order to derive self-consistent limits on DM models with non-standard cross-sections.

Our other motivation is the well-known Solar Abundance Problem: recent photospheric analyses \cite{APForbidO,CtoO,AspIV,AspVI,AGS05,ScottVII,Melendez08,Scott09Ni,AGSS} indicate that the solar metallicity is 10--20\% lower than previously thought, putting predictions of solar evolution models computed with the improved photospheric abundances at stark odds with sound-speed profiles inferred from helioseismology \cite{Bahcall:2004yr, Basu:2004zg, Bahcall06, Yang07, Basu08, Serenelli:2009yc}.  A deal of effort has been devoted to finding solutions to this quandary \cite{Bahcall05, Badnell05, Arnett05, Charbonnel05, Guzik05, Castro07, Christensen09, Guzik10, Serenelli11, Frandsen10, Taoso10, Cumberbatch10, Vincent12}, but so far no proposal has really proven viable.  Upcoming solar neutrino experiments such as SNO+, Hyper-Kamiokande and LENA should allow the solar core metallicity and temperature to be measured even more accurately than can be done with helioseismology alone \cite{Haxton13,Lopes13b,Smirnov13}, hopefully shedding some light on the discrepancy.  One might postulate that the disagreement is in fact caused by enhanced energy transport in the solar core mediated by non-standard WIMP-nucleon couplings; the calculations we carry out here allow this proposition to be tested.

This paper is structured as follows: first, in Section \ref{sec:backbackground} we provide some background on WIMP-nucleon couplings and how one goes about calculating energy transport by WIMPs, introducing the thermal conduction coefficients $\alpha$ and $\kappa$.  In Section \ref{sec:background} we briefly summarise the relevant equations from Ref.\ \cite{GouldRaffelt90a} that allow calculation of  $\alpha$ and $\kappa$. In Section \ref{sec:non-standard} we modify this treatment to account for non-trivial velocity and momentum structure in the scattering cross-section, and show the effect on $\alpha$ and $\kappa$. In Section \ref{sec:luminosity} we provide some examples of the effect on energy transport in the Sun, then finish with some discussion and concluding remarks in Section \ref{sec:conclusion}.

\section{Preliminary theory}
\label{sec:backbackground}

The quantity that describes microscopic interactions between WIMPs and Standard Model nuclei is the total cross-section $\sigma$ for WIMP-nucleon scattering.  Determining the impacts of energy conduction by a population of weakly-interacting particles in a star means working out their influence on bulk stellar properties like the radial temperature, density and pressure profiles.  Initial solutions were obtained by way of simple analytic approximations \cite{Faulkner85,Spergel85,Gilliland86}, but this quickly gave way to the established approach of solving the Boltzmann collision equation (BCE) inside the spherically symmetric potential well of a star \cite{Nauenberg87,GouldRaffelt90a,GouldRaffelt90b}. The formalism and a general solution for standard WIMP-quark couplings were described by Gould and Raffelt \cite{GouldRaffelt90a}. There are three quantities that, together, are sufficient to describe the transport of energy by a diffuse, weakly-interacting massive particle $\chi$ in a dense medium such as a star:\begin{itemize}
\item $l_\chi$, defined as the inverse of the mean number of WIMP-nucleon interactions per unit length; this gives the typical distance travelled by a WIMP between scatterings;
\item $\alpha$, related to the thermal diffusion coefficient; this parameterises the efficiency of a species' diffusion inside the potential well;
\item $\kappa$, the dimensionless thermal conduction coefficient; this describes the efficiency of energy transfer from layer to layer in the star. 
\end{itemize}
Given a form of the WIMP-nucleon cross-section $\sigma$ as a function of the microscopic kinetic variables ($v_{\rm rel}, q$, etc.), the coefficients $\alpha$ and $\kappa$ are quantities that are averaged over the kinetic gas distributions, thus depending only on $\mu$, the ratio between the WIMP mass and the mass of the nucleus with which it scatters.  They can therefore be computed once and for all for each type of WIMP-nucleon coupling.

In the following sections, we compute the thermal conduction coefficients $\alpha$ and $\kappa$ of WIMPs with non-standard couplings to the standard model, going beyond the $\sigma = const.$ case computed explicitly by Ref.\ \cite{GouldRaffelt90a}. Because these computations must be done on a case by case basis, here we focus on the specific velocity- and momentum-dependent cross-sections $\sigma \varpropto v_{\rm rel}^{2n}$ and $\sigma \varpropto q^{2n}$, with $n = \{-1,1,2\}$.\footnote{We do not consider cases where $n \le -2$, as this causes the total momentum transfer to diverge; conduction properties would then intrinsically depend on the chosen cutoff scale.}  Here $v_{\rm rel}$ is the WIMP-nucleus relative velocity, and $q$ is the momentum transferred during a collision. Our choice of these types of couplings is motivated in part by a number of recent direct-detection experiments \cite{Bernabei08,CoGeNTAnnMod11,CRESST11,Agnese:2013rvf}, but they can easily occur theoretically \cite{Fitzpatrick13,Kumar13}.  Mixed couplings (whether linear combinations of these couplings or cross-terms between them) also each require explicit calculation, and cannot be treated using combinations of the results we present here; we leave these for future work.

The dominant operators that arise in the most basic models of interactions between WIMPs $\chi$ and standard model particles (namely quarks, denoted $Q$) are $O_{SI} = \bar \chi \chi \bar Q Q$ and $ O_{SD} = \bar \chi \gamma_\mu \gamma_5 \chi \bar Q \gamma^\mu \gamma_5 Q$. These respectively lead to the regular spin-independent $\ssi$ and spin-dependent $\ssd$ cross-sections, which have no velocity or momentum structure beyond standard kinematic factors, and have been the focus of exclusion efforts in direct experimental searches to date.

There is, however, no guarantee that the dominant coupling between the dark sector and the SM is as simple as $O_{SI}$ or $O_{SD}$. Possibilities include a finite particle radius (the dark sector analogue of the nuclear form factor), a vector coupling to quarks, parity-violating interactions such as $\bar \chi \gamma_5 \chi \bar Q Q$ or $\bar \chi \gamma_\mu \gamma_5 \chi \bar Q \gamma^\mu Q$ or even a small dipole or anapole interaction with the standard model \cite{Pospelov00,Sigurdson04,Chang:2009yt,Feldstein:2009tr,Feldstein10,Chang:2010en,Fitzpatrick10,Barger11,Frandsen:2013cna,DelNobile:2013cva}. If these terms dominate, then the interaction cross-section will be proportional to some power of the momentum transfer $q$:
\begin{equation}
\sigma = \sigma_0 \left(\frac{q}{q_0}\right)^{2n},
\label{pdep}
\end{equation}
where $\sigma_0$ is the cross-section at some normalisation momentum $q_0$. Models charged under multiple gauge fields can display interactions that go as a sum of different powers of $q^2$ \cite{Feldstein:2009tr}.  Even for scattering via standard $O_{SI}$ or $O_{SD}$ couplings, scattering between WIMPs and nuclei can be strongly momentum-dependent due to the nuclear form factor.  This is especially important at high momentum transfer, so has a large impact on rates at which WIMPs can be scattered to below the local escape velocity and captured by a star.  In this paper we neglect the influence of nuclear form factors, mainly because the majority of WIMP scattering in stellar interiors is with hydrogen and helium, at much lower momentum transfer than
successful capture events. In the cases where momentum is exchanged with heavier nuclei (mainly C, N and O), the suppression due to a Helm form factor is only a few percent at relevant values of $q$ -- much smaller, for example, than uncertainties on the the local interstellar WIMP density or velocity distribution. The form factors should in principle be taken into account if the impact of a given WIMP on a particular star is to be calculated in its fullest, goriest detail.
Operators also exist which produce interactions that depend on the relative velocity $\vrel$ between the WIMP and the nucleus:
\begin{equation}
\sigma = \sigma_0 \left(\frac{\vrel}{v_0}\right)^{2n}.
\label{vdep}
\end{equation}
The collisional cross-section $\sigma$ is proportional to the squared matrix element amplitude $|\mathcal M|^2$, which is typically a function of the centre-of-mass energy. This dependence can be expanded as Taylor series of $(\vrel/c)^2$, where the only sizeable contribution in the non-relativistic limit is the constant term. However, in some models the form of $|\mathcal M|^2$ causes a partial or exact cancellation of the constant term, leaving $\sigma \propto \vrel^2$. This is called \textit{p-wave} suppression. The case in which the second term is also suppressed -- leading to $\sigma \propto \vrel^4$, is referred to as \textit{d-wave}.  On the other hand, DM models which scatter through the exchange of a massive force carrier exhibit a Sommerfeld-like enhancement. These can display ``resonant'' behaviour, which leads to a scattering cross-section proportional to $\vrel^{-2}$ \cite{Sommerfeld,Hisano05,AHDM}.

Given a certain WIMP mass, a good choice for the reference velocity $v_0$ in a star is the temperature-dependent velocity 
\begin{equation}
\label{vt}
v_T(m_\chi,r) = [2 k_{\rm B} T(r)/\mx]^{1/2}.
\end{equation}
The typical thermal WIMP velocity can be obtained by multiplying $v_T$ by a further factor of $(3/2)^{1/2}$ \cite{GouldRaffelt90a}.  Here $T(r)$ refers to the stellar temperature profile as function of radius $r$.  In this paper we adopt $v_0=v_T(m_\chi=20\,\mathrm{GeV},r=0)=110 $\,km\,s$^{-1}$ in the Sun, and provide a simple means for rescaling our results to other values of $v_0$.  Our chosen value is comparable to the typical relative velocities in direct detection experiments, where $v_0 \sim 250$\,km\,s$^{-1}$ is often employed.  Similarly for $q_0$, where our default is $q_0=40$\,MeV: this is about the typical momentum transferred both in thermal collisions in the Sun and direct detection.  This value corresponds to nuclear recoil energies of $E_{\rm R} =  q^2/2m_{\rm nuc} \sim 10$\,keV in direct detection experiments.  As for $v_0$, our results can be easily rescaled to any other $q_0$; the process is described in the passage following (\ref{sigmaqm2}).

In this paper we will not focus on specific particle physics models or the WIMP-nucleon couplings they give rise to. Rather, we will take the general forms (\ref{pdep}) and (\ref{vdep}), and compute the effect on thermal conduction by an additional heavy species inside a star. 

\section{Thermal Conduction in a Star}
\label{sec:background}

\subsection{Theory}

The theoretical framework of energy transport by WIMPs was constructed in detail by Gould and Raffelt \cite{GouldRaffelt90a}.  Transport by a weakly-coupled species can occur in two distinct regimes. If the typical inter-scattering distance $l_\chi$ of a particle $\chi$ is much larger than the typical geometric length scale $r_\chi$ (which in this case is the length scale of the WIMP distribution in a star, not the stellar radius), one reaches the Knudsen limit of non-local transport. In the opposite regime, the Knudsen number $K \equiv l_\chi/r_\chi$ is  $\ll 1$, and WIMPs scatter many times per scale height as they travel through the medium.  Here we are interested in the latter regime, where local thermal equilibrium (LTE) holds. 

In \cite{GouldRaffelt90a}, Gould and Raffelt analysed the conduction of energy by a collection of massive particles $\chi$ in the LTE regime. They modelled energy transport using the Boltzmann collision equation:
\begin{equation}
DF = l_\chi^{-1} C F.
\label{BCE}
\end{equation}
$F(\vect u, \vect r, t)$ here is the phase space density for particle velocity $\vect u$ and position $\vect r$ at time $t$.  The typical inter-scattering distance for a WIMP is the inverse of the mean number of interactions it undergoes per unit length 
\begin{equation}
\label{lchi}
l_\chi(\vect u,\vect r) = \left[\sum_i n_i(r) \langle\sigma_i(\vect{u}) \rangle \right]^{-1}, 
\end{equation}
were the index $i$ denotes the individual nuclear species present in the stellar gas, and the angular brackets represent thermal averaging. $n_i(\vect r)$ is simply the number density of species $i$ at location $\vect r$. The relative weighting of different WIMP velocities in the thermal average depends on the local WIMP velocity distribution, which is itself a function of the temperature with height in the star $T(\vect r)$; the thermal averaging is therefore itself understood to be dependent on $\vect r$. The differential operator is $D(\vect u, \vect r, t) = \del_t + \vect u \cdot \nabla_{\vect r} +\vect g(\vect r) \cdot \nabla_{\vect u}  $, where $\vect g(\vect r)$ is the gravitational acceleration. 

To proceed, we make three key approximations: 
\begin{enumerate}
\item The dilute gas approximation: WIMP-WIMP interactions can be neglected because they are much less frequent than WIMP-nucleon interactions.\footnote{This amounts to the restriction on the WIMP self-interaction term $\sigma_{\chi \chi} \ll \sigma_{\chi n} N_{nuc}/N_\chi$. This is not a strong restriction \textit{per se}, but one must be careful when considering certain DM models with strong self-interactions. For example, DM with large enough capture rates and allowed self interaction cross-sections $\sigma_{\chi \chi} \sim$ 1 cm$^2$/g can lead to large WIMP populations in the sun, violating this condition. This was not taken into account, for example, in the analysis of \cite{Taoso10}.}
\item Local isotropy: this allows us to drop any $\theta, \phi$ dependence in the collision operator $C$.
\item The conduction approximation: energy transport is conductive if $l_\chi$ is smaller than the other two length scales in the problem.  That is, \begin{itemize}
  \item $l_\chi \ll r_\chi \implies K \ll 1$, \ie local thermal equilibrium.
  \item $l_\chi \ll |\nabla \ln T(\vect r)|^{-1}$, \ie the typical inter-scattering distance is much smaller than the length scale over which the temperature changes.  This means that $\varepsilon \equiv l(r) |\nabla \ln T(r)| \ll 1$, allowing a perturbative expansion in powers of $\varepsilon$.
  \end{itemize}
\end{enumerate}
Actually, approximation 3 is only strictly necessary for the computation of $\alpha$ and $\kappa$ rather than their application, as correction factors exist for making the conduction treatment approximately applicable even in the non-local regime when $K > 1$ (see e.g.\ \cite{Bottino02,Scott09}).  In practice both conditions in approximation 3. break down for cross-sections $\sigma \simeq 10^{-39}$\,cm$^2$ and smaller -- we will discuss the correction factors explicitly in Section \ref{sec:luminosity}.
The collision operator $C$ has a straightforward intuitive definition: $CF$ is a function of the WIMP velocity $\vect u$, position $\vect r$ and time $t$, and represents the local rate of change of the WIMP phase space distribution. It can be formally written in terms of the components $\Cin$ and $\Cout$,
\begin{eqnarray}
CF &=& \int \ud^3 v\, C'(\vect u,\vect v, \vect r, t)F(\vect v, \vect r, t), \, \mathrm{with} \label{Cdef} \\
C' &\equiv& \Cin (\vect u,\vect v, \vect r, t) - \Cout (\vect v, \vect r, t)\delta^3(\vect v - \vect u). 
\end{eqnarray}
$\Cin(\vect u,\vect v, \vect r, t)$ is the rate at which particles with velocity $\vect v$ are scattered to velocity $\vect u$.  $\Cout(\vect v,\vect r,t)$ is the rate at which particles with velocity $\vect v$ are scattered to any other velocity.  Performing the integral of $\Cin$ in (\ref{Cdef}) over incoming velocities $\vect v$ gives the total rate of scattering \textit{to} velocity $\vect u$ from any velocity, whereas the integral over $\Cout$ gives the total rate of scattering \textit{from} velocity $\vect u$ to any other velocity.  The difference in these two rates is therefore the net rate of change in the WIMP phase space distribution.  

The BCE was solved by \cite{GouldRaffelt90a} perturbatively, with the expansion 
\begin{equation}
F = F_0 + \varepsilon F_1 + ... 
\label{fperturb}
\end{equation}
where $\varepsilon\ll1$ is an expansion parameter (as explained above),\footnote{Note that in the case where the temperature gradient in a stellar core is so steep that $\varepsilon$ is not small, the entire treatment of energy transport by WIMPs presented here and in \cite{GouldRaffelt90a} breaks down because a perturbative expansion is not possible, even though $K$ may still be much less than 1.  In such interesting cases the phase-space distribution would be strongly non-Boltzmannian, and an entirely different solution to the BCE would be required.} and $F_0$ is a regular Boltzmann distribution.  Such a distribution is in thermal equilibrium by definition, so $CF_0 = 0$.  Using this expansion to first order, dropping the term $\varepsilon D F_1$ because it is much smaller than $DF_0$, the first order BCE is 
\begin{equation}
DF_0 = \frac{\varepsilon}{l_\chi}CF_1.
\label{firstorderBCE}
\end{equation}
Assuming that the thermal timescale in a star is sufficiently long compared to the timescale for conductive energy transport by multiple WIMP scatterings, the solution of the BCE will be time-independent (stationary), such that the time-derivative in $D$ can be ignored.  In this case, the left-hand side of (\ref{firstorderBCE}) is
\begin{equation}
DF_0(\vect u,r) = \frac{\varepsilon}{l_\chi} \left[ \vect \alpha(r)  + \frac{\mx u^2}{2 T(r)}\frac{\nabla \ln T(r)}{|\nabla \ln T(r) | } \right]  \vect u F_0(\vect u, r),
\end{equation}
where $\vect \alpha$ is a separation constant. Assuming the stellar temperature gradient to be spherically symmetric, both $DF_0$ and $F_1$ are pure dipoles. We use the standard notation for spherical harmonics: $Y_j^m(\theta,\varphi)$, where $j$ corresponds to the degree (related to total angular momentum) and $m$ is the order (angular momentum solely in the $\hat r$ direction). We can therefore express the BCE entirely in terms of monopole ($j=m=0$) and dipole ($j = 1, m = 0,\pm 1)$ components of the various quantities; this is the Rosseland approximation.

It is convenient to transform our working variables to dimensionless quantities. We define the WIMP-to-nucleus mass ratio
\begin{equation}
\mu \equiv \mx/ m_{\rm nuc},
\label{mudef}
\end{equation}
and divide velocities by the local temperature-dependent velocity $v_T$ (\ref{vt})
\begin{eqnarray}
\vect x &\equiv& \vect v/v_T, \label{xdef} \\
\vect y &\equiv& \vect u/v_T, \label{ydef} \\ 
\vect z &\equiv& \vect v_{\rm nuc} / v_T, \label{zdef}
\end{eqnarray}
with $v_{\rm nuc}$ denoting the velocity of a nucleus.  

We also define a dimensionless (angle-dependent, differential) cross-section $\shat$, and a total dimensionless (angle-independent, non-differential) cross-section $\shat_{\rm tot}$.  The total dimensionless cross-section is simply $\shat$ integrated over the centre-of-mass scattering angle $\theta_{\rm CM}$:
\begin{equation}
\label{stotdef}
\shat_{\rm tot}(v_{\rm rel}) \equiv \int_{-1}^1 \ud\cos\theta_{\rm CM}\, \shat(v_{\rm rel},q),
\end{equation}
remembering that $q$ is itself a function of both $v_\mathrm{rel}$ and $\cos\theta_{\rm CM}$.  The original dimensionless differential cross-section ($\shat$) is defined by the requirement that the \textit{total} dimensionless cross-section be unity at $v_{\rm rel} = v_T$, \ie 
\begin{equation}
\label{stotdef2}
\shat_{\rm tot}(v_T) = 1.
\end{equation}

Similarly, we define the normalised phase space distribution functions
\begin{equation}
f_\nu^{j,m}(x,\vect r)\,\ud x \equiv \frac{1}{n_{\chi, \mathrm{LTE}}(r)} F_\nu^{j,m}(v,\vect r)\,\ud v, 
\label{fnormeq}
\end{equation}
where $n_{\chi, \mathrm{LTE}}(r)$ is the WIMP number density in the LTE (conductive) approximation, $\nu$ is the expansion order in $\varepsilon$, $j$ is the degree and $m$ is the order of the spherical harmonic expansion.

The first order equation (\ref{firstorderBCE}) can then be expressed as the combination of
\begin{equation}
\left(\alpha_m y - \delta_{m0} y^3 \right) f_0^{0,0}(y,r) = \int \ud x\, C(y,x,r)f_1^{1,m}(x,r)
\label{firstorderBCEv2}
\end{equation}
and the stationarity condition, expressed as the absence of a net WIMP flux across any given surface inside the star:
\begin{equation}
\int \ud x\, x f_1^{1,m} (x,r) = 0.
\label{stationarity}
\end{equation}
Here $\alpha_m$ are thermal diffusivity coefficients corresponding to the dipole coefficients in the spherical harmonic expansion of $\hat{\vect u}\cdot\vect \alpha (r)$.  In the Dirac notation of \cite{GouldRaffelt90a}, where
\begin{eqnarray}
\ket{f} &\equiv& f(y),\\
Q\ket{f} &\equiv& \int \ud x\, Q(y,x)f(x),\\
\braket{g}{f} &\equiv& \int \ud x\, g(x)f(x),\\
\bra{g}Q\ket{f} &\equiv& \int \ud y \ud x\, g(y)Q(y,x)f(x),
\end{eqnarray}
we see that (\ref{firstorderBCEv2}) and (\ref{stationarity}) become quite compact:
\begin{eqnarray}
\alpha_m (r) \ket{y f_0^{0,0}} -\delta_{m0} \ket{y^3 f_0^{0,0}} &=& C\ket{f_1^{1,m}}, \label{keteq}\\
\braket{y}{f_1^{1,m}} &=& 0.
\end{eqnarray}

In terms of the inverse of the collisional operator $C$, we can write the solution to the first-order BCE as
\begin{equation}
\ket{f_1^{1,m}} = \alpha_m C^{-1} \ket{y f_0^{0,0}} - \delta_{m0}C^{-1}\ket{y^3 f_0^{0,0}}.
\label{keteq2}
\end{equation}
The diffusivity coefficients are then fully specified by multiplying (\ref{keteq2}) by $\bra{y}$:
\begin{eqnarray}
\alpha_{\pm 1} &=& 0, \\
\alpha_0 &=& \frac{\bra{y} C^{-1} \ket{y^3 f_0^{0,0}}}{\bra{y} C^{-1} \ket{y f_0^{0,0}}}.
\label{eqn:alpha}
\end{eqnarray}
This can be used to obtain the stationary WIMP density profile \cite{GouldRaffelt90a} 
\begin{eqnarray}
\label{LTEdens}
    n_{\chi,{\rm LTE}}(r) &=& n_{\chi,\mathrm{LTE}}(0)\left[\frac{T(r)}{T(0)}\right]^{3/2} \\
    &\times&\exp\left[-\int^r_0 \ud r'\,\frac{k_{\rm B}\alpha(r')\frac{\ud T(r')}{\ud r'} + 
      m_\chi\frac{\ud \phi(r')}{\ud r'}}{k_{\rm B}T(r')}\right],\nonumber
\end{eqnarray}
where $\phi(r)$ refers to the gravitational potential at height $r$ in the star. Together with the thermal conductivity $\kappa$
\begin{equation}
\kappa = \frac{\sqrt{2}}{3} \braket{y^3}{f_1^{1,0}}, 
\end{equation}
one can then use (\ref{LTEdens}) to finally obtain the luminosity carried by WIMP scattering,
\begin{eqnarray}
\label{LTEtransport}
    L_{\chi,{\rm LTE}}(r) &=& 4\pi r^2 \zeta^{2n}(r) \kappa(r)n_{\chi,{\rm LTE}}(r)l_\chi(r) \nonumber \\
    &\times& \left[\frac{k_\mathrm{B}T(r)}{m_\chi}\right]^{1/2}k_\mathrm{B}\frac{\ud T(r)}{\ud r},
\end{eqnarray}
and the corresponding energy injection rate per unit mass of stellar material:
\begin{equation}
\label{epsLTE}
    \epsilon_{\chi,{\rm LTE}}(r) = \frac{1}{4\pi r^2 \rho(r)}\frac{\ud L_{\chi,{\rm LTE}}(r)}{\ud r}.
\end{equation}
Here $\rho(r)$ is the stellar density, $\zeta(r)= v_0/v_T(r)$ for velocity-dependent scattering and $\zeta(r)= q_0/[m_\chi v_T(r)]$ for momentum-dependent scattering.  The factors of $\zeta$ in (\ref{LTEtransport}) connect the velocity (momentum) scale at which the reference cross-section $\sigma_0$ is defined, to the thermal scale $v_T$ ($m_\chi v_T$) at which the dimensionless conductivity $\kappa$ is computed.  Note that as per Ref.\ \cite{Scott09}, our sign convention is such that positive $L$ and $\epsilon$ refer to energy injection rather than evacuation, which differs from Ref.\ \cite{GouldRaffelt90a}.  

The problem of calculating the rate of energy transport by WIMPs is now one of explicitly calculating $\alpha$ and $\kappa$ for the WIMP-nucleus mass ratio and cross-section of interest.  The only position dependence in the dimensionless BCE arises through the temperature; the transformation to dimensionless variables $x, y$, etc., means that the collision operator $C$, as well as $\alpha$ and $\kappa$, contain no $r$-dependence at all.  This allows $\alpha$ and $\kappa$ to be expressed solely in terms of the mass ratio $\mu$. In a real star, more than one species $i$ of gas is present however, in different ratios at different $r$.  Thus, to compute (\ref{LTEdens}) and (\ref{LTEtransport}) $\alpha$ and $\kappa$ must be computed for the specific mixture of gases at each height \cite{Scott09}:
\begin{equation}
\alpha(r,t) = \sum_i \frac{\sigma_i n_i(r,t)}{\sum_j \sigma_j n_j(r,t)} \alpha_i(\mu_i)
\end{equation}
and
\begin{equation}
\kappa(r,t) = \left[l_\chi(r,t)\sum_i\left\{\kappa_i(\mu_i)l_i(r,t)\right\}^{-1} \right]^{-1}. 
\label{kappaTotal}
\end{equation}
Here $n_i(r,t)$ is the number density of species $i$, while $\sigma_i$ and $l_i(r,t) \equiv (n_i(r)\langle\sigma_i(\vect u) \rangle )^{-1}$ are respectively the scattering cross-section and the typical interscattering distance of WIMPs with species $i$; this is in contrast with the complete sum in (\ref{lchi}). Henceforth we denote $\alpha \equiv \alpha(\mu)$ and $\kappa \equiv \kappa(\mu)$. 

Gould and Raffelt \cite{GouldRaffelt90a} calculated these quantities for a range of mass ratios, but only the case of velocity- and momentum-independent cross-sections; here we extend the treatment to more general cross-sections.

\subsection{Calculations of $\alpha$ and $\kappa$}

Because we plan on evaluating the collision operator $C$ numerically, it makes sense to express the functions $\ket{f(y)}$ as one-dimensional ``vectors'' and the operators as two-dimensional matrices. This allows direct calculation of a well-defined inverse operator $C^{-1}$, simply by inverting the corresponding matrix for $C$.  Although the integrals of the form $Q \ket{f(y)}$ are from zero to infinity, the exponential behaviour of $f(y)$ means that the $y > 3$ contribution to each integral is less than a fraction of a percent. We therefore truncate the vectors and matrices at $y = 5$. 

\begin{figure}[tbp]
\includegraphics[width=.5\textwidth]{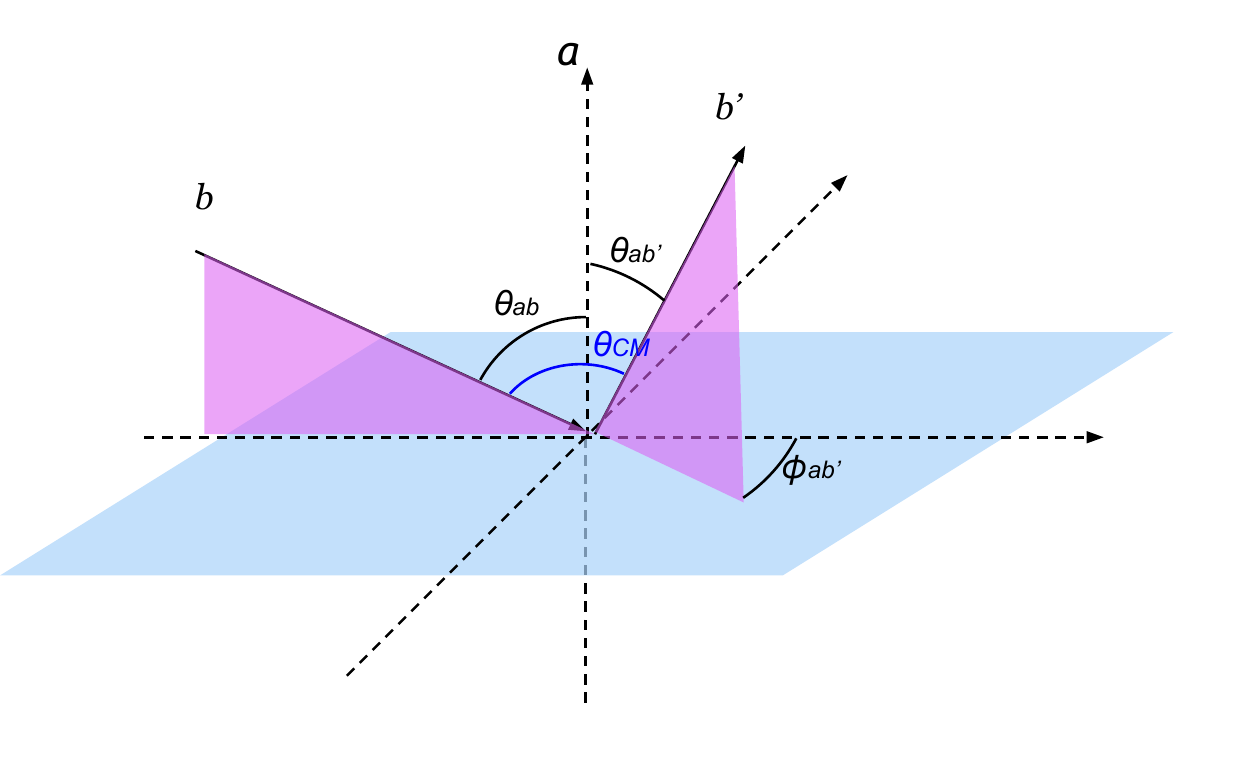}
\caption{\textit{Scattering kinematics in the centre of mass frame, for the computation of $\Cin$. The velocity of the CM with respect to the lab frame is $a$.  The incoming and outgoing WIMP velocities are $b$ and $b'$ respectively. Nucleon velocities are not shown, as they are simply related to the WIMP quantities in this frame.}}
\label{kinfig}
\end{figure}

$\Cout$ is relatively straightforward to construct:
\begin{equation}
\Cout(x, r) = \int \ud^3 z\, |\vect x - \vect z| \hat \sigma_{\rm tot}(v_T | \vect x - \vect z |) F_{\rm {nuc}}(\vect z),
\label{CoutDefinition}
\end{equation}
where $F_{\rm {nuc}}(\vect z)$ is the velocity distribution of the nuclei,
\begin{equation}
F_{\rm {nuc}}(\vect z) = (\pi \mu)^{-3/2} e^{-|\vect z|^2/\mu}.
\label{fnucdef}
\end{equation}
An expression for $\Cin$ is not as obvious. We transform to centre of mass (CM) coordinates, where $\vect a$ is the velocity of the CM frame, and $\vect b$ and $\vect b'$ are respectively the incoming and outgoing WIMP velocities in that frame.\footnote{In the language of Ref.~\cite{GouldRaffelt90a} $\vect a$ and $\vect b$ correspond to $\vect s$ and $\vect t$, respectively.} These quantities are illustrated in Figure \ref{kinfig}.

\begin{table*}[t!]
\caption{Analytic forms of the integral (\ref{CoutDefinition}), given as a function of $w \equiv y/\sqrt{\mu}$.  Expressions are identical for velocity-dependent and momentum-dependent cross-sections of the same order $2n$.}
\label{tab:coutv}
\vspace{2mm}
\begin{tabular}{l@{\hspace{1cm}}l}
\hline\hline\vspace{1mm}
$\sigma=const.$& $\Cout(y) = \mu^{1/2} \left[ \left(w + \frac{1}{2w}\right)\mathrm{erf} (w) + \frac{1}{\sqrt{\pi}}\exp (-w^2)\right]$\\\vspace{1mm}
$\sigma\varpropto v_\mathrm{rel}^2, q^2$ & $\Cout(y) = \mu^{3/2} \left[\frac{3 + 12w^2 + 4 w^4}{4w} \mathrm{erf}(w) +\frac{5 + 2 w^2}{2 \sqrt{\pi}}\exp(-w^2)   \right]$ \\\vspace{1mm}
$\sigma\varpropto v_\mathrm{rel}^4, q^4$& $\Cout(y) = \mu^{5/2} \left[\frac{ \left(15+8 w^6+60  w^4+90 w^2\right)}{8w}\mathrm{erf}(w)+\frac{ \left(33 +4  w^4+28  w^2\right)}{4\sqrt{\pi}}\exp(-w^2)\right]$\\\vspace{1mm}
$\sigma\varpropto v_\mathrm{rel}^{-2}, q^{-2}$ & $\Cout(y) = \mu^{-1/2} w^{-1} \mathrm{erf}(w) $\\
\hline\hline
\end{tabular}
\end{table*}

Ref.\ \cite{GouldRaffelt90a} then finds
\begin{eqnarray}
\label{cinint}
\Cin^j(y,x,r) &=& (1+\mu)^4 \frac{y}{x}\int_0^\infty \ud a\, \int_0^\infty \ud b\, F_{\rm nuc}(\vect z) 2 \pi b \langle P_j \hat \sigma \rangle \nonumber \\
&\times& \Theta(y-|a-b|)\Theta(a+b - y)\nonumber\\
&\times& \Theta(x-|a-b|)\Theta(a+b-x).
\end{eqnarray}
$\langle P_j (\cos \theta_{\rm lab}) \hat \sigma \rangle$ is the angle-averaged product of the differential cross-section with the $j$-th Legendre polynomial, as expanded around transverse scattering angles in the lab frame.  The angle averaging is performed in the azimuthal direction around the $\vect a$ axis in the CM frame. The integral over the zenith angle is performed implicitly when integrating over $a$ and $b$. Note that in this angle-average, $P_j$ is a function of $\theta_{\rm lab}$, whereas $ \shat $ is a function of $\vrel$ and $\theta_{\rm CM}$. The two angles are related as
\begin{eqnarray}
\cos \theta_{\rm CM} &=& A + B\cos \phi_{a b'}, \\
\cos \theta_{\rm lab} &=& G + B\frac{b^2}{xy} \cos \phi_{a b'}, \\
\end{eqnarray}
with 
\begin{eqnarray}
A &\equiv& \cos \theta_{a b'} \cos \theta_{a b} \label{Adef} \\ 
  &=& \frac{(x^2-a^2-b^2)(y^2-a^2-b^2)}{4a^2 b^2}, \nonumber\\
B &\equiv& \sin \theta_{a b'} \sin \theta_{a b}, \label{Bdef} \\
G &\equiv& \frac{(x^2+a^2-b^2)(y^2+a^2-b^2)}{4a^2 xy}.\label{Gdef}
\end{eqnarray}
As we are only solving the BCE to first order, the only relevant term in (\ref{cinint}) is $\langle P_1 \hat \sigma \rangle$.  Noting that \mbox{$\vrel=(1+\mu)bv_T$,} we then have
\begin{eqnarray}
\langle P_j \shat \rangle = \langle P_1 \shat \rangle &=& \frac{1}{2\pi}\int_0^{2\pi} d \phi_{a b'} \left(G + B\frac{b^2}{xy} \cos \phi_{a b'}\right) \nonumber\\
&\times& \shat \left[(1+\mu)bv_T, A + B\cos \phi_{a b'}\right].
\end{eqnarray}

\section{conduction coefficients for non-standard WIMPs}
\label{sec:non-standard}
We write the cross-sections of interest to us explicitly in terms of the dimensionless variables. We will find  $\Cout$ will usually have a tractable analytic form, but that the evaluation of the kinematic integrals for $\Cin$ (\ref{cinint}) must be done numerically, which we do using the CUBPACK multi-dimensional adaptive cubature integration package \cite{CUBPACK}. In every case, we produce an $N \times N$ linearly-spaced matrix for $C(y,x)$, which we explicitly invert to find $\alpha$, $\ket{f_1^{1,0}(y)}$ and $\kappa$.  We find $N=500$ to be sufficient: both $\alpha$ and $\kappa$ vary by less than one part in $10^{5}$ when going from $N = 400$ to $N = 500$, across the full range of $\mu$ values we consider.  For the standard $\sigma = const.$ case, our method of explicit matrix inversion yields identical results to the values of $\alpha(\mu)$ and $\kappa(\mu)$ presented in Table I of Ref.\ \cite{GouldRaffelt90a}, where a slightly different iterative method was used. These values are shown in the figures in this section and are given explicitly in Table \ref{aktable} at the end of the text. Our values of $\alpha$ agree exactly within given precision with Gould and Raffet for small $\mu$, and to within a part in $\sim 10^4$ as $\mu \rightarrow 100$.  Our $\kappa$ results agree within $10^{-3}$ at small $\mu$, up to $\sim 1\%$ as $\mu$ approaches 100. These differences are small enough not to have any appreciable impact on modelled energy transport. We have furthermore cross-checked our results for the collision operator $\Cin$ and $\Cout$ as well as the first-order solution $f_1^{1,0}(y)$ presented in the figures of Ref.\ \cite{GouldRaffelt90a}.

\subsection{Velocity-dependent scattering}
\label{subsec:vdepscat}
\begin{figure*}[tbp]
\begin{tabular}{c@{\hspace{0.04\textwidth}}c}
\includegraphics[width=0.48\textwidth]{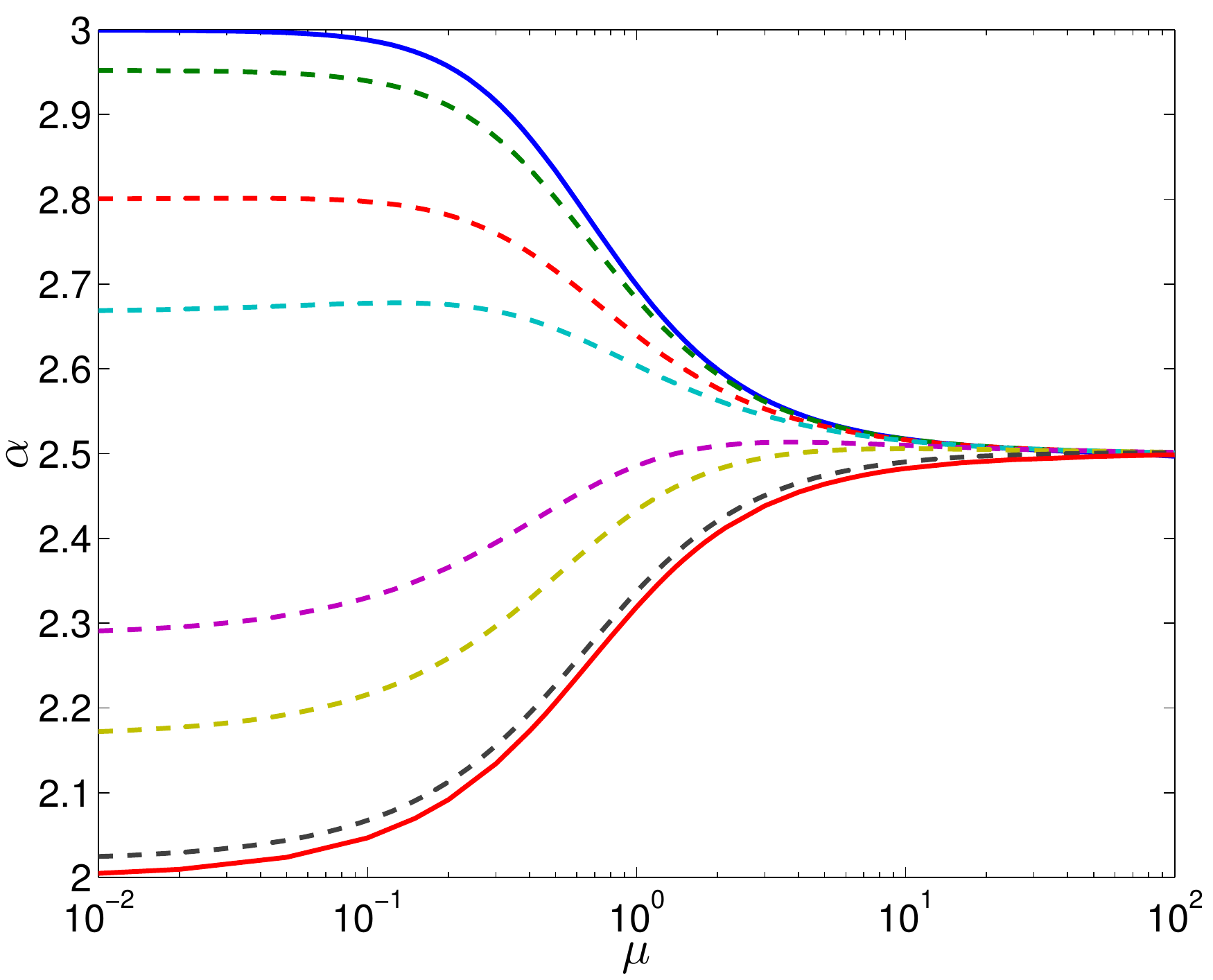} & \includegraphics[width=0.48\textwidth]{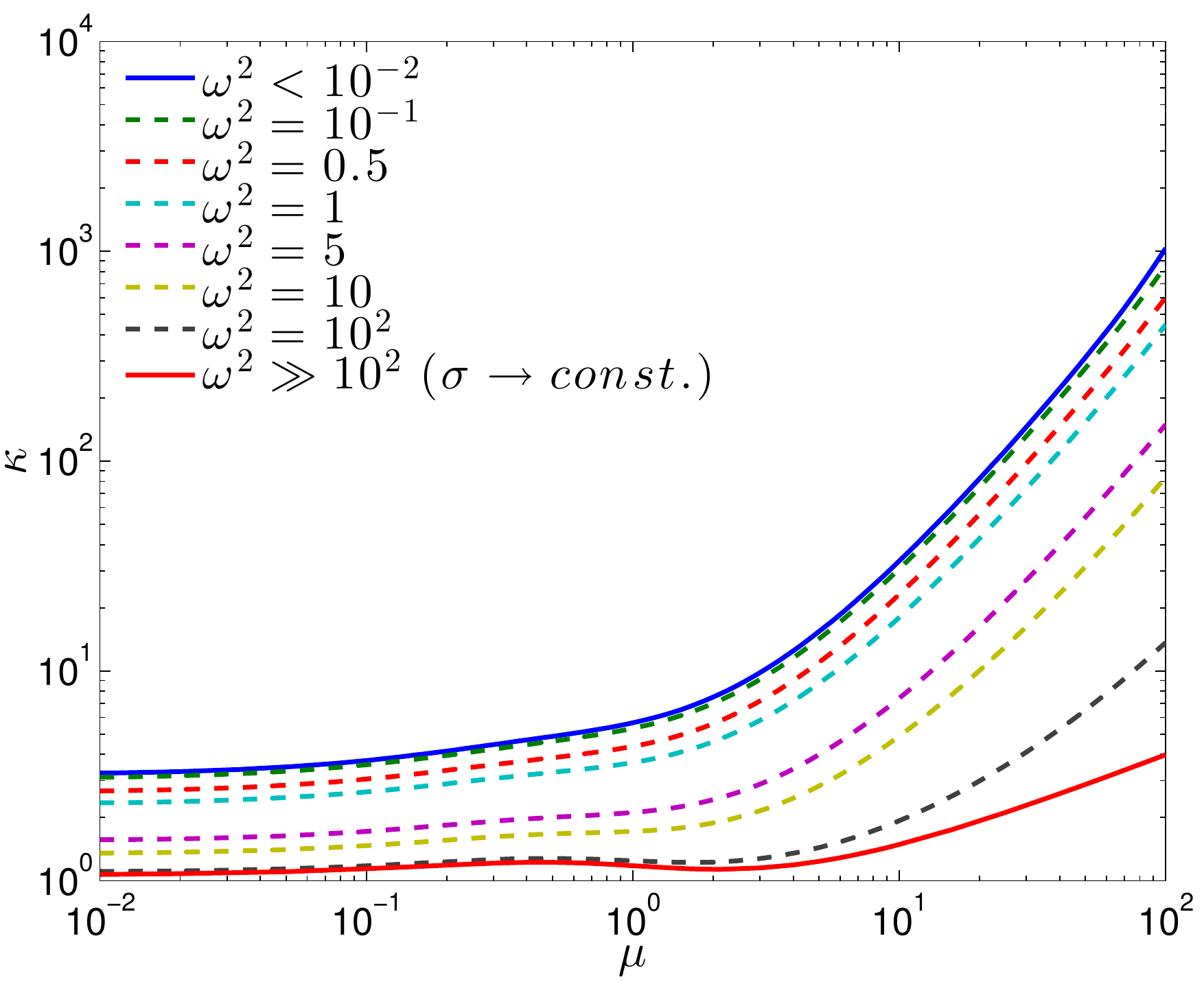} \\
\end{tabular}
\caption{\textit{The effect of a cutoff $\omega$ to the enhancement on $\alpha$ and $\kappa$, as defined in (\ref{v2cutoff}). Increasing the cutoff from $\omega = 0$ to $\omega \gg 1$ smoothly transforms $\alpha$ and $\kappa$ from the $\sigma \propto \vrel^{-2}$ case to the $\sigma = const.$ case. This type of behaviour could occur, for example, in models of Sommerfeld-enhanced DM scattering, which corresponds to an attractive force with a massive force carrier. }}
\label{fig:vm2cutoff}
\end{figure*}

The relative velocity is the difference between the incoming WIMP speed $x$ and the nucleus speed $z$; expressed in terms of these velocities, the dimensionless differential cross-section is just
\begin{equation}
\hat \sigma_{v^{2n}} = \frac12|\vect x  - \vect z | ^{2n}.
\end{equation}
$\Cout$ is then modified by extra powers of the relative velocity. The angular and radial integrals can be performed analytically, and the results are given in Table \ref{tab:coutv}. $\Cin$ is also modified in a simple way. The angular dependence drops out in the centre of mass frame where we compute $\Cin$ and $\vect x - \vect z = (1 + \mu) \vect b$, so that 
\begin{equation}
\hat \sigma_{v^{2n}} = \frac12(1+\mu)^{2n} b^{2n}.
\label{cinv2}
\end{equation}
These are related to (\ref{vdep}) via:
\begin{equation} 
\sigma = 2 \sigma_0 \zeta^{-2n} \shat_{v^{2n}},
\end{equation}
where $\zeta$ is defined below (\ref{epsLTE}). There is one final complication in the computation of $\Cin$ (\ref{cinint}) for the $\vrel^{-2}$ case: the expression (\ref{cinv2}) diverges for $b \rightarrow 0$, complicating the numerical integration of $\Cin$, even though the integral itself is finite. We address this by imposing a cutoff velocity $\om$ such that:
\begin{equation}
\shat_{v^{-2}} \rightarrow \frac{1 + \om^2}{2\left(|\vect x - \vect z |^2 + \om^2\right) } = \frac{1 + \om^2}{2(1+\mu)^2 b^2 + 2\om^2},
\label{v2cutoff}
\end{equation}
where the factor in the numerator ensures that $\hat \sigma_{\rm tot} (v_T) = 1$. Both $\alpha$ and $\kappa$ converge to finite values as $\om^2 \rightarrow 0$. This convergence is illustrated in Figure \ref{fig:vm2cutoff}. As $\omega^2$ is varied from $10^{-3}$ to $10^{-2}$, $\alpha(\mu)$ varies by less than one part in $10^4$, whereas $\kappa(\mu)$ changes by less than $1\%$ at low $\mu$, up to $\sim 4\%$ for $\mu = 100$.  The thermally-averaged cross-section, necessary for computing a particle's typical inter-scattering distance $l_\chi$, can potentially depend on non-zero $\om$. A large value is reasonable, for example, in the case of resonant Sommerfeld-enhanced scattering, where the cross-section can reach a saturation value for $\vrel/c \approx 10^{-5}-10^{-3}$, below which $\sigma$ becomes constant.  
 
\begin{figure*}[p]
\begin{tabular}{c@{\hspace{0.04\textwidth}}c}
\multicolumn{2}{c}{\includegraphics[height = 0.32\textwidth]{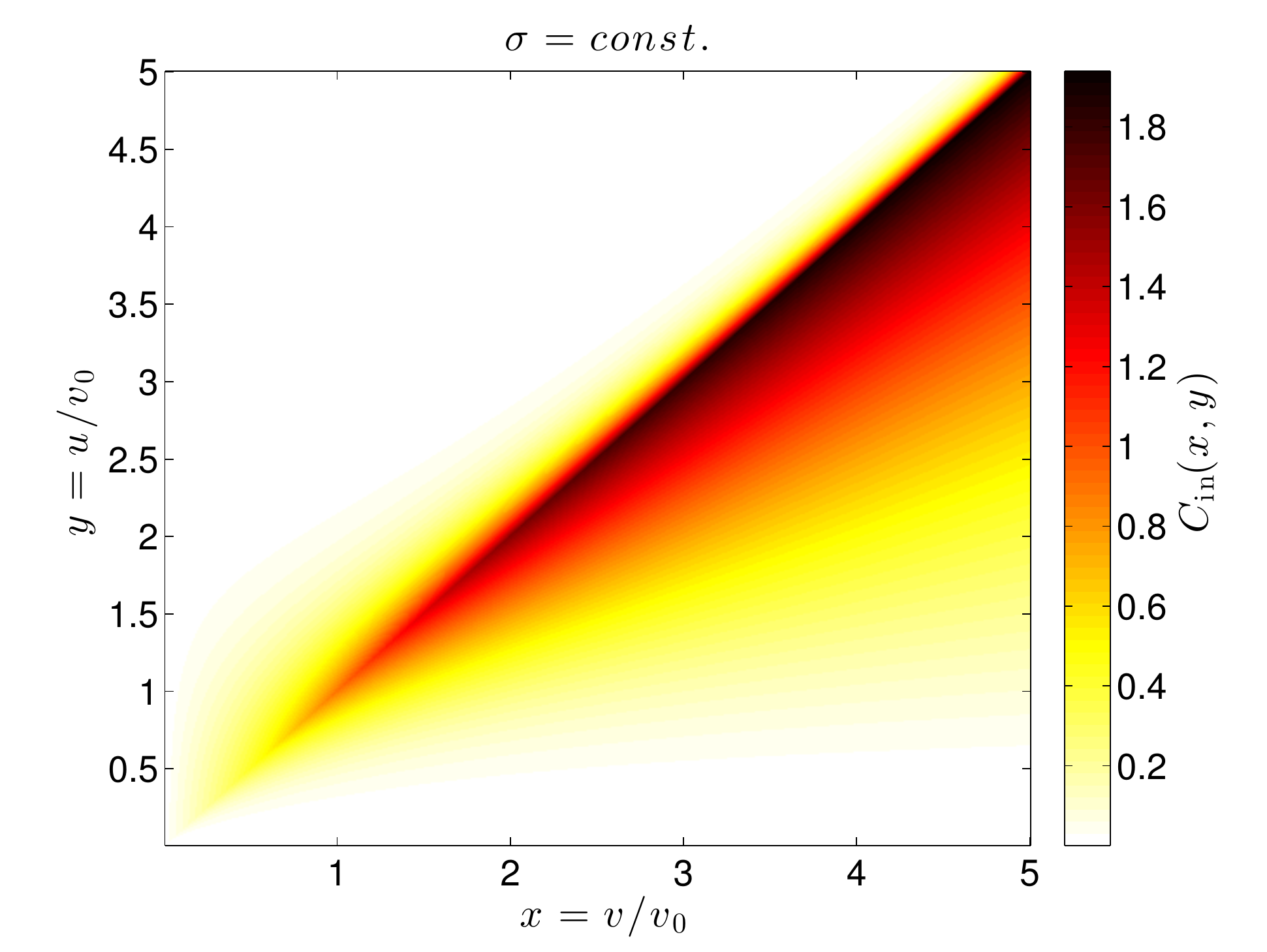}} \\
\includegraphics[height=0.32\textwidth]{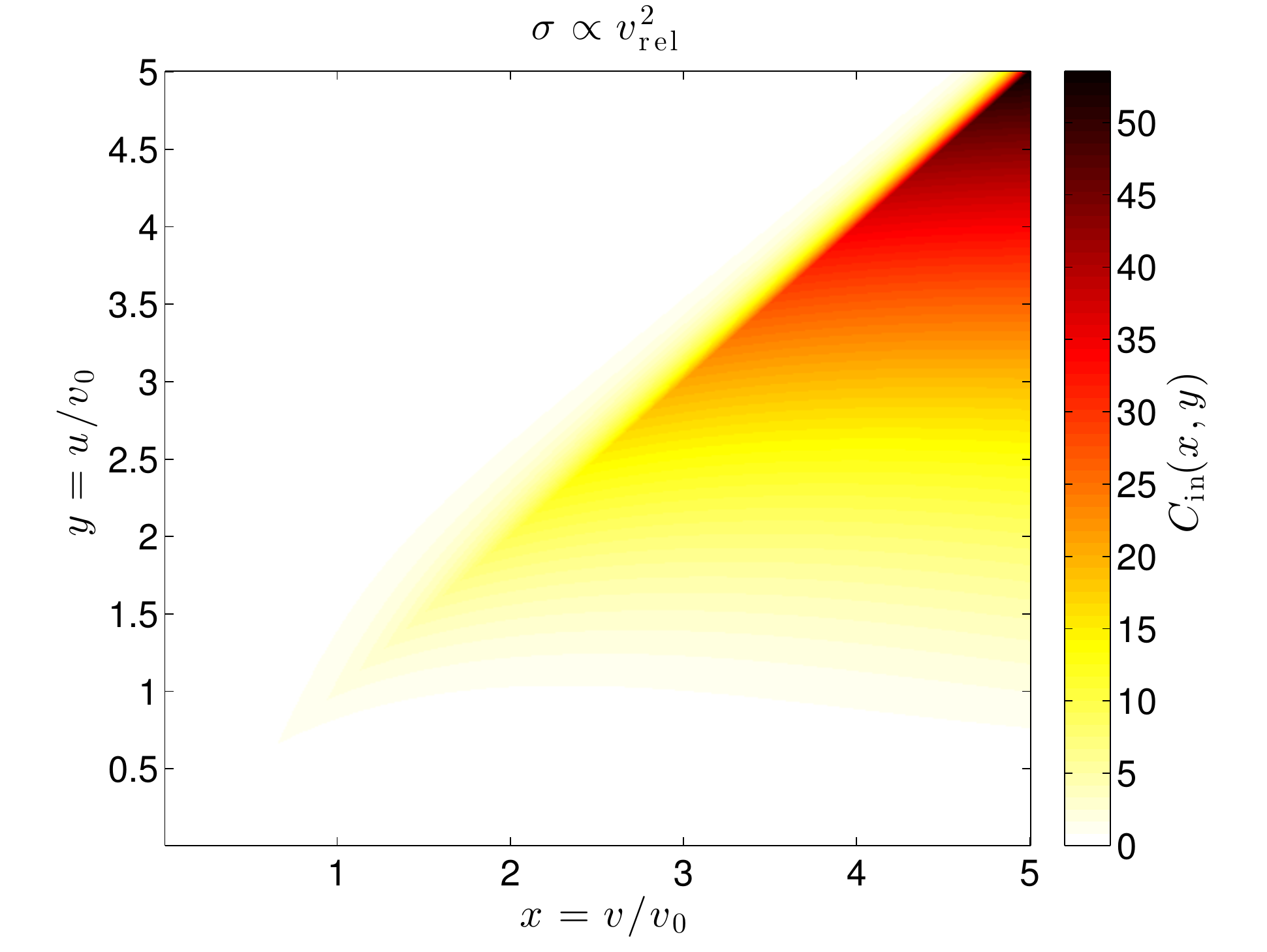} & \includegraphics[height=0.32\textwidth]{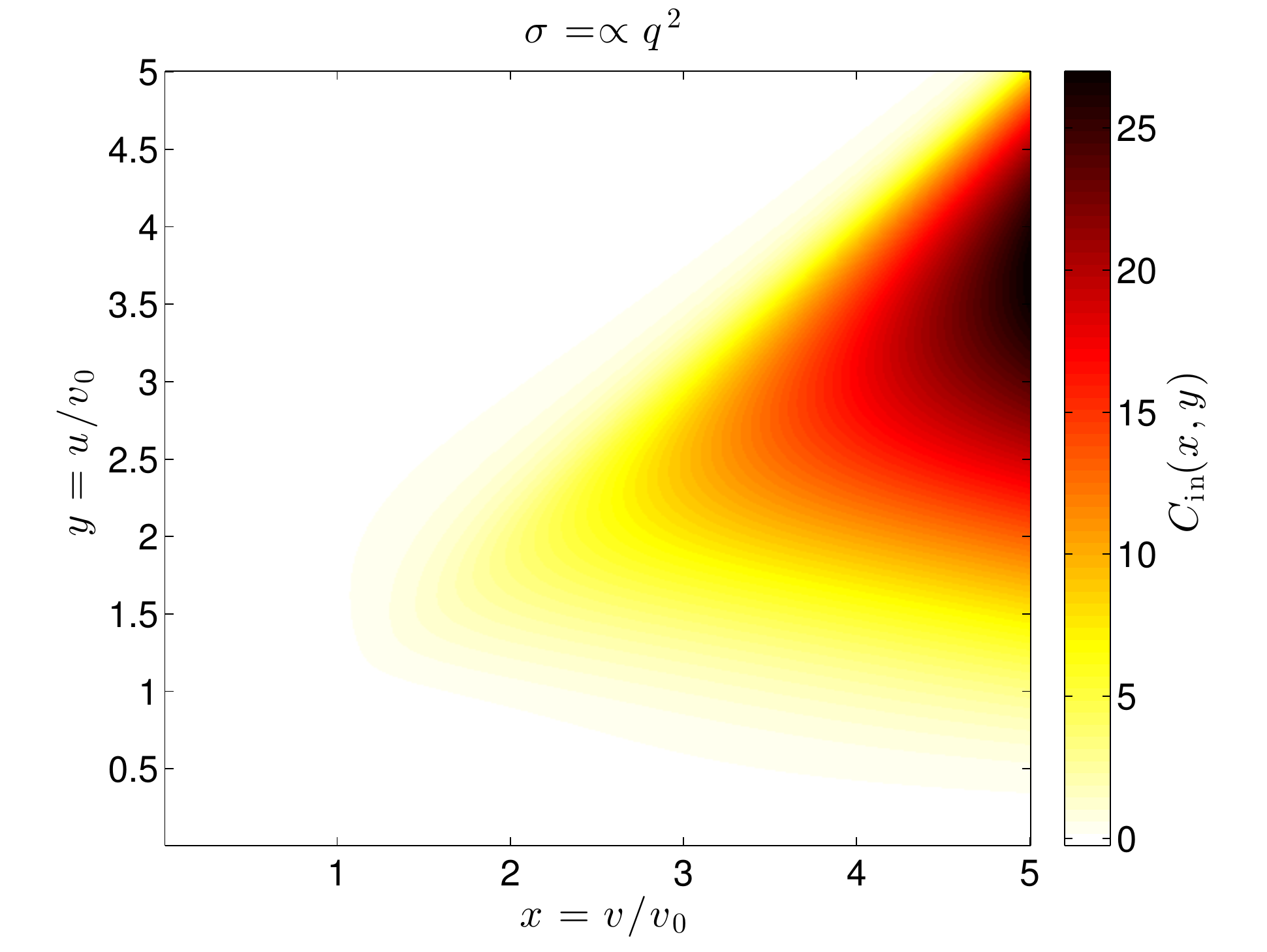} \\
 \includegraphics[height=0.32\textwidth]{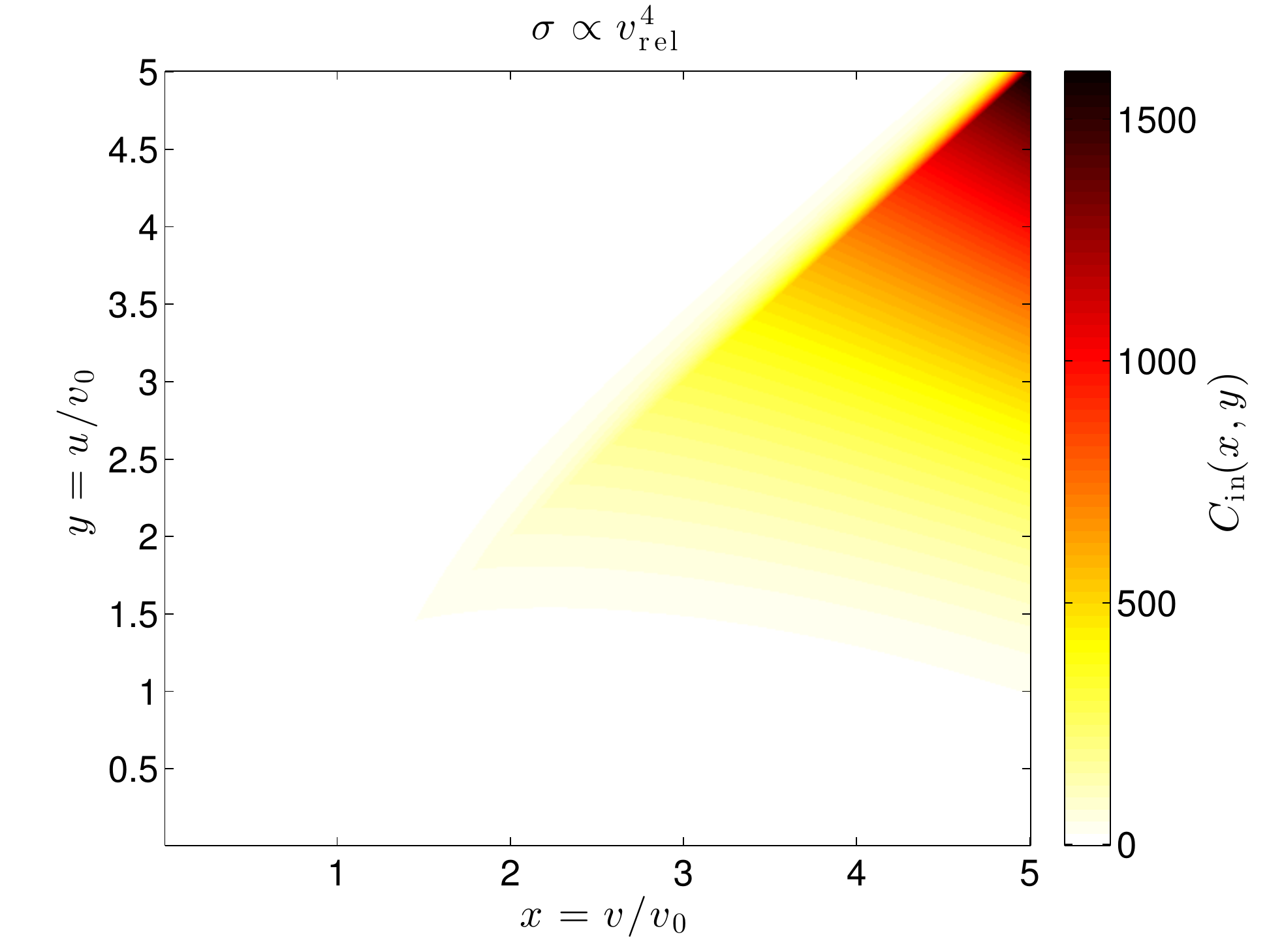} & \includegraphics[height=0.32\textwidth]{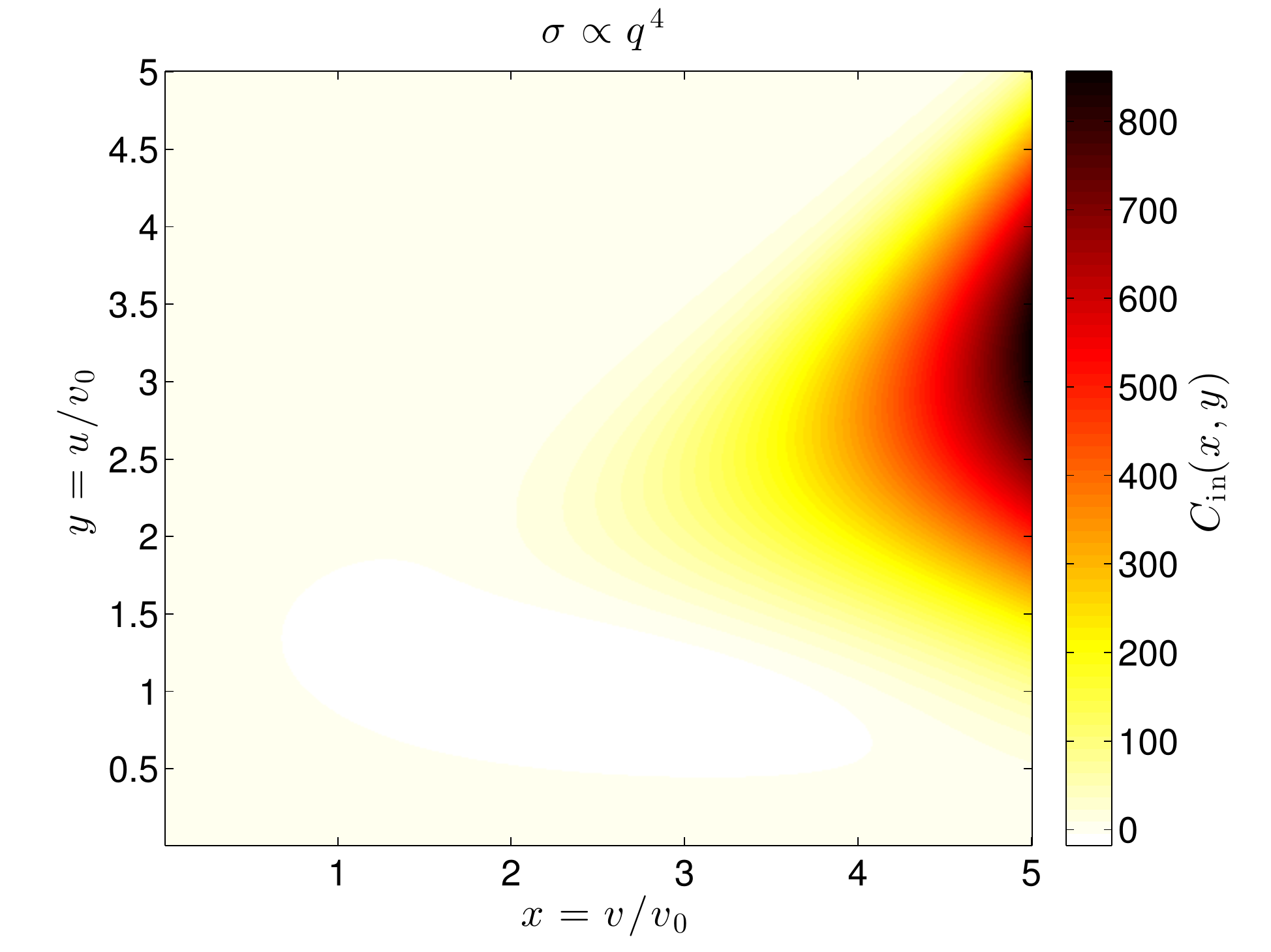} \\
  \includegraphics[height = 0.32\textwidth]{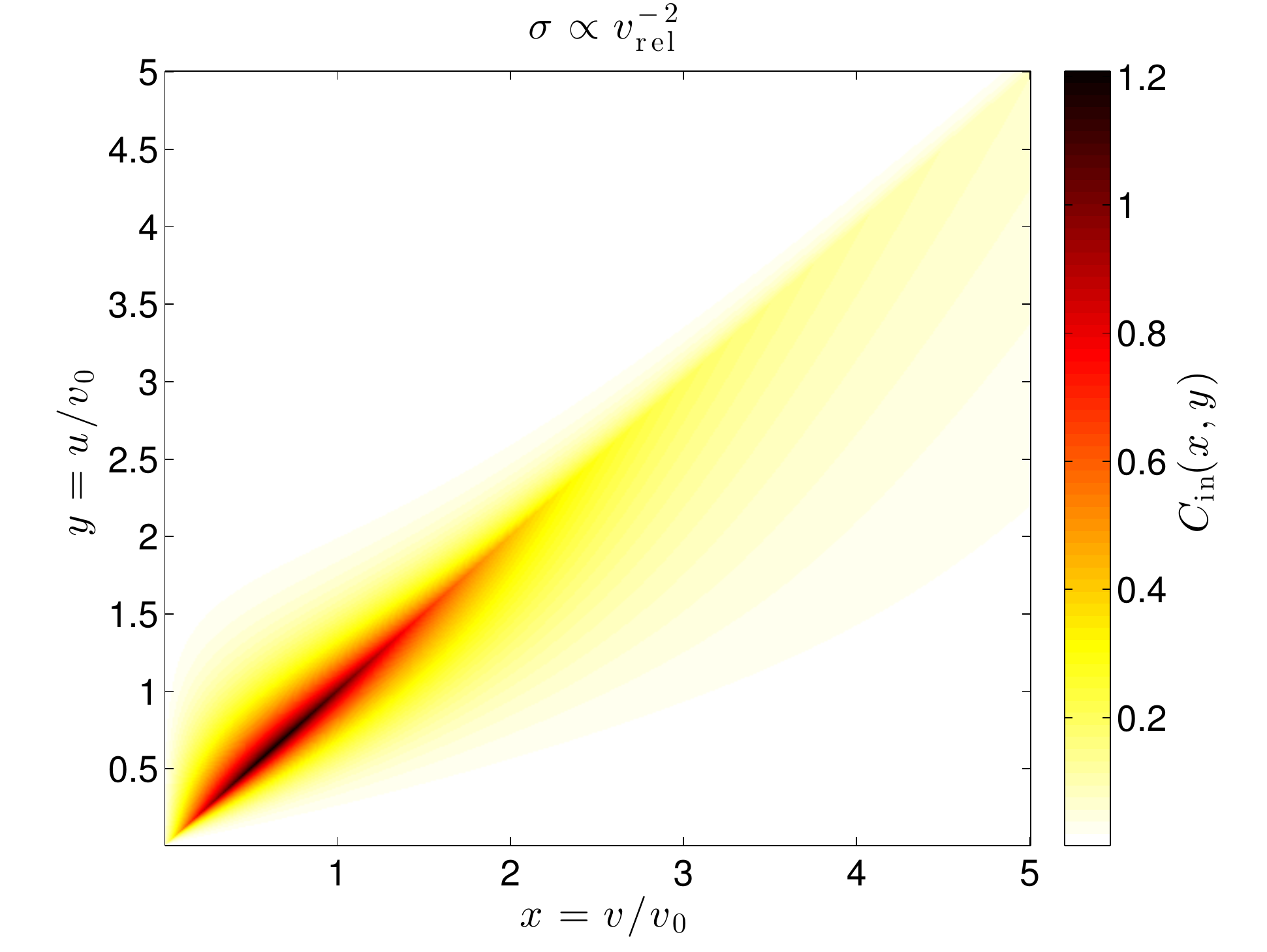} & \includegraphics[height = 0.32\textwidth]{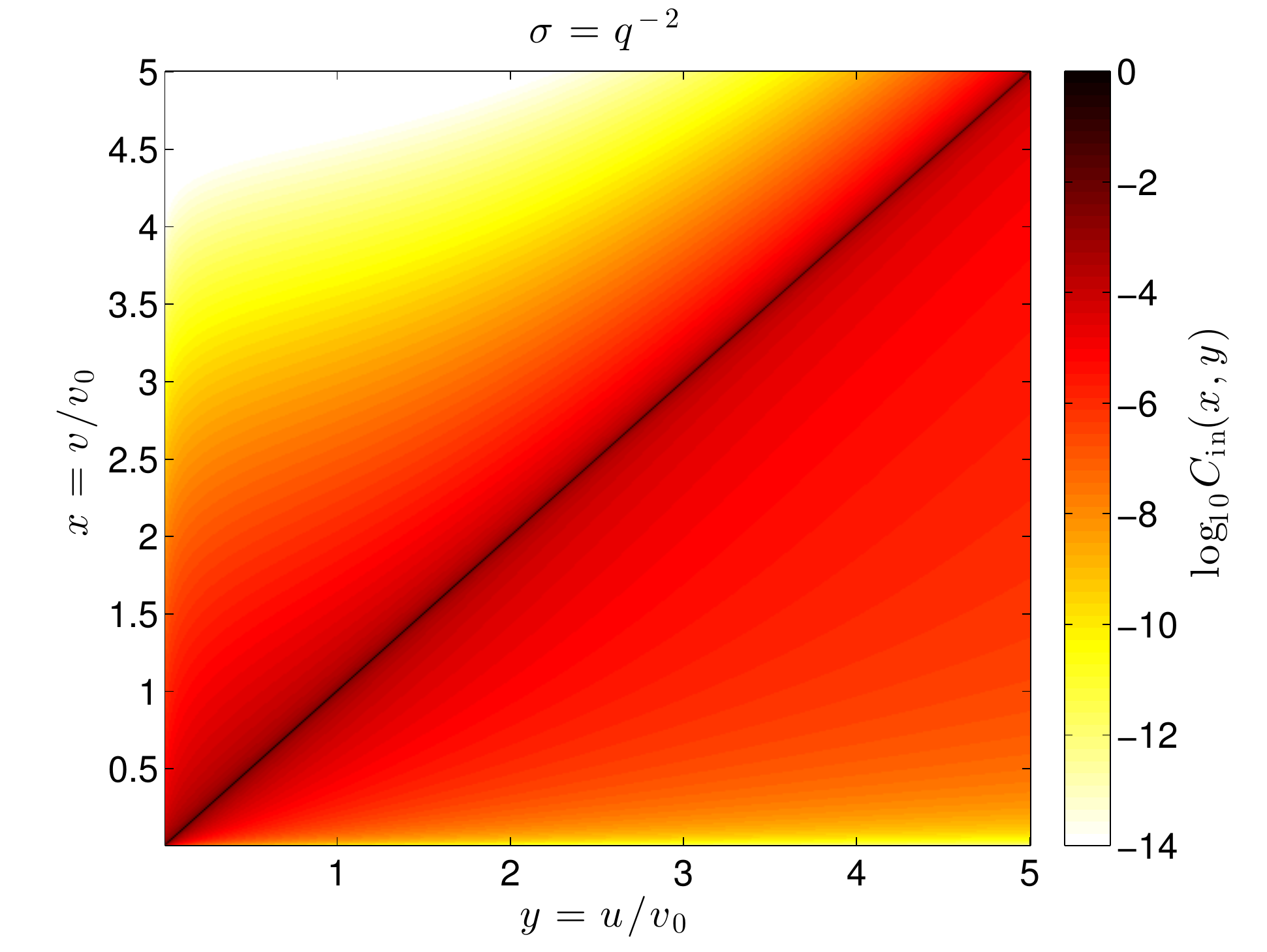} \\
\end{tabular}
\caption{\it Partial dimensionless collision operator $\Cin(y,x)$ with $\mu = \mx/m_{\rm nuc} = 1$ for the various types of cross-section, as a function of the WIMP's dimensionless incoming velocity $x$ and outgoing velocity $y$. The angular integral in the $q^{-2}$ case (bottom-right panel) is regulated with $\xi = 10^{-8}$ (\ref{P1qm2}) to remove the divergence at $x = y$. Note also the log scale in this panel. }
\label{Cinfig}
\end{figure*}

Values of $\Cin(y,x)$ for $\mu = 1$ are illustrated on the left-hand side of Figure \ref{Cinfig}. For positive powers of $\vrel$, this results in enhanced scattering of particles with high incoming velocity $v$; conversely this enhancement is present a low incoming velocities for $\sigma \propto \vrel^{-2}$.

\begin{figure*}[tbp]
\begin{tabular}{c@{\hspace{0.04\textwidth}}c}
\includegraphics[width=0.48\textwidth]{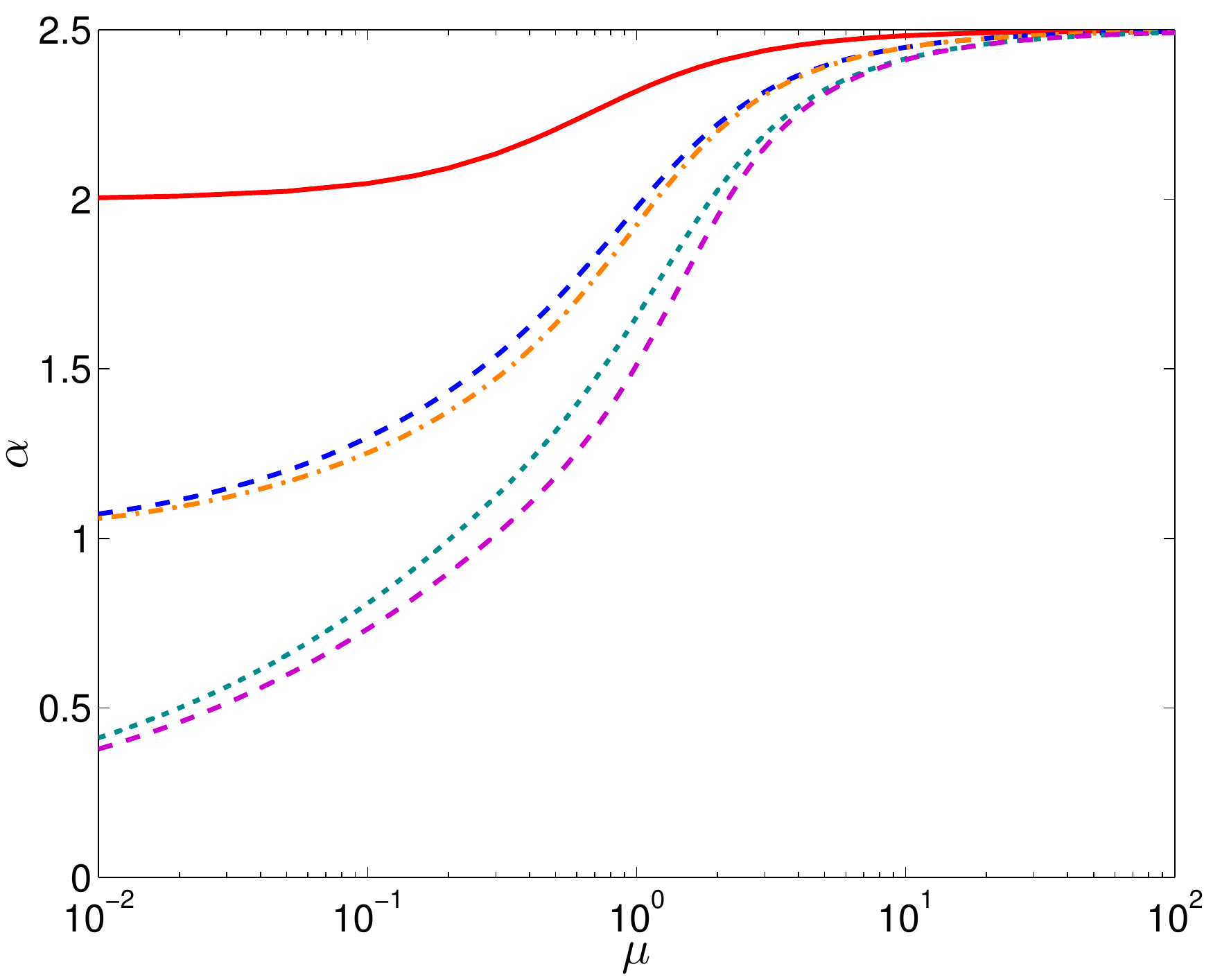} & \includegraphics[width=0.48\textwidth]{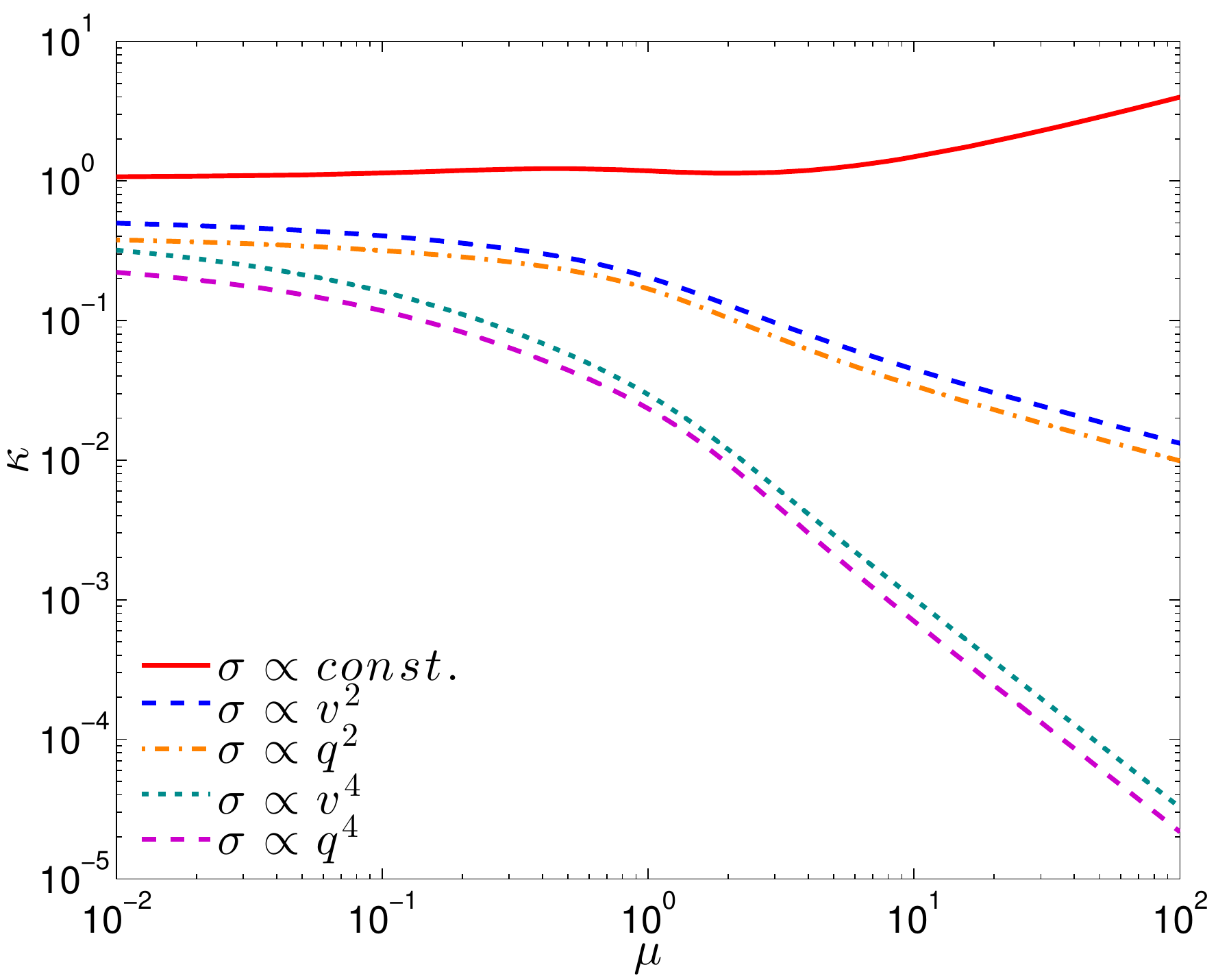} \\
\includegraphics[width=0.48\textwidth]{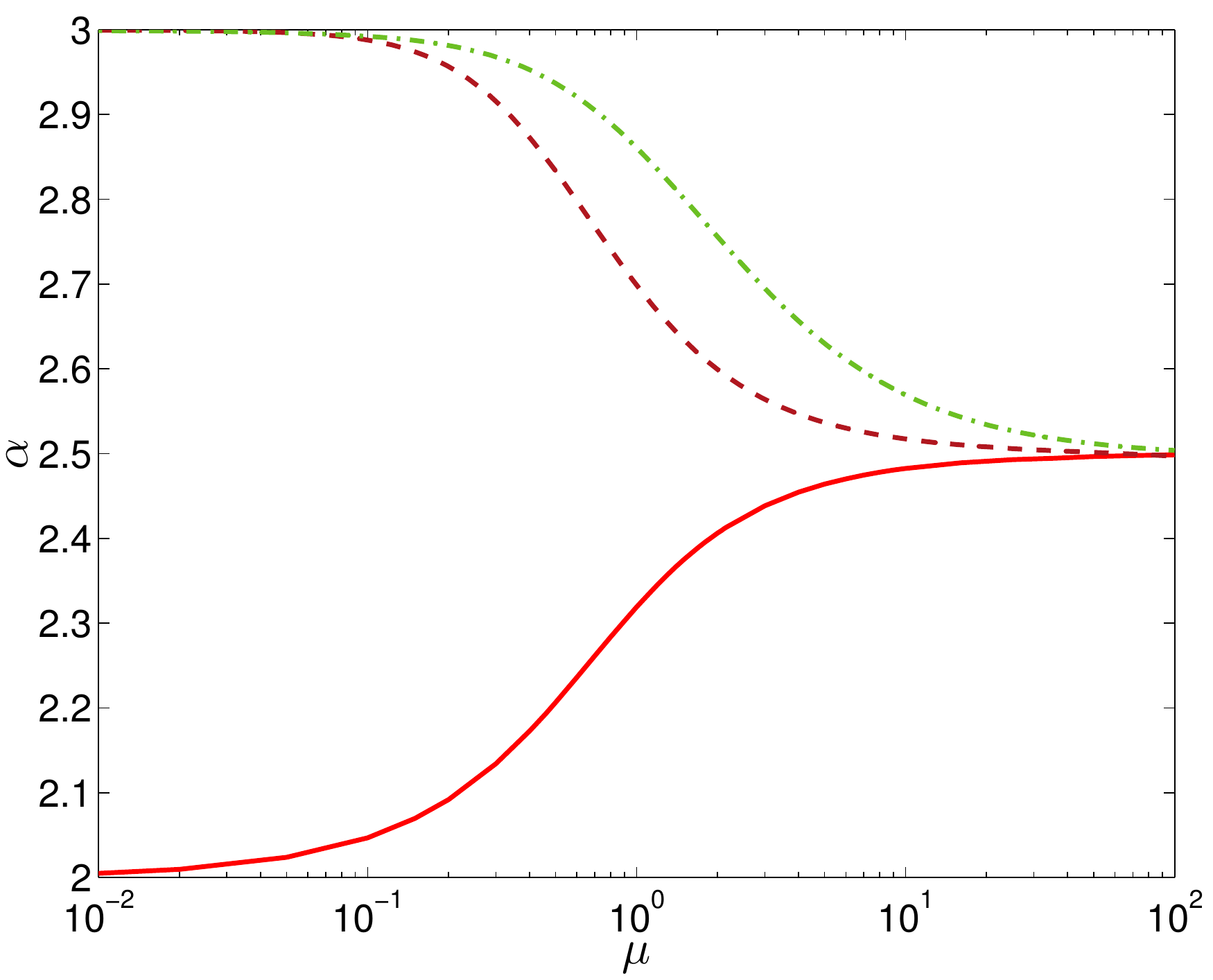} & \includegraphics[width=0.48\textwidth]{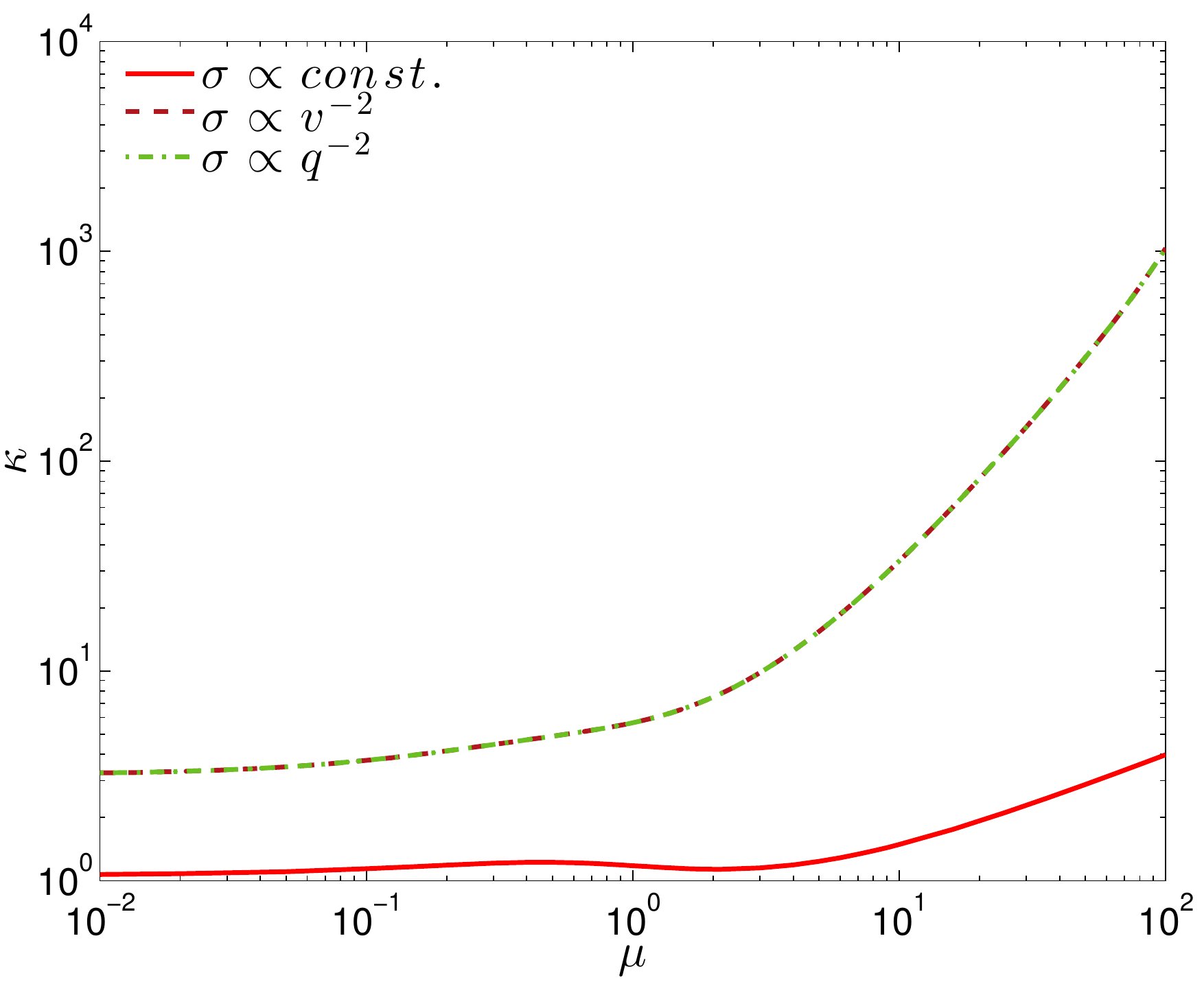} \\
\end{tabular}
\caption{\textit{ Dimensionless thermal diffusivity $\alpha$ (left) and conductivity $\kappa$ (right) as a function of $\mu = \mx/\mn$.  Curves are shown for velocity-dependent scattering cross-sections $\sigma \propto \vrel^{2n}$ and for momentum-dependent cross-sections $\sigma \propto q^{2n}$. The upper panels show curves for positive values of $n$,  and bottom panels show results for $n = -1$. In every case the thick red line is the fiducial case $\sigma = const.$, which was calculated explicitly by Gould and Raffelt \cite{GouldRaffelt90a}, and is identical to the result presented in their Figure 5 (panel b); this is in contrast to the results obtained by \cite{Nauenberg87}, who did not include the $\Cin$ contribution . Note that, because we regulate the cross-section for $q^{-2}$-dependent scattering using the momentum transfer cross-section (\ref{sTdef}), the $\kappa$ curves for $\vrel^{-2}$ and $q^{-2}$ are identical.}}
\label{v2p2fig}
\end{figure*}

In Figure \ref{v2p2fig}, the thermal conduction coefficients $\alpha$ and $\kappa$ are shown for $\vrel^{2}$ scattering (blue dashed lines), and for $\vrel^4$ (cyan dotted lines). For the $\vrel^{-2}$ case, $\kappa$ and $\alpha$ converge to well-defined curve as $\om^2 \rightarrow 0$; these are the values we show in the bottom panels of Figure \ref{v2p2fig} (dark red dashed lines), and later use to compute $L(r)$. Numerical values are provided in Table \ref{aktable}. 

The effect of positive powers of $\vrel$ is to suppress $\alpha$ at low values of $\mu$, and to suppress $\kappa$ for large values of $\mu$. $\alpha$ is a measure of how well WIMPs can diffuse outward in the potential well of a star. This suppression ultimately means that the distribution $n_\chi(r)$ will be more compact than in the standard case. The effect of a $\kappa$ suppression, meanwhile, implies that the WIMP population will be less efficient at heat transport via scattering with light nuclei such as hydrogen. In the lower panel of Figure \ref{v2p2fig} we show that the opposite behaviour is true for $\sigma \propto \vrel^{-2}$: a ``fluffier'' WIMP distribution around the stellar core, with enhanced heat transport via collisions. 

In the case of a large cutoff velocity $\omega$, the thermal conduction coefficients lie in between the $\sigma = const.$ and $\sigma \propto \vrel^{-2}$ cases. We illustrate this in Figure \ref{fig:vm2cutoff}, for several values of the dimensionless cutoff velocity $\omega$. Sommerfeld-enhanced DM models, therefore, would benefit from some of the enhanced conduction of the $\vrel^{-2}$ case.

\subsection{Momentum-dependent scattering}
The momentum transferred during a collision, $q = \sqrt{2\mn E_{\rm R}}$ is equal to the total momentum gained or lost by the WIMP:
\begin{eqnarray}
q^2 &=& \mx^2 |\vect u - \vect v|^2 \nonumber \\
 	&=& q_0^2 \zeta^{-2} |\vect x - \vect y|^2 \nonumber \\
	&=& 2 b^2 \zeta^{-2} q_0^2 (1-\cos \theta_{\rm CM}).
\label{pdepeq}
\end{eqnarray}
The latter expression allows us to write the cross-section in terms of the centre-of-mass velocity of the incoming WIMP, $b$, and the CM scattering angle $\theta_{\rm CM}$, as the scattering event does not change the WIMP's speed in the CM frame, only its direction. 

Then 
\begin{equation}
\shat_{q^{2n}} = 2^{-n}H_n|\vect x - \vect y|^{2n} = H_nb^{2n} (1-\cos \theta_{\rm CM})^{n},
\label{sigmaofq}
\end{equation}
where the normalisation factor required to ensure that $\hat\sigma_{\rm tot}(v_T) = 1$ is
\begin{equation}
\label{H_n}
H_n = \frac{(1+\mu)^{2n}}{\int_{-1}^1  (1-\cos \theta')^{n}\, \ud \cos \theta' }.
\end{equation}
The structure of (\ref{sigmaofq}) shows us that 
\begin{equation}
\shat_{q^{2n}} = 2\shat_{v^{2n}}\frac{(1-\cos \theta_{\rm CM})^{n}}{\int_{-1}^1 (1-\cos \theta')^{n}\, \ud \cos \theta' },
\end{equation}
so 
\begin{equation}
\shat_{\mathrm{tot},q^{2n}} = \shat_{\mathrm{tot},v^{2n}} = (1+\mu)^{2n} b^{2n},
\end{equation}
as we would expect, given that $\shat_\mathrm{tot}$ is independent of $\theta_{\rm CM}$.  This also means that $C_{\mathrm{out},q^{2n}} =  C_{\mathrm{out},v^{2n}}$. 
In terms of the dimensionful cross-section (\ref{pdep}),
\begin{equation}
\sigma = \frac{2^n}{H_n} \sigma_0 \zeta^{-2n} \shat_{q^{2n}}.
\end{equation}

The component $\Cin$ requires an explicit evaluation of $\langle P_1 \shat (\cos \theta_{\rm CM}) \rangle$:
\begin{eqnarray}
\langle P_1 \shat_0 \rangle &=& \frac12 G, \label{P1q0} \\
\langle P_1 \shat_{q^{2}} \rangle &=& \frac12b^2(1+\mu)^2 \left[G(1-A) - \frac{b^2B^2}{2xy}\right], \label{P1q2} \\
\langle P_1 \shat_{q^{4}} \rangle &=&    \frac38b^4(1+\mu)^4 \Bigg[ \left\{G(1-A) - \frac{b^2B^2}{xy}\right\}(1-A) \nonumber\\
                                  &+& \frac{GB^2}{2} \Bigg], \label{P1q4} \\
\langle P_1 \shat_{q^{-2}} \rangle &=& \frac{1}{\ln 2 - \ln \xi}b^{-2}(1+\mu)^{-2} \nonumber \\
                                   &\times&\left[\frac{(1-A+\xi)b^2 +  Gxy}{xy\sqrt{(1-A+\xi)^2-B^2}} - \frac{b^2}{xy}\right], \label{P1qm2}
\end{eqnarray}
where $A$, $B$ and $G$ are defined in (\ref{Adef}--\ref{Gdef}). The parameter $\xi$ comes from the replacement $\cos \theta_{\rm CM} \rightarrow \cos \theta_{\rm CM} - \xi$. We make this substitution because for $n = -1$, the factor $(1 - \cos \theta)^{-1}$ in (\ref{sigmaofq}) causes the cross-section to diverge for forward scattering at small angles; likewise for the integral (\ref{stotdef}). For the computation of $\alpha$ in (\ref{eqn:alpha}), the powers of $C^{-1}$ allow the divergence to cancel.  However, $\kappa$, which governs momentum transfer, formally diverges. Fortunately, the divergence at $\theta_{\rm CM}=0$ corresponds to forward scattering, in which no momentum is actually transferred. We regulate this divergence with the ``momentum transfer cross-section,'' which comes from plasma physics \cite{Krstic} but has also been applied to parameterise transport by WIMPs \cite{Tulin:2013teo}:
%\footnote{a second quantity, the viscosity cross-section can also be defined $\sigma_V = \int d \Omega \sin^2 \theta d\sigma/d\Omega$ which regulates both forward and backward-scattering. This is useful for conduction by self-scattering of a single species, but would defeat our purpose by erasing back-scattering, which is most efficient at momentum transfer. }:
\begin{align}
\label{sTdef}
\shat_{\rm tot,T}(v_{\rm rel}) &\equiv \int_{-1}^1 \ud\cos\theta_{\rm CM}\, \shat_{\rm T}(v_{\rm rel},q), \\
\shat_{\rm T}(v_{\rm rel},q) &\equiv (1-\cos\theta_{\rm CM})\shat(v_{\rm rel},q),
\end{align}
where $\shat$ is as per (\ref{sigmaofq}), but with the replacement $n\to n+1$ in the denominator of (\ref{H_n}).  This gives
\begin{equation}
\label{P1qm2reg}
 \langle P_1 \shat_{\mathrm{T},q^{-2}} \rangle  =  \frac12 \frac{G}{b^2 (1+ \mu)^2},
\end{equation}
leading to a finite, well-defined value of $\kappa$. 

The behaviour of $\Cin$ as a function of $x$ and $y$ for momentum-dependent cross-sections is shown in the right-hand panels of figure $\ref{Cinfig}$. The two upper right-hand panels show an enhancement in the scattering rate when the momentum transfer is large, over both the constant and velocity-dependent cases, which favour collisions closer to the $x = y$ line. The lower-right panel shows the $q^{-2}$ case computed with (\ref{P1qm2}) and $\xi = 10^{-8}$, illustrating the divergent behaviour in the forward-scattering rate, and a general suppression of collisions which lead to appreciable momentum exchange.

The values of $\alpha$ and $\kappa$ are shown in dot-dashed orange for the $q^{2}$ case and in dot-dashed green for $q^{-2}$ in Figure \ref{v2p2fig}.  The  $q^4$ case is also shown (dashed purple lines). The behaviour of $\alpha$ and $\kappa$ for a $q^{2n}$ cross-section is qualitatively similar to the $\vrel^{2n}$ case.  Indeed, $\kappa$ is identical for the $q^{-2}$ and $\vrel^{-2}$ cases, thanks to the momentum-transfer cross-section, which essentially deletes the angular dependence in the $q^{-2}$ case and makes $C$ identical for momentum and velocity-dependent cross-sections when $n=-1$.  In the next section we will show that, once this behaviour is combined with the different typical inter-scattering distances in models with momentum-dependent and velocity-dependent scattering, the conduction of energy can be greatly modified with respect to the fiducial constant cross-section case.

\section{Effect on energy transport in the Sun}
\label{sec:luminosity}
We now illustrate the effect of a non-standard WIMP-nucleon cross-section on the conduction of energy within the Sun. Our discussion applies equally well to any star, but modelling and measurements of the Sun and its properties are far more precise than for other stars. We use the temperature, density and elemental composition profiles computed with the AGSS09ph solar model\footnote{publicly available at \url{http://www.mpa-garching.mpg.de/~aldos/}} \cite{AGSS,Serenelli:2009yc}. 

Scattering cross-sections $\sigma_0 \lesssim 10^{-38}$\,cm$^2$ allowed by bounds from direct detection experiments can easily lead to energy transport outside the LTE regime, \ie non-local transport. To account for this, based on the results of Gould \& Raffelt \cite{GouldRaffelt90a,GouldRaffelt90b}, Refs.\ \cite{Bottino02,Scott09} introduce the quantities:
\begin{eqnarray}
\mathfrak{h}(r) &=& \left(\frac{r - r_\chi}{r_\chi}\right)^3 + 1, \label{hr}\\
\mathfrak f(K) &=& \frac{1}{1+ \left(\frac{K}{K_0}\right)^{1/\tau}}, \label{fK}
\end{eqnarray}
where $K$ is the Knudsen number, $K_0 \simeq 0.4$ and $\tau \simeq 0.5$. The WIMP scale height $r_\chi$ as a function of the central temperature $T_c$ and density $\rho_c$ is
\begin{equation}
r_\chi = \left(\frac{3 k_B T_c}{2\pi G \rho_c \mx}\right)^{1/2}.
\label{scaleheight}
\end{equation}
The WIMP distribution becomes a combination of the isothermal and LTE distributions:
\begin{equation}
n_\chi(r) = \mathfrak f(K) n_{\chi,{\rm LTE}} + \left[1 - \mathfrak f(K) \right] n_{\chi,{\rm iso}},
\end{equation}
where
\begin{equation}
\label{isodens}
n_{\chi,{\rm iso}}(r,t) = N(t)\frac{e^{-\frac{r^2}{r_\chi^2}}}{\pi^{3/2}r_\chi^3}.
\end{equation}
The total luminosity can be finally written:
\begin{equation}
L_{\chi,\rm total}(r,t) = \mathfrak f(K) \mathfrak h(r,t)L_{\chi,{\rm LTE}}(r,t).
\end{equation}
The expressions (\ref{hr}, \ref{fK}) are fits to the results of direct Monte Carlo solutions to the Boltzmann Equation for $n=0$ in (\ref{pdep}, \ref{vdep}).  In principle similar solutions should be obtained and fitted for $n\ne0$, but carrying out such Monte Carlo simulations is well beyond the scope of this paper.  We expect that the form of the suppression functions will be broadly similar to (\ref{hr}, \ref{fK}) for all $n$, however.  Conductive transport is generically expected to be most efficient for some $K$ close to but less than 1, because the larger the inter-scattering distance $l_\chi$ of a WIMP relative to the scale height $r_\chi$, the further it will have travelled (in an effective sense) in the star, and therefore the greater impact it will have on the local luminosity function, when it deposits its energy -- but beyond $K\sim1$, the conduction approximation breaks down, and transport must be less efficient.  Transport is also enhanced in the surface regions of a star due to the enhancement of (\ref{LTEdens}) at large $r$ compared to (\ref{isodens}), and suppressed in the stellar core generically, as the radial height is actually less than $l_\chi$, causing the impacts of conductive transport to be predominantly felt `downstream' (i.e.\ at radii larger than $l\kappa(\mu)/\kappa(0)$; see \cite{GouldRaffelt90a} for details).

For the calculation of WIMP conductive luminosities, we require the typical inter-scattering distance (\ref{lchi}), which depends on the local velocity-averaged scattering cross-section.  Taking a Maxwell-Boltzmann distribution for the velocities of the WIMPs and nuclei once more, the distribution of their relative velocities, in normalised units, is
\begin{equation}
F_{\chi-n}(\vect x - \vect z) = \left[\pi(1+\mu)\right]^{-3/2}e^{-\frac{|\vect x - \vect z|^2}{1+\mu}}.
\end{equation}
This then gives:
\begin{eqnarray}
\left\langle\sigma_{v^2}\right\rangle &= \sigma_0 \zeta^{-2} \left\langle \frac{\vrel^2}{v_T^2} \right\rangle &= \frac32\sigma_0\zeta^{-2}(1+\mu), \label{sigmav2}\\
\left\langle\sigma_{v^4}\right\rangle &= \sigma_0 \zeta^{-4} \left\langle \frac{\vrel^4}{v_T^4} \right\rangle &= \frac{15}{4}\sigma_0\zeta^{-4}(1+\mu)^2, \\
\left\langle\sigma_{v^{-2}}\right\rangle &= \sigma_0 \zeta^2 \left\langle \frac{v_T^2}{\vrel^2} \right\rangle &= 2\sigma_0\zeta^{2}(1+\mu)^{-1}, 
\end{eqnarray}
where $\zeta \equiv v_0/v_T$. The averaged $q$-dependent cross-sections follow from these, as $\langle b^{2n} \rangle = \langle (\vrel/v_0)^{2n} \rangle/(1+\mu)^{2n}$:
\begin{eqnarray}
\left\langle\sigma_{q^2}\right\rangle &=& 6 \sigma_0 \zeta^{-2} (1+\mu)^{-1}, \\
\left\langle\sigma_{q^4}\right\rangle &=& 40 \sigma_0 \zeta^{-4} (1+\mu)^{-2}, \\
\left\langle\sigma_{q^{-2}}\right\rangle &=& \sigma_0 \zeta^2 (1+\mu), \label{sigmaqm2}
\end{eqnarray}
where for $q$-dependent cross-sections $\zeta \equiv q_0/ (\mx v_T)$. We have again regulated the divergence in the $q^{-2}$ case using the momentum transfer cross-section.

Here we take the opportunity to point out that our results for $\alpha$ and $\kappa$ are fully independent of $\zeta$, and therefore $v_0$ and $q_0$.  To apply them in cases where $v_0$ or $q_0$ differ from the canonical values we have assumed in this paper, one need only use the desired $v_0$ or $q_0$ when implementing (\ref{LTEtransport}) and the appropriate one of (\ref{sigmav2})--(\ref{sigmaqm2}).

To properly compute the effect of WIMPs in the Sun, we will need to include the conduction coefficients in a full solar evolution code that also includes a detailed computation of the velocity/momentum-dependent capture, evaporation and annihilation rates (if any); that work will appear soon \cite{Vincent14}. For the purpose of this paper, we simply assume a steady state with a set number of WIMPs per baryon in the present Sun ($n_\chi/n_\mathrm{b}$), in order to directly compare the conductive efficiencies of cross-sections with different velocity and momentum scalings.  Likewise, we simply choose a few benchmark values of $\mx$, $\sigma_0$, $v_0$ and $q_0$ at which to make these comparisons, and assume for the sake of example that $\sigma_0$ scales with nuclear mass as it does for isospin-conserving spin-independent couplings (\ie proportional to the square of the atomic number).
 
\begin{figure*}[htbp]
\begin{tabular}{c@{\hspace{0.04\textwidth}}c}
\includegraphics[width=0.48\textwidth]{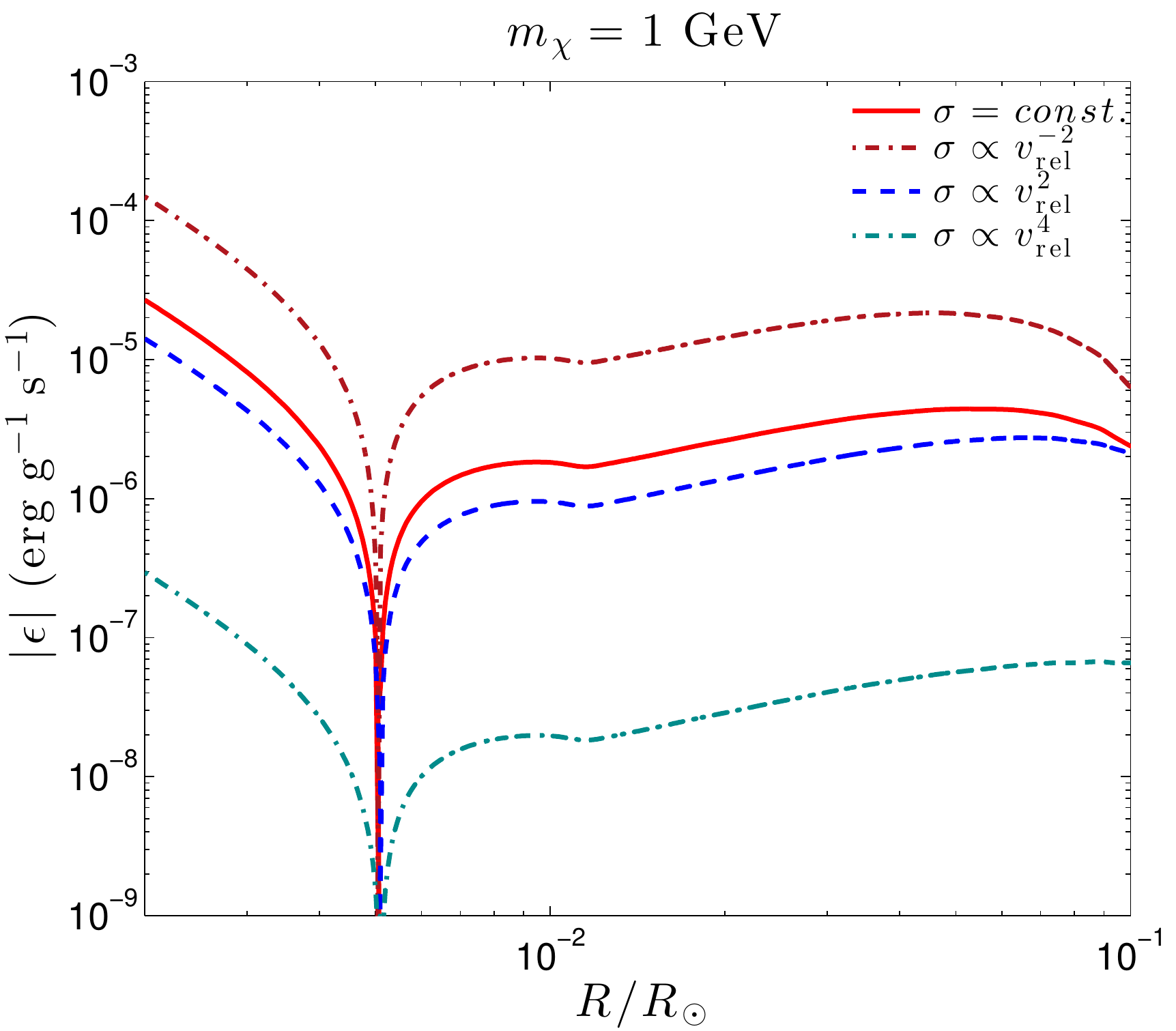} & \includegraphics[width=0.48\textwidth]{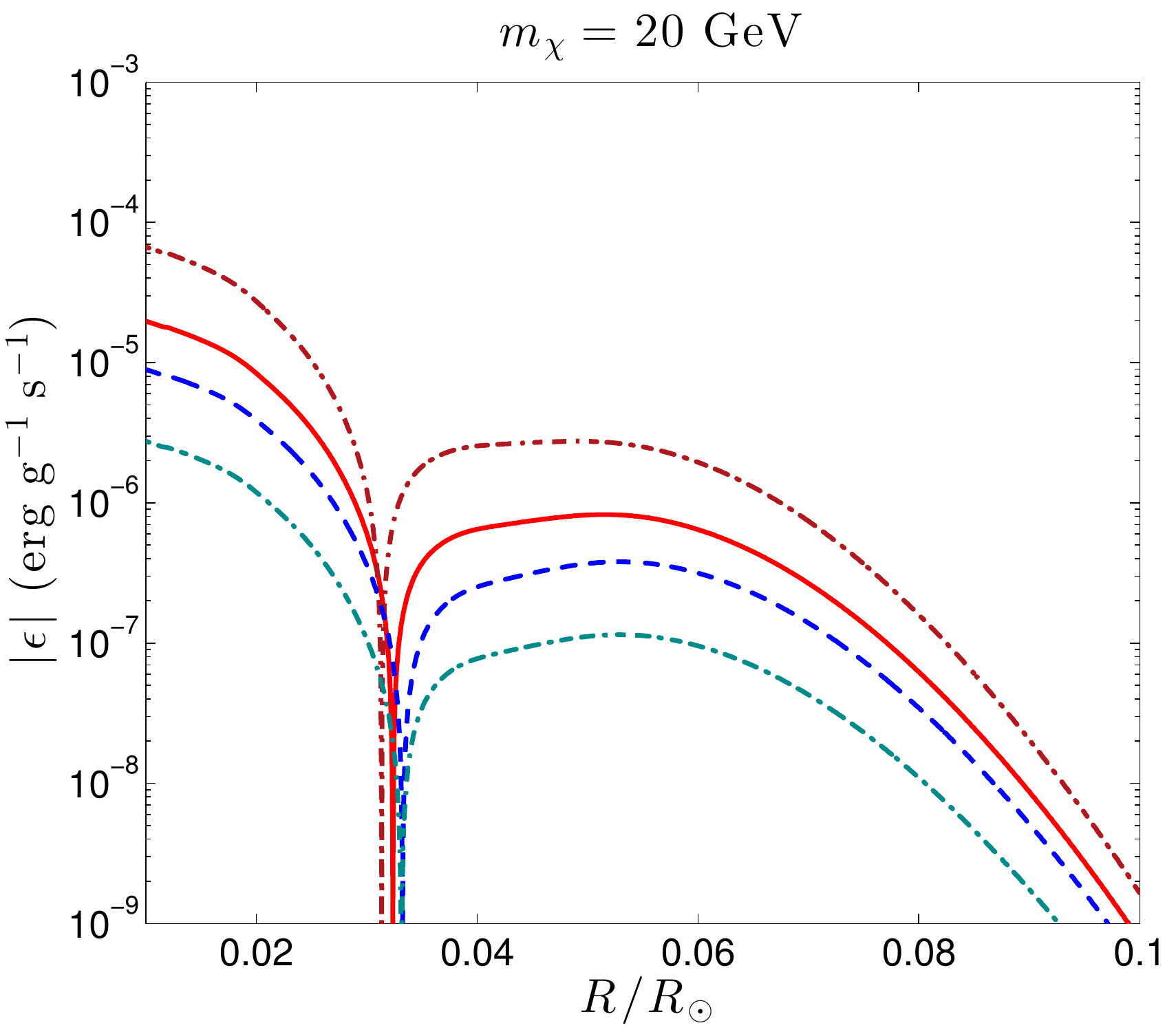} \\
\end{tabular}
\caption{\textit{ Absolute energy transported $|\epsilon|$ by WIMPs of mass $\mx = 1$ GeV (left) and $\mx = 20$ GeV (right) with a velocity dependent cross-section. Transported energy is negative at small $R$ (below the dip) and becomes positive at larger $R$, representing transport from the solar core into outer layers. We have taken the AGSS09ph solar model \cite{Serenelli:2009yc} with $n_\chi/n_\mathrm{b}=10^{-16}$ WIMPs per baryon, $v_0 = 110$ km/s, and $\sigma_0 = 10^{-39}$ cm$^{2}$.}} 
\label{fig:etransv}
\end{figure*}

In Figure \ref{fig:etransv} we illustrate the effect on energy transport by 1\,GeV\footnote
{
Although the tendency in solar dark matter literature at present is to assume that evaporation is efficient below 4\,GeV and inefficient above it, this is a simplification of a result obtained by Gould \cite{Gould87a} in the context of now-defunct models designed to explain the solar neutrino problem.  The reality is far more nuanced than this naive approximation would suggest: the efficiency of evaporation depends sensitively upon mass-matching between different nuclei in the Sun and the WIMP, as well as the thermal velocity widths of the nuclei and their radial stratification, and the actual Knudsen number of the system, which depends directly on the cross-section itself.  This sensitivity will be further modified by any velocity or momentum-dependence to the cross-section.  These effects are ignored in even the best modern analyses, so the true impact of evaporation for modern dark matter models, both at low and high WIMP masses, still remains to be reliably calculated.
}
 and 20\,GeV WIMPs in the Sun, using $n_\chi/n_\mathrm{b} = 10^{-16}$ WIMPs per baryon, $v_0 = 110$ km/s, and $\sigma_0 = 10^{-39}$ cm$^{2}$.  These roughly correspond to results of \cite{Taoso10}, which provide some small effects on solar structure.  The combined effect of $\alpha$, $\kappa$ and $l_\chi$ gives a suppression of energy transport in the case of a $\vrel^2$ or $\vrel^4$ cross-section, but an enhancement for the $\vrel^{-2}$ case, due to the increased scattering at low velocity.

\begin{figure*}[tbp]
\begin{tabular}{c@{\hspace{0.04\textwidth}}c}
\includegraphics[width=0.48\textwidth]{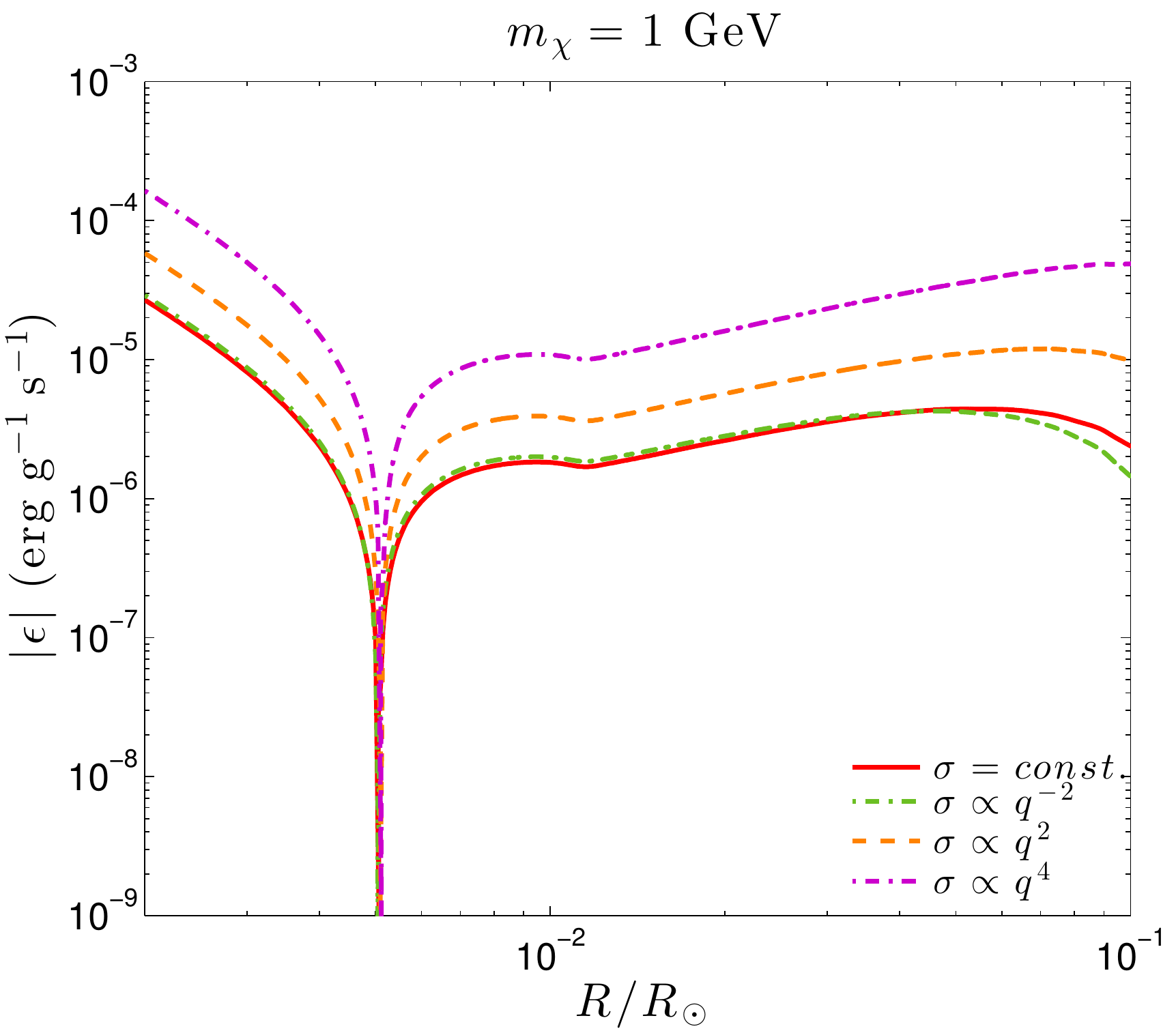} & \includegraphics[width=0.48\textwidth]{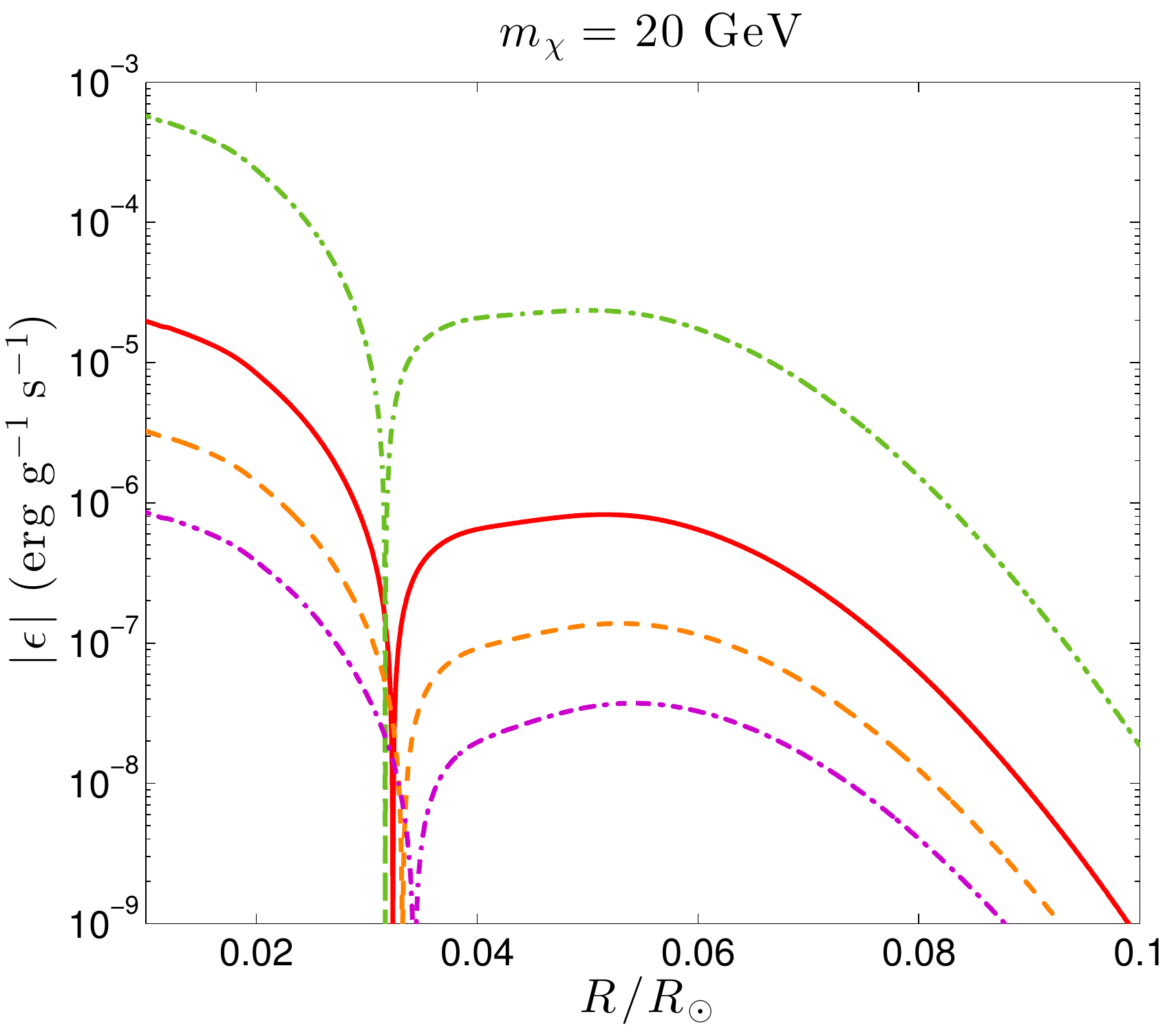} \\
\end{tabular}
\caption{\textit{ Absolute energy transported $|\epsilon|$ by WIMPs of mass $\mx = 1$ GeV (left) and $\mx = 20$ GeV (right) with a momentum-dependent cross-section and $q_0= 40$ MeV. We have taken the AGSS09ph solar model \cite{Serenelli:2009yc} with $n_\chi/n_\mathrm{b}=10^{-16}$ WIMPs per baryon, and $\sigma_0 = 10^{-39}$ cm$^{2}$.}} 
\label{fig:etransq}
\end{figure*}

Figure \ref{fig:etransq} shows the absolute energy transported by WIMPs with a momentum-dependent cross-section for $\mx = 1$\,GeV and $\mx = 20$\,GeV, again with $n_\chi/n_\mathrm{b} = 10^{-16}$ WIMPs per baryon and $\sigma_0 = 10^{-39}$ cm$^{2}$. Here we use $q_0 = 40$\,MeV, which corresponds to a nuclear recoil energy of $E_R = 11$\,keV in a germanium direct detection experiment, and 29\,keV in a silicon detector. It is apparent that at higher masses, a $q^{2n}$ cross-section with negative $n$ can result in heat transport that is enhanced by many orders of magnitude relative to standard WIMPs with a constant cross-section.  At lower masses, the opposite is true: negative $n$ suppresses heat transport, whereas positive $n$ can greatly enhance it.

\begin{figure*}[tbp]
\begin{tabular}{c@{\hspace{0.04\textwidth}}c}
\includegraphics[width=0.48\textwidth]{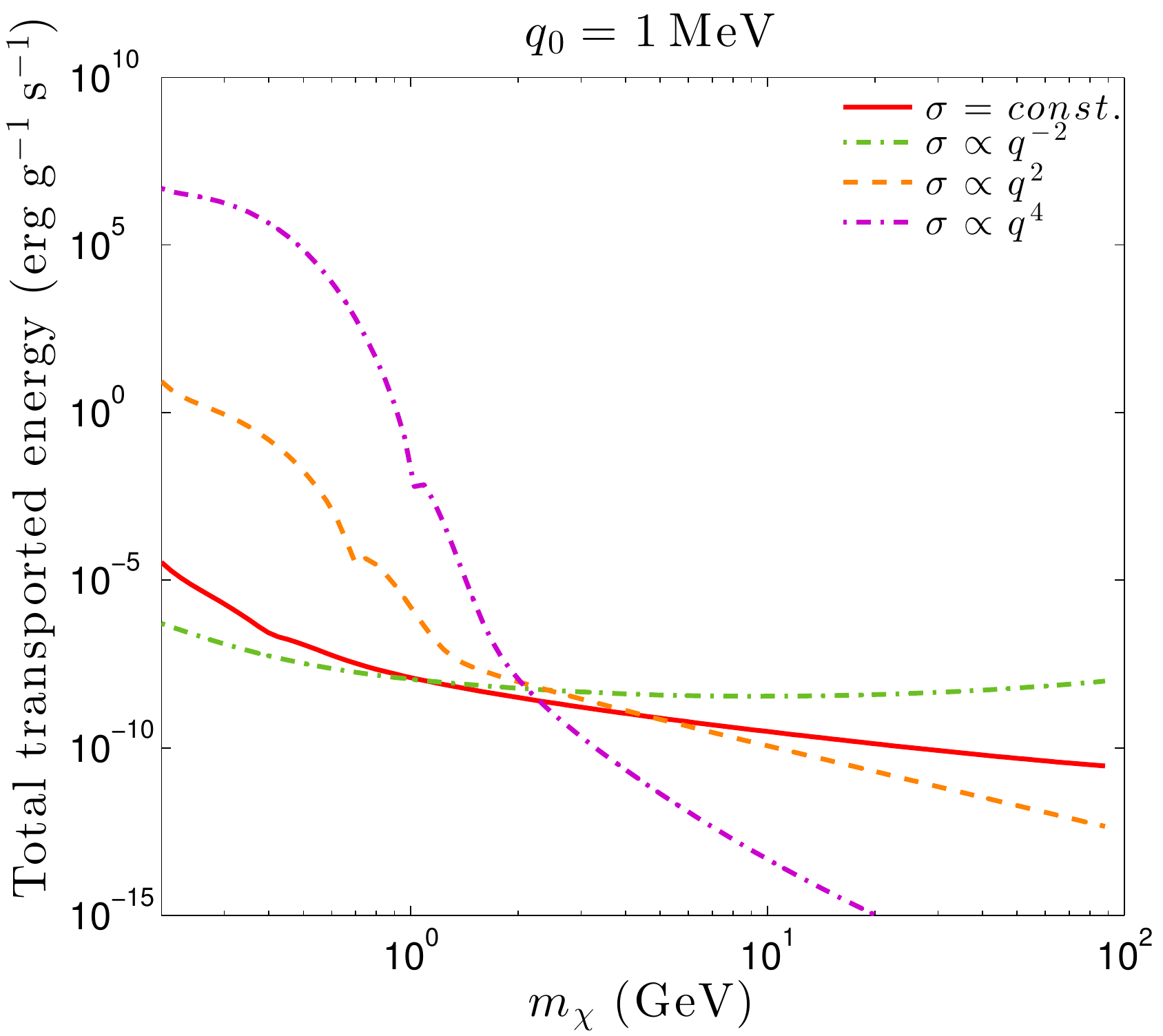} & \includegraphics[width=0.48\textwidth]{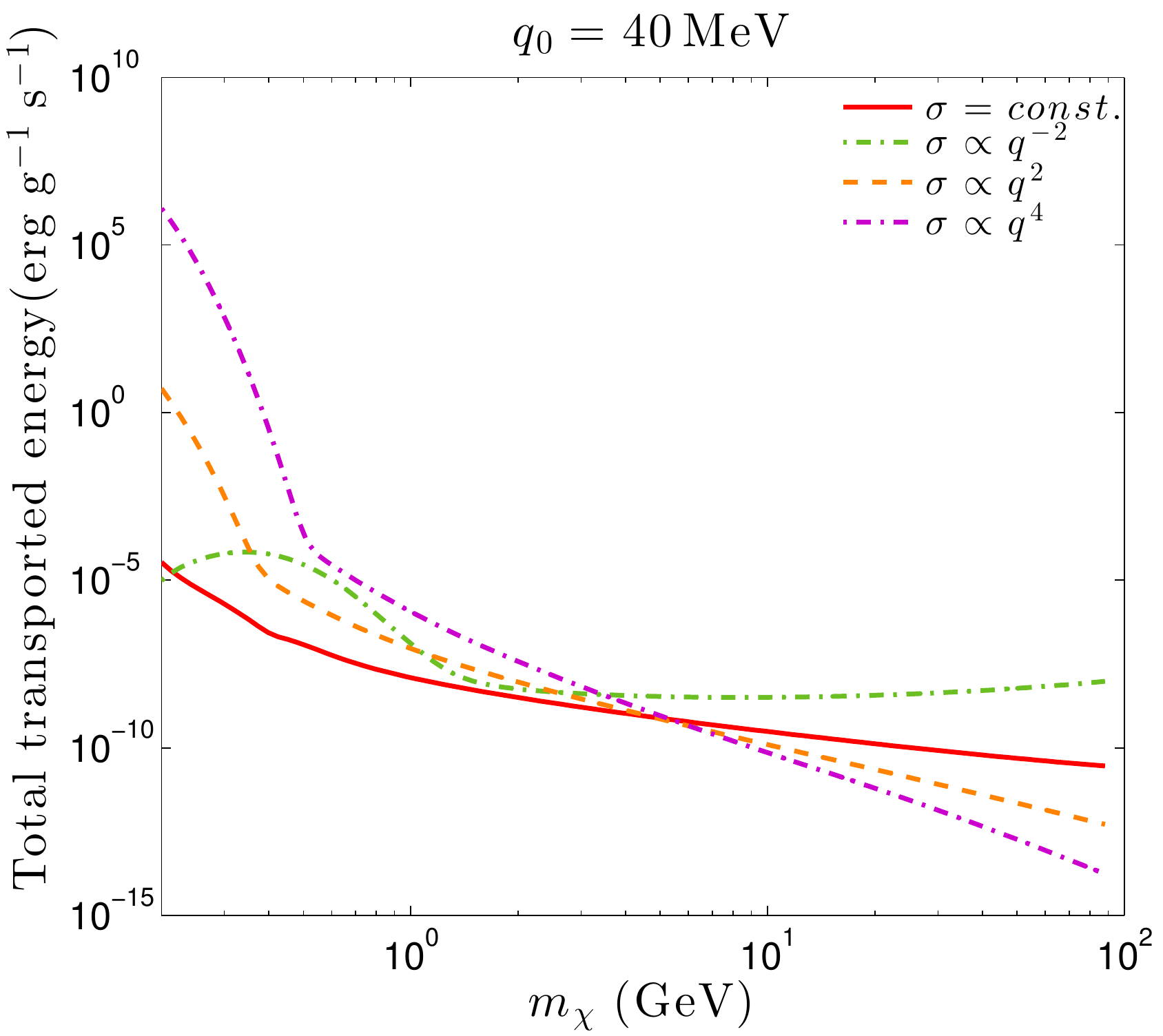}\\
\end{tabular}
\caption{\textit{Total transferred energy $ r_\odot^{-3}\int_0^{r_\odot} |\epsilon(r)|r^2 \ud r$ as a function of the WIMP mass for momentum-dependent scattering.  Here $\sigma_0 = 10^{-39}$ cm$^2$, $n_\chi/n_b = 10^{-16}$ WIMPs per baryon and $q_0 = 1$\,MeV (left) or $q_0 = 40$\,MeV (right). These reference momenta can be compared with nuclear recoil energies in direct detection experiments using the expression $q_0 = \sqrt{2 m_{\rm nuc} E_R}$. For recoils on Germanium, for instance, they correspond to nuclear recoil energies of 7 eV and 11 keV, respectively. These show that, depending on the $q$ region probed by direct detection experiments, solar observations may be far more, or far less sensitive than underground detectors.}}
\label{fig:eps_mdep}
\end{figure*}

Figure \ref{fig:eps_mdep} illustrates this point in more detail.  Here we show the full dependence of the total transported energy on the WIMP mass $\mx$, for each type of $q$-dependent scaling.  We also compare this behaviour for two different example values of $q_0$. For $q_0 = 1$\,MeV -- corresponding to recoil energies in detectors of only a few eV -- the behaviour of $\epsilon$ as a function of the sign of $n$ is reversed below $\mx \sim 2$\,GeV. The value of $q_0 = 40$\,MeV that we choose for the right-hand figure is closer to the region probed by underground direct detection experiments, and shows a reversal at slightly higher masses, around $\mx \sim 4$\,GeV. This shows that for the mass region probed by such experiments, the effects of solar heat transport can be much larger than previously computed in the $\sigma = const.$ case and that for such models, the comparison between experiments and solar effects should be made on a case-by-case basis. 

In contrast with the behaviour shown in Figure \ref{fig:eps_mdep} for $q$-dependent scattering, the total energy transported by $\vrel$-dependent scattering does not show a reversal with respect to $n$ when the WIMP mass is changed. This can also be seen by comparing Figures \ref{fig:etransv} and \ref{fig:etransq}: for $q$, the impact depends on WIMP mass, whereas for $\vrel$-dependent scattering, negative $n$ enhances energy transport and positive $n$ decreases it, independent of the mass.  The qualitative difference between the impacts of the WIMP mass for $\vrel$ and $q$-dependent cross-sections is due to the additional factor of $(1+\mu)^2$ in the thermally-averaged cross-section for $q$-dependent scattering.  This adds an additional dependence on WIMP mass to the typical inter-scattering distance, Knudsen number, and therefore the total $K$-dependent luminosity suppression.

Again, we stress that to compute an accurate profile of energy transport, one should include the feedback on solar structure itself during stellar evolution.

\section{Discussion and Conclusion}
\label{sec:conclusion}
Figures \ref{fig:etransv}--\ref{fig:eps_mdep} contrast the potential impacts on energy transport in the Sun of nuclear scattering cross-sections with different momentum and velocity scalings.  The differences we see suggest that solar physics will be a very useful complement to direct detection experiments in ruling out or confirming specific forms of $\sigma(\vrel,q)$.  

It is worth noting that the cross-sections which give rise to enhanced energy transport will also enhance or suppress the capture of DM by the Sun in the first place. In order to become gravitationally bound to the Sun, a free WIMP must lose enough kinetic energy to fall below the local escape velocity. A scattering cross-section that goes as $\vrel^{-2}$ would boost the chance of a free WIMP at the low end of the Galactic DM velocity distribution scattering off a nucleus, whereas a $q^{2n}$ cross-section with $n > 0$ would enhance the rate of collisions leading to large enough energy losses for the WIMP's speed to fall below $v_{\rm esc}$.  The combination of enhanced capture and heat transport is therefore expected to lead to observable effects on the solar structure for some models, even for very small scattering cross-sections.

Finally we point out the possible application of this enhanced heat transport to the so-called Solar Abundance Problem.  It has been suggested \cite{Frandsen10,Taoso10} that at intermediate radii, $\mx = 5$\,GeV DM can partly reconcile the discrepancy between the results of solar models computed with the latest abundances, and the sound speed profile inferred from helioseismology. In this scenario, DM accumulation is usually maximised by taking it to be asymmetric (non-annihilating). Simulations \cite{Cumberbatch10} indicate that the effects of energy transported from the core can propagate to higher radii and affect the sound speed up to the base of the convection zone.  However, to do so without the effects of momentum- or velocity-dependent couplings, the scattering cross-section must be extremely high ($\sim$$10^{-35}$\,cm$^2$), in order to accumulate sufficient quantities of DM to have an observable effect. The enhancement provided to capture and conduction by $\vrel^{2n}$ or $q^{2n}$ scattering of DM on nuclei has the potential to finally make DM a viable solution to the abundance problem. There is thus a strong possibility that models such as Sommerfeld-enhanced or dipole DM, whose motivation stem from different areas of DM phenomenology, are particularly suited to resolving this outstanding problem in solar physics.

The framework developed above -- an extension of the process established by Gould and Raffelt \cite{GouldRaffelt90a} -- allows the computation of the thermal conduction parameters $\alpha$ and $\kappa$ for DM particles whose interactions with nuclei are velocity or momentum-dependent. We have computed these parameters explicitly for $\sigma \propto \vrel^{2n}$ and $q^{2n}$, for $n = \{-1,1,2\}$. Finally, we have shown the impact on heat transfer in the Sun with respect to the fiducial constant $\sigma$ case, for parameters currently allowed by direct detection experiments.  Our results indicate that this impact could be rather substantial.  In a follow-up paper \cite{Vincent14}, we plan to incorporate these results into state-of-the-art solar simulation software to include realistic DM capture and the feedback on solar structure itself.

\textit{Note added: } As we were finalising this paper for resubmission, ref.\ \cite{Lopes14} appeared on the arXiv, investigating the impacts on the solar structure of a model with a massive force carrier, similar to the $\vrel^{-2}$ case with a cuttoff $\omega$ that we studied.  That paper employed the original formalism of \cite{GouldRaffelt90a}, which we have shown here to be invalid for such dark matter models, and considered only mean WIMP velocities rather than full velocity distributions as we do here.  Nonetheless, the authors did employ a full solar evolutionary model, calculate the capture and annihilation rates for their model, and find promising results with regards to the impacts on the Sun and the potential for solving the Solar Abundance Problem.  This makes an analysis of solar models with velocity- and momentum-dependent energy transport correctly implemented on the basis of the results in this paper \cite{Vincent14}, all the more interesting a prospect.

\section*{Acknowledgements}
We thank Aldo Serenelli for useful discussions. AV acknowledges support from FQRNT and European contracts  FP7--PEOPLE--2011--ITN and PITN--GA--2011--289442--INVISIBLES. PS is supported by a Banting Fellowship, administered by the Natural Science and Engineering Research Council of Canada.

\appendix

\begin{table*}[t]
\caption{List of variables used in this article, with their description, the location in the text where they are first defined and their respective typical units. A dash (---) entry in the unit column corresponds to dimensionless quantities.}
\begin{tabular}{| l | l | c| c |}
\hline
Parameter & Description &Definition & Typ.\ Units \\ \hline
%Lowercase
$\vect a$ & Velocity of CM frame &Fig. \ref{kinfig} & --- \\
$\vect b$ & Incoming WIMP velocity in CM frame & Fig. \ref{kinfig} & --- \\
$c$ (subscript) & Value at $r = 0$ (solar centre) & & --- \\
$f(x,\vect r)$& Dimensionless WIMP phase space distribution & (\ref{fnormeq}) & --- \\
$j$ (index) & Degree of spherical harmonic $Y_j^m$ & & --- \\
$l_\chi$ & Typical interscattering distance & Sec. \ref{sec:backbackground} & cm \\
%$l_i$ & Interscattering distance with species $i$ & (\ref{kappaTotal}) & cm \\
$m$ (index) & Order of spherical harmonic $Y_j^m$ & & --- \\
$m_\chi$ & DM particle mass & Sec. \ref{sec:backbackground} & GeV \\
$m_{\rm nuc}$ & Nucleus mass & Sec. \ref{sec:backbackground}  & GeV \\
$n$ & Exponent of $q$ or $\vrel$ dependence of $\sigma$ & (\ref{pdep}-\ref{vdep}) & --- \\
$n_i(r)$ & Number density of species $i$ & (\ref{fnormeq}) & cm$^{-3}$ \\
$q$ & Transferred momentum & (\ref{pdep}) & GeV \\
$q_0$ & Reference scale for $q$  & (\ref{pdep})  & GeV \\
$\vect r$ & Position within the Sun & & cm \\
$r_\chi$ & WIMP ``scale height'' & (\ref{scaleheight}) & cm \\
t & time & & s \\
$\vect u$ & WIMP velocity & (\ref{BCE}) & cm s$^{-1}$ \\
$\vect v$ & Aux. WIMP velocity variable & (\ref{Cdef}) & cm s$^{-1}$ \\
$\vect v_{nuc}$ & Nucleon velocity & (\ref{zdef}) & cm s$^{-1}$ \\
$\vrel$ & Relative WIMP-nucleon velocity before scattering  & (\ref{vdep}) & cm s$^{-1}$ \\
$v_0$ & Reference scale for $\vrel$  & (\ref{vdep}) & cm s$^{-1}$ \\
$v_T$ & Typical thermal velocity & (\ref{vt}) & cm s$^{-1}$ \\
$w$ & $y/\sqrt{\mu}$ & Tab. \ref{tab:coutv} & --- \\
$\vect x$ & Dimensionless WIMP velocity $\vect v / v_T$ & (\ref{xdef}) & ---\\
$\vect y$ & Dimensionless WIMP velocity $\vect u / v_T$ & (\ref{xdef}) & ---\\
$\vect z$ & Dimensionless nucleus velocity $\vect v_{nuc} / v_T$ & (\ref{xdef}) & ---\\
%Uppercase
$A,B,G$ & Aux. functions of ($x,y,a,b$) & (\ref{Adef}-\ref{Gdef}) & ---\\
$C$ & Collision operator & (\ref{BCE}), (\ref{Cdef}) & --- \\
$D$ & Differential (Laplace) operator & Sec. \ref{sec:background} & s$^{-1}$ \\
$F(\vect u,\vect r,t)$ &WIMP phase space density & (\ref{BCE}) & cm$^{-4}$s \\
$F_{\rm nuc} (z, \vect r)$ & Dimensionless nucleus phase space distribution & (\ref{fnucdef}) & --- \\
$H_n$ & Normalization factor for $\hat \sigma_{\mathrm{tot}}$, for exponent $n$ & (\ref{H_n}) & --- \\
$K$ & Knudsen number $K = l_\chi/r_\chi$ & Sec. \ref{sec:background} & --- \\
$L_\chi(r)$ & Luminosity carried by WIMP scattering & (\ref{LTEtransport}) & erg s$^{-1}$ \\
$P_j$ & Legendre polynomial of degree $j$ & (\ref{cinint}) & --- \\
$T(r)$ & Local temperature & (\ref{vt}) &K \\
%Greek
$\alpha$ & Dimensionless thermal diffusion coefficient & Sec. \ref{sec:backbackground} & --- \\

$\epsilon$ & Energy transport rate by WIMPs & (\ref{epsLTE}) & erg g$^{-1}$ s$^{-1}$ \\
$\varepsilon$ & BCE expansion parameter $l_\chi | \nabla \ln T(r)|$ & Sec. \ref{sec:background} & --- \\
$\zeta$ & rescaling parameter  & (\ref{epsLTE}) & --- \\
$\kappa$ & Dimensionless thermal conduction coefficient &Sec. \ref{sec:backbackground}  & --- \\
$\mu$ & mass ratio $\mx / m_{\rm nuc}$ & (\ref{mudef}) & --- \\
$\nu$ (index) & Order in $\varepsilon$ expansion & (\ref{fperturb}) & ---\\
$\xi$ & Regulator for angular integral & (\ref{P1qm2}) & --- \\
$\rho_i$ & Mass density of species $i$ & & GeV cm$^{-3}$ \\ 
$\sigma$ & WIMP-nucleon scattering cross-section & Sec. \ref{sec:backbackground} & cm$^2$ \\
$\sigma_0$ &Reference scale for $\sigma$ &Sec. \ref{sec:backbackground} & cm$^2$ \\
$\hat \sigma$ & Dimensionless cross-section & (\ref{stotdef})& --- \\
$\hat \sigma_{\mathrm{tot}} $& Angle-integrated dimensionless cross-section & (\ref{stotdef})& --- \\
$\theta_{CM}$ & Scattering angle in CM frame &  Fig. \ref{kinfig}& rad \\
$\chi$ & Dark matter particle label & Sec. \ref{sec:backbackground} & --- \\
$\omega$ & Dimensionless velocity cutoff & (\ref{v2cutoff}) & --- \\
$\Theta(x)$ & Heaviside step function & & --- \\
\hline
\end{tabular}
\end{table*}

\newpage
\setlength{\tabcolsep}{3.5pt}
\begin{table*}[t]
\caption{Tabulated values of $\alpha$ (left) and  $\log_{10}\kappa$ (right) for the different cross-sections studied in this paper, at selected logarithmically-spaced values of $\mu = \mx/m_{\rm nuc}$.}
\label{aktable}
\begin{tabular}{| c | c c c c c c c | c c c c c c c |}
\hline
\multirow{2}{*}[2.4pt]{\backslashbox{\hspace{2mm}$\mu$}{$\sigma \propto$}} & $const.$ & $\vrel^{-2}$ & $\vrel^2$ & $\vrel^4$ & $q^{-2}$ &$q^2$ & $q^4$ & $const.$ & $\vrel^{-2}$ & $\vrel^2$ & $\vrel^4$ & $q^{-2}$ &$q^2$ & $q^4$ \\
& & & & $\alpha$   & &  &   & &  & & $\log_{10} \kappa$  & & & \\
\hline 
    0.0100&    2.0049&    2.9994&    1.0721&    0.4117&    2.9990&    1.0588&    0.3789&    0.0317&    0.5139&   -0.3024&   -0.4960&    0.5139&   -0.4225&   -0.6543\\
    0.0120&    2.0059&    2.9993&    1.0814&    0.4331&    2.9988&    1.0665&    0.3981&    0.0317&    0.5148&   -0.3070&   -0.5107&    0.5148&   -0.4262&   -0.6675\\
    0.0145&    2.0071&    2.9993&    1.0920&    0.4562&    2.9987&    1.0754&    0.4189&    0.0323&    0.5167&   -0.3114&   -0.5264&    0.5167&   -0.4298&   -0.6817\\
    0.0175&    2.0085&    2.9992&    1.1040&    0.4810&    2.9985&    1.0855&    0.4410&    0.0333&    0.5190&   -0.3163&   -0.5433&    0.5190&   -0.4339&   -0.6971\\
    0.0210&    2.0102&    2.9990&    1.1173&    0.5075&    2.9983&    1.0967&    0.4647&    0.0345&    0.5218&   -0.3217&   -0.5617&    0.5218&   -0.4384&   -0.7139\\
    0.0254&    2.0122&    2.9988&    1.1320&    0.5358&    2.9980&    1.1091&    0.4900&    0.0359&    0.5251&   -0.3277&   -0.5818&    0.5251&   -0.4434&   -0.7322\\
    0.0305&    2.0146&    2.9985&    1.1483&    0.5659&    2.9977&    1.1230&    0.5169&    0.0376&    0.5290&   -0.3343&   -0.6036&    0.5290&   -0.4489&   -0.7522\\
    0.0368&    2.0176&    2.9980&    1.1662&    0.5980&    2.9973&    1.1383&    0.5456&    0.0396&    0.5336&   -0.3415&   -0.6273&    0.5336&   -0.4550&   -0.7741\\
    0.0443&    2.0211&    2.9973&    1.1860&    0.6321&    2.9968&    1.1552&    0.5761&    0.0419&    0.5390&   -0.3494&   -0.6530&    0.5390&   -0.4617&   -0.7979\\
    0.0534&    2.0253&    2.9962&    1.2077&    0.6684&    2.9961&    1.1739&    0.6085&    0.0446&    0.5452&   -0.3581&   -0.6810&    0.5452&   -0.4691&   -0.8239\\
    0.0643&    2.0304&    2.9947&    1.2315&    0.7069&    2.9953&    1.1945&    0.6429&    0.0477&    0.5524&   -0.3676&   -0.7113&    0.5524&   -0.4771&   -0.8522\\
    0.0774&    2.0365&    2.9926&    1.2575&    0.7477&    2.9943&    1.2172&    0.6792&    0.0513&    0.5605&   -0.3779&   -0.7441&    0.5605&   -0.4859&   -0.8830\\
    0.0933&    2.0438&    2.9894&    1.2860&    0.7908&    2.9929&    1.2421&    0.7176&    0.0553&    0.5698&   -0.3892&   -0.7796&    0.5698&   -0.4955&   -0.9164\\
    0.1123&    2.0525&    2.9849&    1.3172&    0.8365&    2.9912&    1.2696&    0.7582&    0.0597&    0.5801&   -0.4014&   -0.8178&    0.5801&   -0.5060&   -0.9526\\
    0.1353&    2.0630&    2.9787&    1.3512&    0.8847&    2.9889&    1.2999&    0.8009&    0.0644&    0.5915&   -0.4146&   -0.8591&    0.5915&   -0.5173&   -0.9917\\
    0.1630&    2.0755&    2.9699&    1.3885&    0.9355&    2.9859&    1.3334&    0.8458&    0.0693&    0.6038&   -0.4290&   -0.9033&    0.6038&   -0.5297&   -1.0339\\
    0.1963&    2.0904&    2.9581&    1.4293&    0.9893&    2.9820&    1.3705&    0.8932&    0.0742&    0.6170&   -0.4447&   -0.9507&    0.6170&   -0.5433&   -1.0792\\
    0.2364&    2.1078&    2.9423&    1.4741&    1.0462&    2.9768&    1.4117&    0.9431&    0.0789&    0.6308&   -0.4619&   -1.0014&    0.6308&   -0.5582&   -1.1277\\
    0.2848&    2.1282&    2.9222&    1.5232&    1.1066&    2.9702&    1.4576&    0.9960&    0.0829&    0.6450&   -0.4808&   -1.0554&    0.6450&   -0.5747&   -1.1794\\
    0.3430&    2.1514&    2.8972&    1.5771&    1.1710&    2.9616&    1.5091&    1.0523&    0.0860&    0.6594&   -0.5017&   -1.1129&    0.6594&   -0.5932&   -1.2344\\
    0.4132&    2.1774&    2.8677&    1.6362&    1.2401&    2.9508&    1.5667&    1.1130&    0.0876&    0.6738&   -0.5253&   -1.1739&    0.6738&   -0.6142&   -1.2927\\
    0.4977&    2.2058&    2.8342&    1.7005&    1.3146&    2.9373&    1.6311&    1.1794&    0.0875&    0.6883&   -0.5519&   -1.2390&    0.6883&   -0.6386&   -1.3544\\
    0.5995&    2.2359&    2.7982&    1.7698&    1.3955&    2.9210&    1.7023&    1.2534&    0.0855&    0.7033&   -0.5824&   -1.3085&    0.7033&   -0.6671&   -1.4201\\
    0.7221&    2.2667&    2.7611&    1.8431&    1.4835&    2.9016&    1.7797&    1.3370&    0.0816&    0.7192&   -0.6173&   -1.3833&    0.7192&   -0.7006&   -1.4906\\
    0.8697&    2.2971&    2.7247&    1.9186&    1.5785&    2.8794&    1.8615&    1.4321&    0.0763&    0.7373&   -0.6571&   -1.4645&    0.7373&   -0.7399&   -1.5676\\
    1.0476&    2.3261&    2.6906&    1.9941&    1.6793&    2.8544&    1.9449&    1.5390&    0.0702&    0.7586&   -0.7020&   -1.5530&    0.7586&   -0.7853&   -1.6527\\
    1.2619&    2.3528&    2.6596&    2.0669&    1.7831&    2.8274&    2.0264&    1.6557&    0.0641&    0.7845&   -0.7516&   -1.6497&    0.7845&   -0.8366&   -1.7477\\
    1.5199&    2.3767&    2.6324&    2.1345&    1.8859&    2.7989&    2.1027&    1.7770&    0.0589&    0.8160&   -0.8052&   -1.7547&    0.8160&   -0.8929&   -1.8536\\
    1.8307&    2.3975&    2.6090&    2.1950&    1.9831&    2.7697&    2.1710&    1.8957&    0.0556&    0.8542&   -0.8615&   -1.8674&    0.8542&   -0.9528&   -1.9700\\
    2.2051&    2.4153&    2.5894&    2.2477&    2.0711&    2.7407&    2.2302&    2.0049&    0.0548&    0.8997&   -0.9193&   -1.9863&    0.8997&   -1.0146&   -2.0953\\
    2.6561&    2.4302&    2.5731&    2.2923&    2.1476&    2.7125&    2.2798&    2.0999&    0.0572&    0.9528&   -0.9773&   -2.1093&    0.9528&   -1.0766&   -2.2265\\
    3.1993&    2.4426&    2.5597&    2.3295&    2.2119&    2.6858&    2.3208&    2.1788&    0.0631&    1.0133&   -1.0345&   -2.2346&    1.0133&   -1.1376&   -2.3606\\
    3.8535&    2.4528&    2.5487&    2.3601&    2.2648&    2.6610&    2.3540&    2.2423&    0.0727&    1.0812&   -1.0903&   -2.3604&    1.0812&   -1.1968&   -2.4948\\
    4.6416&    2.4612&    2.5397&    2.3851&    2.3079&    2.6384&    2.3810&    2.2928&    0.0857&    1.1559&   -1.1444&   -2.4857&    1.1559&   -1.2539&   -2.6276\\
    5.5908&    2.4681&    2.5324&    2.4056&    2.3428&    2.6181&    2.4027&    2.3326&    0.1022&    1.2369&   -1.1967&   -2.6102&    1.2369&   -1.3087&   -2.7582\\
    6.7342&    2.4737&    2.5265&    2.4223&    2.3710&    2.6002&    2.4204&    2.3641&    0.1218&    1.3236&   -1.2472&   -2.7336&    1.3236&   -1.3613&   -2.8867\\
    8.1113&    2.4783&    2.5216&    2.4360&    2.3939&    2.5844&    2.4346&    2.3892&    0.1444&    1.4154&   -1.2961&   -2.8560&    1.4154&   -1.4121&   -3.0133\\
    9.7701&    2.4822&    2.5177&    2.4472&    2.4126&    2.5708&    2.4463&    2.4094&    0.1695&    1.5117&   -1.3437&   -2.9779&    1.5117&   -1.4611&   -3.1384\\
    11.768&    2.4854&    2.5145&    2.4564&    2.4278&    2.5590&    2.4558&    2.4256&    0.1968&    1.6119&   -1.3901&   -3.0992&    1.6119&   -1.5088&   -3.2624\\
    14.175&    2.4881&    2.5118&    2.4640&    2.4404&    2.5490&    2.4635&    2.4389&    0.2262&    1.7156&   -1.4355&   -3.2201&    1.7156&   -1.5552&   -3.3855\\
    17.074&    2.4903&    2.5096&    2.4703&    2.4507&    2.5404&    2.4699&    2.4496&    0.2573&    1.8224&   -1.4800&   -3.3408&    1.8224&   -1.6006&   -3.5081\\
    20.565&    2.4920&    2.5077&    2.4755&    2.4592&    2.5332&    2.4752&    2.4584&    0.2898&    1.9321&   -1.5239&   -3.4615&    1.9321&   -1.6453&   -3.6303\\
    24.771&    2.4934&    2.5060&    2.4798&    2.4663&    2.5271&    2.4795&    2.4657&    0.3237&    2.0444&   -1.5671&   -3.5820&    2.0444&   -1.6891&   -3.7521\\
    29.837&    2.4946&    2.5046&    2.4833&    2.4721&    2.5220&    2.4831&    2.4716&    0.3586&    2.1594&   -1.6099&   -3.7027&    2.1594&   -1.7325&   -3.8738\\
    35.938&    2.4956&    2.5034&    2.4862&    2.4769&    2.5177&    2.4860&    2.4765&    0.3946&    2.2771&   -1.6522&   -3.8233&    2.2771&   -1.7753&   -3.9953\\
    43.288&    2.4964&    2.5023&    2.4886&    2.4809&    2.5141&    2.4884&    2.4806&    0.4315&    2.3978&   -1.6941&   -3.9439&    2.3978&   -1.8176&   -4.1166\\
    52.140&    2.4972&    2.5012&    2.4906&    2.4843&    2.5111&    2.4904&    2.4839&    0.4690&    2.5224&   -1.7357&   -4.0646&    2.5224&   -1.8597&   -4.2380\\
    62.803&    2.4978&    2.5001&    2.4924&    2.4870&    2.5085&    2.4921&    2.4867&    0.5073&    2.6519&   -1.7770&   -4.1853&    2.6519&   -1.9015&   -4.3593\\
    75.646&    2.4984&    2.4990&    2.4939&    2.4894&    2.5064&    2.4935&    2.4890&    0.5462&    2.7884&   -1.8181&   -4.3060&    2.7884&   -1.9431&   -4.4806\\
    91.116&    2.4989&    2.4976&    2.4952&    2.4914&    2.5045&    2.4946&    2.4909&    0.5859&    2.9354&   -1.8589&   -4.4267&    2.9354&   -1.9844&   -4.6018\\
    100.00&    2.4991&    2.4969&    2.4957&    2.4923&    2.5037&    2.4951&    2.4917&    0.6061&    3.0146&   -1.8792&   -4.4870&    3.0146&   -2.0050&   -4.6624\\ \hline
  \end{tabular}
\end{table*}
\bibliographystyle{JHEP_pat}
\bibliography{solarDM,DMbiblio,AbuGen,CandO,CObiblio,solarAxions}

\providecommand{\href}[2]{#2}\begingroup\raggedright\begin{thebibliography}{100}

\bibitem{GouldRaffelt90a}
A.~{Gould} and G.~{Raffelt}, {\it {Thermal conduction by massive particles}},
  {\em \apj} {\bf 352} (1990) 654--668.

\bibitem{Press85}
W.~H. {Press} and D.~N. {Spergel}, {\it {Capture by the sun of a galactic
  population of weakly interacting, massive particles}},  {\em \apj} {\bf 296}
  (1985) 679--684.

\bibitem{Griest87}
K.~{Griest} and D.~{Seckel}, {\it {Cosmic asymmetry, neutrinos and the Sun.}},
  {\em \nphysb} {\bf 283} (1987) 681--705.

\bibitem{Gould87b}
A.~{Gould}, {\it {Resonant enhancements in weakly interacting massive particle
  capture by the earth}},  {\em \apj} {\bf 321} (1987) 571--585.

\bibitem{Steigman78}
G.~{Steigman}, H.~{Quintana}, C.~L. {Sarazin}, and J.~{Faulkner}, {\it
  {Dynamical interactions and astrophysical effects of stable heavy
  neutrinos}},  {\em \aj} {\bf 83} (1978) 1050--1061.

\bibitem{Spergel85}
D.~N. {Spergel} and W.~H. {Press}, {\it {Effect of hypothetical, weakly
  interacting, massive particles on energy transport in the solar interior}},
  {\em \apj} {\bf 294} (1985) 663--673.

\bibitem{Faulkner85}
J.~{Faulkner} and R.~L. {Gilliland}, {\it {Weakly interacting, massive
  particles and the solar neutrino flux}},  {\em \apj} {\bf 299} (1985)
  994--1000.

\bibitem{Scott09}
P.~{Scott}, M.~{Fairbairn}, and J.~{Edsj{\"o}}, {\it {Dark stars at the
  Galactic Centre - the main sequence}},  {\em \mnras} {\bf 394} (2009)
  82--104, [\href{http://xxx.lanl.gov/abs/{arXiv:0809.1871}}{{\tt
  {arXiv:0809.1871}}}].

\bibitem{Turck12}
S.~{Turck-Chi{\`e}ze} and I.~{Lopes}, {\it {Solar-stellar astrophysics and dark
  matter}},  {\em Research in Astronomy and Astrophysics} {\bf 12} (2012)
  1107--1138.

\bibitem{Zurek13}
K.~M. {Zurek}, {\it {Asymmetric Dark Matter: Theories, Signatures, and
  Constraints}},  \href{http://xxx.lanl.gov/abs/1308.0338}{{\tt
  arXiv:1308.0338}}.

\bibitem{Krauss86}
L.~M. {Krauss}, M.~{Srednicki}, and F.~{Wilczek}, {\it {Solar System
  constraints and signatures for dark-matter candidates}},  {\em \prd} {\bf 33}
  (1986) 2079--2083.

\bibitem{Gaisser86}
T.~K. {Gaisser}, G.~{Steigman}, and S.~{Tilav}, {\it {Limits on
  cold-dark-matter candidates from deep underground detectors}},  {\em \prd}
  {\bf 34} (1986) 2206--2222.

\bibitem{Gandhi:1993ce}
R.~Gandhi, J.~L. Lopez, D.~V. Nanopoulos, K.-j. Yuan, and A.~Zichichi, {\it
  {Scrutinizing supergravity models through neutrino telescopes}},  {\em
  Phys.Rev.} {\bf D49} (1994) 3691--3703,
  [\href{http://xxx.lanl.gov/abs/astro-ph/9309048}{{\tt astro-ph/9309048}}].

\bibitem{Bottino:1994xp}
A.~Bottino, N.~Fornengo, G.~Mignola, and L.~Moscoso, {\it {Signals of
  neutralino dark matter from earth and sun}},  {\em Astropart.Phys.} {\bf 3}
  (1995) 65--76, [\href{http://xxx.lanl.gov/abs/hep-ph/9408391}{{\tt
  hep-ph/9408391}}].

\bibitem{Bergstrom98b}
L.~{Bergstr{\"o}m}, J.~{Edsj{\"o}}, and P.~{Gondolo}, {\it {Indirect detection
  of dark matter in km-size neutrino telescopes}},  {\em \prd} {\bf 58} (1998)
  103519, [\href{http://xxx.lanl.gov/abs/hep-ph/9806293}{{\tt
  hep-ph/9806293}}].

\bibitem{Barger02}
V.~{Barger}, F.~{Halzen}, D.~{Hooper}, and C.~{Kao}, {\it {Indirect search for
  neutralino dark matter with high energy neutrinos}},  {\em \prd} {\bf 65}
  (2002) 075022, [\href{http://xxx.lanl.gov/abs/hep-ph/0105182}{{\tt
  hep-ph/0105182}}].

\bibitem{Desai04}
Super-Kamiokande Collaboration: S.~{Desai}, Y.~{Ashie}, {\em et.~al.}, {\it
  {Search for dark matter WIMPs using upward through-going muons in
  Super-Kamiokande}},  {\em \prd} {\bf 70} (2004) 083523,
  [\href{http://xxx.lanl.gov/abs/hep-ex/0404025}{{\tt hep-ex/0404025}}].

\bibitem{Desai08}
Super-Kamiokande Collaboration: S.~{Desai}, K.~{Abe}, {\em et.~al.}, {\it
  {Study of TeV neutrinos with upward showering muons in Super-Kamiokande}},
  {\em \app} {\bf 29} (2008) 42--54,
  [\href{http://xxx.lanl.gov/abs/0711.0053}{{\tt arXiv:0711.0053}}].

\bibitem{IceCube09}
IceCube Collaboration: R.~{Abbasi}, Y.~{Abdou}, {\em et.~al.}, {\it {Limits on
  a Muon Flux from Neutralino Annihilations in the Sun with the IceCube
  22-String Detector}},  {\em \prl} {\bf 102} (2009) 201302,
  [\href{http://xxx.lanl.gov/abs/0902.2460}{{\tt arXiv:0902.2460}}].

\bibitem{IceCube09_KK}
IceCube Collaboration: R.~{Abbasi}, Y.~{Abdou}, {\em et.~al.}, {\it {Limits on
  a muon flux from Kaluza-Klein dark matter annihilations in the Sun from the
  IceCube 22-string detector}},  {\em \prd} {\bf 81} (2010) 057101,
  [\href{http://xxx.lanl.gov/abs/0910.4480}{{\tt arXiv:0910.4480}}].

\bibitem{IC40DM}
IceCube Collaboration: R.~{Abbasi}, Y.~{Abdou}, {\em et.~al.}, {\it {Multiyear
  search for dark matter annihilations in the Sun with the AMANDA-II and
  IceCube detectors}},  {\em \prd} {\bf 85} (2012) 042002,
  [\href{http://xxx.lanl.gov/abs/1112.1840}{{\tt arXiv:1112.1840}}].

\bibitem{SuperK11}
Super-Kamiokande Collaboration: T.~{Tanaka}, K.~{Abe}, {\em et.~al.}, {\it {An
  Indirect Search for Weakly Interacting Massive Particles in the Sun Using
  3109.6 Days of Upward-going Muons in Super-Kamiokande}},  {\em \apj} {\bf
  742} (2011) 78, [\href{http://xxx.lanl.gov/abs/1108.3384}{{\tt
  arXiv:1108.3384}}].

\bibitem{IC22Methods}
P.~{Scott}, C.~{Savage}, J.~{Edsj{\"o}}, and {the IceCube Collaboration:
  R.~Abbasi et al.}, {\it {Use of event-level neutrino telescope data in global
  fits for theories of new physics}},  {\em \jcap} {\bf 11} (2012) 57,
  [\href{http://xxx.lanl.gov/abs/1207.0810}{{\tt arXiv:1207.0810}}].

\bibitem{Silverwood12}
H.~{Silverwood}, P.~{Scott}, {\em et.~al.}, {\it {Sensitivity of
  IceCube-DeepCore to neutralino dark matter in the MSSM-25}},  {\em \jcap}
  {\bf 3} (2013) 27, [\href{http://xxx.lanl.gov/abs/1210.0844}{{\tt
  arXiv:1210.0844}}].

\bibitem{IC79}
IceCube Collaboration: M.~G. {Aartsen}, R.~{Abbasi}, {\em et.~al.}, {\it
  {Search for Dark Matter Annihilations in the Sun with the 79-String IceCube
  Detector}},  {\em \prl} {\bf 110} (2013) 131302,
  [\href{http://xxx.lanl.gov/abs/1212.4097}{{\tt arXiv:1212.4097}}].

\bibitem{SalatiSilk89}
P.~{Salati} and J.~{Silk}, {\it {A stellar probe of dark matter annihilation in
  galactic nuclei}},  {\em \apj} {\bf 338} (1989) 24--31.

\bibitem{BouquetSalati89a}
A.~{Bouquet} and P.~{Salati}, {\it {Life and death of cosmions in stars}},
  {\em \aap} {\bf 217} (1989) 270--282.

\bibitem{Moskalenko07}
I.~V. {Moskalenko} and L.~L. {Wai}, {\it {Dark Matter Burners}},  {\em \apjl}
  {\bf 659} (2007) L29--L32,
  [\href{http://xxx.lanl.gov/abs/astro-ph/0702654}{{\tt astro-ph/0702654}}].

\bibitem{Spolyar08}
D.~{Spolyar}, K.~{Freese}, and P.~{Gondolo}, {\it {Dark Matter and the First
  Stars: A New Phase of Stellar Evolution}},  {\em \prl} {\bf 100} (2008)
  051101, [\href{http://xxx.lanl.gov/abs/0705.0521}{{\tt arXiv:0705.0521}}].

\bibitem{Bertone07}
G.~{Bertone} and M.~{Fairbairn}, {\it {Compact stars as dark matter probes}},
  {\em \prd} {\bf 77} (2008) 043515,
  [\href{http://xxx.lanl.gov/abs/0709.1485}{{\tt arXiv:0709.1485}}].

\bibitem{Fairbairn08}
M.~{Fairbairn}, P.~{Scott}, and J.~{Edsj{\"o}}, {\it {The zero age main
  sequence of WIMP burners}},  {\em \prd} {\bf 77} (2008) 047301,
  [\href{http://xxx.lanl.gov/abs/{arXiv:0710.3396}}{{\tt {arXiv:0710.3396}}}].

\bibitem{Scott08a}
P.~{Scott}, J.~{Edsj{\"o}}, and M.~{Fairbairn}, {\it {Low mass stellar
  evolution with WIMP capture and annihilation}},  in {\em Dark Matter in
  Astroparticle and Particle Physics: Dark 2007} (H.~K. {Klapdor-Kleingrothaus}
  and G.~F. {Lewis}, eds.), World Scientific, Singapore (2008) 387--392,
  [\href{http://xxx.lanl.gov/abs/{arXiv:0711.0991}}{{\tt {arXiv:0711.0991}}}].

\bibitem{Iocco08a}
F.~{Iocco}, {\it {Dark Matter Capture and Annihilation on the First Stars:
  Preliminary Estimates}},  {\em \apjl} {\bf 677} (2008) L1--L4,
  [\href{http://xxx.lanl.gov/abs/0802.0941}{{\tt arXiv:0802.0941}}].

\bibitem{Iocco08b}
F.~{Iocco}, A.~{Bressan}, {\em et.~al.}, {\it {Dark matter annihilation effects
  on the first stars}},  {\em \mnras} {\bf 390} (2008) 1655--1669,
  [\href{http://xxx.lanl.gov/abs/0805.4016}{{\tt arXiv:0805.4016}}].

\bibitem{Casanellas09}
J.~{Casanellas} and I.~{Lopes}, {\it {The Formation and Evolution of Young
  Low-mass Stars within Halos with High Concentration of Dark Matter
  Particles}},  {\em \apj} {\bf 705} (2009) 135--143,
  [\href{http://xxx.lanl.gov/abs/0909.1971}{{\tt arXiv:0909.1971}}].

\bibitem{Ripamonti10}
E.~{Ripamonti}, F.~{Iocco}, {\em et.~al.}, {\it {First star formation with dark
  matter annihilation}},  {\em \mnras} {\bf 406} (2010) 2605--2615,
  [\href{http://xxx.lanl.gov/abs/1003.0676}{{\tt arXiv:1003.0676}}].

\bibitem{Zackrisson10a}
E.~{Zackrisson}, P.~{Scott}, {\em et.~al.}, {\it {Finding High-redshift Dark
  Stars with the James Webb Space Telescope}},  {\em \apj} {\bf 717} (2010)
  257--267, [\href{http://xxx.lanl.gov/abs/{arXiv:1002.3368}}{{\tt
  {arXiv:1002.3368}}}].

\bibitem{Zackrisson10b}
E.~{Zackrisson}, P.~{Scott}, {\em et.~al.}, {\it {Observational constraints on
  supermassive dark stars}},  {\em \mnras} {\bf 407} (2010) L74--L78,
  [\href{http://xxx.lanl.gov/abs/{arXiv:1006.0481}}{{\tt {arXiv:1006.0481}}}].

\bibitem{Scott11}
P.~{Scott}, A.~{Venkatesan}, {\em et.~al.}, {\it {Impacts of Dark Stars on
  Reionization and Signatures in the Cosmic Microwave Background}},  {\em \apj}
  {\bf 742} (2011) 129, [\href{http://xxx.lanl.gov/abs/1107.1714}{{\tt
  arXiv:1107.1714}}].

\bibitem{Rott13}
C.~{Rott}, J.~M. {Siegal-Gaskins}, and J.~F. {Beacom}, {\it {New sensitivity to
  solar WIMP annihilation using low-energy neutrinos}},  {\em \prd} {\bf 88}
  (2013) 055005, [\href{http://xxx.lanl.gov/abs/1208.0827}{{\tt
  arXiv:1208.0827}}].

\bibitem{Bernal:2012qh}
N.~Bernal, J.~Mart'n-Albo, and S.~Palomares-Ruiz, {\it {A novel way of
  constraining WIMPs annihilations in the Sun: MeV neutrinos}},  {\em JCAP}
  {\bf 1308} (2013) 011, [\href{http://xxx.lanl.gov/abs/1208.0834}{{\tt
  arXiv:1208.0834}}].

\bibitem{Gilliland86}
R.~L. {Gilliland}, J.~{Faulkner}, W.~H. {Press}, and D.~N. {Spergel}, {\it
  {Solar models with energy transport by weakly interacting particles}},  {\em
  \apj} {\bf 306} (1986) 703--709.

\bibitem{Renzini87}
A.~{Renzini}, {\it {Effects of cosmions in the sun and in globular cluster
  stars}},  {\em \aap} {\bf 171} (1987) 121.

\bibitem{Spergel88}
D.~N. {Spergel} and J.~{Faulkner}, {\it {Weakly interacting, massive particles
  in horizontal-branch stars}},  {\em \apjl} {\bf 331} (1988) L21--L24.

\bibitem{Faulkner88}
J.~{Faulkner} and F.~J. {Swenson}, {\it {Main-sequence evolution with efficient
  central energy transport}},  {\em \apjl} {\bf 329} (1988) L47--L50.

\bibitem{Bouquet89}
A.~{Bouquet}, J.~{Kaplan}, and F.~{Martin}, {\it {Weakly interacting massive
  particles and stellar structure}},  {\em \aap} {\bf 222} (1989) 103--116.

\bibitem{BouquetSalati89b}
A.~{Bouquet} and P.~{Salati}, {\it {Dark matter and the suppression of stellar
  core convection}},  {\em \apj} {\bf 346} (1989) 284--288.

\bibitem{Salati90}
P.~{Salati}, {\it {Dark matter particles as inhibitors of the solar core
  pulsations}},  {\em \apj} {\bf 348} (1990) 738--747.

\bibitem{Dearborn90b}
D.~{Dearborn}, G.~{Raffelt}, P.~{Salati}, J.~{Silk}, and A.~{Bouquet}, {\it
  {Dark matter and the age of globular clusters}},  {\em \nat} {\bf 343} (1990)
  347.

\bibitem{Dearborn90}
D.~{Dearborn}, G.~{Raffelt}, P.~{Salati}, J.~{Silk}, and A.~{Bouquet}, {\it
  {Dark matter and thermal pulses in horizontal-branch stars}},  {\em \apj}
  {\bf 354} (1990) 568--582.

\bibitem{GH90}
Y.~{Giraud-Heraud}, J.~{Kaplan}, F.~M. {de Volnay}, C.~{Tao}, and
  S.~{Turck-Chieze}, {\it {WIMPs and solar evolution code}},  {\em \solphys}
  {\bf 128} (1990) 21--33.

\bibitem{CDalsgaard92}
J.~{Christensen-Dalsgaard}, {\it {Solar models with enhanced energy transport
  in the core}},  {\em \apj} {\bf 385} (1992) 354--362.

\bibitem{Faulkner93}
J.~{Faulkner} and F.~J. {Swenson}, {\it {Sub-giant branch evolution and
  efficient central energy transport}},  {\em \apj} {\bf 411} (1993) 200--206.

\bibitem{Iocco12}
F.~{Iocco}, M.~{Taoso}, F.~{Leclercq}, and G.~{Meynet}, {\it {Main Sequence
  Stars with Asymmetric Dark Matter}},  {\em \prl} {\bf 108} (2012) 061301,
  [\href{http://xxx.lanl.gov/abs/1201.5387}{{\tt arXiv:1201.5387}}].

\bibitem{Lopes02a}
I.~P. {Lopes}, J.~{Silk}, and S.~H. {Hansen}, {\it {Helioseismology as a new
  constraint on supersymmetric dark matter}},  {\em \mnras} {\bf 331} (2002)
  361--368, [\href{http://xxx.lanl.gov/abs/astro-ph/0111530}{{\tt
  astro-ph/0111530}}].

\bibitem{Lopes02b}
I.~P. {Lopes}, G.~{Bertone}, and J.~{Silk}, {\it {Solar seismic model as a new
  constraint on supersymmetric dark matter}},  {\em \mnras} {\bf 337} (2002)
  1179--1184, [\href{http://xxx.lanl.gov/abs/astro-ph/0205066}{{\tt
  astro-ph/0205066}}].

\bibitem{Bottino02}
A.~{Bottino}, G.~{Fiorentini}, {\em et.~al.}, {\it {Does solar physics provide
  constraints to weakly interacting massive particles?}},  {\em \prd} {\bf 66}
  (2002) 053005, [\href{http://xxx.lanl.gov/abs/hep-ph/0206211}{{\tt
  hep-ph/0206211}}].

\bibitem{Frandsen10}
M.~T. {Frandsen} and S.~{Sarkar}, {\it {Asymmetric Dark Matter and the Sun}},
  {\em \prl} {\bf 105} (2010) 011301,
  [\href{http://xxx.lanl.gov/abs/1003.4505}{{\tt arXiv:1003.4505}}].

\bibitem{Taoso10}
M.~{Taoso}, F.~{Iocco}, G.~{Meynet}, G.~{Bertone}, and P.~{Eggenberger}, {\it
  {Effect of low mass dark matter particles on the Sun}},  {\em \prd} {\bf 82}
  (2010) 083509, [\href{http://xxx.lanl.gov/abs/1005.5711}{{\tt
  arXiv:1005.5711}}].

\bibitem{Cumberbatch10}
D.~T. {Cumberbatch}, J.~A. {Guzik}, J.~{Silk}, L.~S. {Watson}, and S.~M.
  {West}, {\it {Light WIMPs in the Sun: Constraints from helioseismology}},
  {\em \prd} {\bf 82} (2010) 103503,
  [\href{http://xxx.lanl.gov/abs/1005.5102}{{\tt arXiv:1005.5102}}].

\bibitem{Lopes10}
I.~{Lopes} and J.~{Silk}, {\it {Neutrino Spectroscopy Can Probe the Dark Matter
  Content in the Sun}},  {\em Science} {\bf 330} (2010) 462.

\bibitem{Lopes:2012}
I.~{Lopes} and J.~{Silk}, {\it {Solar Constraints on Asymmetric Dark Matter}},
  {\em \apj} {\bf 757} (2012) 130,
  [\href{http://xxx.lanl.gov/abs/1209.3631}{{\tt arXiv:1209.3631}}].

\bibitem{Lopes:2013}
I.~{Lopes} and J.~{Silk}, {\it {Solar Neutrino Physics: Sensitivity to Light
  Dark Matter Particles}},  {\em \apj} {\bf 752} (2012) 129,
  [\href{http://xxx.lanl.gov/abs/1309.7573}{{\tt arXiv:1309.7573}}].

\bibitem{Casanellas:2013}
J.~{Casanellas} and I.~{Lopes}, {\it {First Asteroseismic Limits on the Nature
  of Dark Matter}},  {\em \apjl} {\bf 765} (2013) L21,
  [\href{http://xxx.lanl.gov/abs/1212.2985}{{\tt arXiv:1212.2985}}].

\bibitem{Lopes:2014}
I.~{Lopes}, K.~{Kadota}, and J.~{Silk}, {\it {Constraint on Light Dipole Dark
  Matter from Helioseismology}},  \href{http://xxx.lanl.gov/abs/1310.0673}{{\tt
  arXiv:1310.0673}}.

\bibitem{Bernabei08}
DAMA Collaboration: R.~{Bernabei}, P.~{Belli}, {\em et.~al.}, {\it {First
  results from DAMA/LIBRA and the combined results with DAMA/NaI}},  {\em
  \epjc} (2008) 167, [\href{http://xxx.lanl.gov/abs/0804.2741}{{\tt
  arXiv:0804.2741}}].

\bibitem{CoGeNTAnnMod11}
CoGeNT Collaboration: C.~E. {Aalseth}, P.~S. {Barbeau}, {\em et.~al.}, {\it
  {Search for an Annual Modulation in a p-Type Point Contact Germanium Dark
  Matter Detector}},  {\em \prl} {\bf 107} (2011) 141301,
  [\href{http://xxx.lanl.gov/abs/1106.0650}{{\tt arXiv:1106.0650}}].

\bibitem{CRESST11}
CRESST Collaboration: G.~{Angloher}, M.~{Bauer}, {\em et.~al.}, {\it {Results
  from 730 kg days of the CRESST-II Dark Matter search}},  {\em \epjc} {\bf 72}
  (2012) 1971, [\href{http://xxx.lanl.gov/abs/1109.0702}{{\tt
  arXiv:1109.0702}}].

\bibitem{Agnese:2013rvf}
CDMS Collaboration: R.~Agnese {\em et.~al.}, {\it {Dark Matter Search Results
  Using the Silicon Detectors of CDMS II}},  {\em \prl} (2013) in press,
  [\href{http://xxx.lanl.gov/abs/1304.4279}{{\tt arXiv:1304.4279}}].

\bibitem{Angle:2011th}
XENON10 Collaboration: J.~Angle {\em et.~al.}, {\it {A search for light dark
  matter in XENON10 data}},  {\em \prl} {\bf 107} (2011) 051301,
  [\href{http://xxx.lanl.gov/abs/1104.3088}{{\tt arXiv:1104.3088}}].

\bibitem{XENON2013}
XENON100 Collaboration: E.~{Aprile}, M.~{Alfonsi}, {\em et.~al.}, {\it {Dark
  Matter Results from 225 Live Days of XENON100 Data}},  {\em \prl} {\bf 109}
  (2012) 181301, [\href{http://xxx.lanl.gov/abs/1207.5988}{{\tt
  arXiv:1207.5988}}].

\bibitem{COUPP12}
COUPP Collaboration: E.~{Behnke}, J.~{Behnke}, {\em et.~al.}, {\it {First dark
  matter search results from a 4-kg CF$_{3}$I bubble chamber operated in a deep
  underground site}},  {\em \prd} {\bf 86} (2012) 052001,
  [\href{http://xxx.lanl.gov/abs/1204.3094}{{\tt arXiv:1204.3094}}].

\bibitem{SIMPLE11}
SIMPLE Collaboration: M.~{Felizardo}, T.~A. {Girard}, {\em et.~al.}, {\it
  {Final Analysis and Results of the Phase II SIMPLE Dark Matter Search}},
  {\em \prl} {\bf 108} (2012) 201302,
  [\href{http://xxx.lanl.gov/abs/1106.3014}{{\tt arXiv:1106.3014}}].

\bibitem{Bozorgnia:2013hsa}
N.~{Bozorgnia}, J.~{Herrero-Garcia}, T.~{Schwetz}, and J.~{Zupan}, {\it
  {Halo-independent methods for inelastic dark matter scattering}},  {\em
  \jcap} {\bf 7} (2013) 49, [\href{http://xxx.lanl.gov/abs/1305.3575}{{\tt
  arXiv:1305.3575}}].

\bibitem{Mao:2013nda}
Y.-Y. Mao, L.~E. Strigari, and R.~H. Wechsler, {\it {Connecting Direct Dark
  Matter Detection Experiments to Cosmologically Motivated Halo Models}},
  \href{http://xxx.lanl.gov/abs/1304.6401}{{\tt arXiv:1304.6401}}.

\bibitem{Masso09}
E.~{Mass{\'o}}, S.~{Mohanty}, and S.~{Rao}, {\it {Dipolar dark matter}},  {\em
  \prd} {\bf 80} (2009) 036009, [\href{http://xxx.lanl.gov/abs/0906.1979}{{\tt
  arXiv:0906.1979}}].

\bibitem{Feldstein:2009tr}
B.~Feldstein, A.~L. Fitzpatrick, and E.~Katz, {\it {Form Factor Dark Matter}},
  {\em \jcap} {\bf 1001} (2010) 020,
  [\href{http://xxx.lanl.gov/abs/0908.2991}{{\tt arXiv:0908.2991}}].

\bibitem{Chang:2009yt}
S.~Chang, A.~Pierce, and N.~Weiner, {\it {Momentum Dependent Dark Matter
  Scattering}},  {\em \jcap} {\bf 1001} (2010) 006,
  [\href{http://xxx.lanl.gov/abs/0908.3192}{{\tt arXiv:0908.3192}}].

\bibitem{Feldstein10}
B.~{Feldstein}, A.~L. {Fitzpatrick}, E.~{Katz}, and B.~{Tweedie}, {\it {A
  simple explanation for DAMA with moderate channeling}},  {\em \jcap} {\bf 3}
  (2010) 29, [\href{http://xxx.lanl.gov/abs/0910.0007}{{\tt arXiv:0910.0007}}].

\bibitem{Chang10}
S.~{Chang}, J.~{Liu}, A.~{Pierce}, N.~{Weiner}, and I.~{Yavin}, {\it {CoGeNT
  interpretations}},  {\em \jcap} {\bf 8} (2010) 18,
  [\href{http://xxx.lanl.gov/abs/1004.0697}{{\tt arXiv:1004.0697}}].

\bibitem{An10}
H.~{An}, S.-L. {Chen}, R.~N. {Mohapatra}, S.~{Nussinov}, and Y.~{Zhang}, {\it
  {Energy dependence of direct detection cross section for asymmetric mirror
  dark matter}},  {\em \prd} {\bf 82} (2010) 023533,
  [\href{http://xxx.lanl.gov/abs/1004.3296}{{\tt arXiv:1004.3296}}].

\bibitem{Chang:2010en}
S.~Chang, N.~Weiner, and I.~Yavin, {\it {Magnetic Inelastic Dark Matter}},
  {\em \prd} {\bf 82} (2010) 125011,
  [\href{http://xxx.lanl.gov/abs/1007.4200}{{\tt arXiv:1007.4200}}].

\bibitem{Barger11}
V.~{Barger}, W.-Y. {Keung}, and D.~{Marfatia}, {\it {Electromagnetic properties
  of dark matter: Dipole moments and charge form factor}},  {\em Physics
  Letters B} {\bf 696} (2011) 74--78,
  [\href{http://xxx.lanl.gov/abs/1007.4345}{{\tt arXiv:1007.4345}}].

\bibitem{Fitzpatrick10}
A.~L. {Fitzpatrick} and K.~M. {Zurek}, {\it {Dark moments and the DAMA-CoGeNT
  puzzle}},  {\em \prd} {\bf 82} (2010) 075004,
  [\href{http://xxx.lanl.gov/abs/1007.5325}{{\tt arXiv:1007.5325}}].

\bibitem{Frandsen:2013cna}
M.~T. {Frandsen}, F.~{Kahlhoefer}, C.~{McCabe}, S.~{Sarkar}, and
  K.~{Schmidt-Hoberg}, {\it {The unbearable lightness of being: CDMS versus
  XENON}},  {\em \jcap} {\bf 7} (2013) 23,
  [\href{http://xxx.lanl.gov/abs/1304.6066}{{\tt arXiv:1304.6066}}].

\bibitem{LUX13}
LUX Collaboration: D.~Akerib {\em et.~al.}, {\it {First results from the LUX
  dark matter experiment at the Sanford Underground Research Facility}},
  \href{http://xxx.lanl.gov/abs/1310.8214}{{\tt arXiv:1310.8214}}.

\bibitem{Gresham13}
M.~I. {Gresham} and K.~M. {Zurek}, {\it {Light Dark Matter Anomalies After
  LUX}},  {\em \prd} {\bf 89} (2014)
  [\href{http://xxx.lanl.gov/abs/1311.2082}{{\tt arXiv:1311.2082}}].

\bibitem{Cirelli13}
M.~{Cirelli}, E.~{Del Nobile}, and P.~{Panci}, {\it {Tools for
  model-independent bounds in direct dark matter searches}},  {\em \jcap} {\bf
  10} (2013) 19, [\href{http://xxx.lanl.gov/abs/1307.5955}{{\tt
  arXiv:1307.5955}}].

\bibitem{APForbidO}
C.~{Allende Prieto}, D.~L. {Lambert}, and M.~{Asplund}, {\it {The Forbidden
  Abundance of Oxygen in the Sun}},  {\em \apjl} {\bf 556} (2001) L63--L66,
  [\href{http://xxx.lanl.gov/abs/astro-ph/0106360}{{\tt astro-ph/0106360}}].

\bibitem{CtoO}
C.~{Allende Prieto}, D.~L. {Lambert}, and M.~{Asplund}, {\it {A Reappraisal of
  the Solar Photospheric C/O Ratio}},  {\em \apjl} {\bf 573} (2002) L137--L140,
  [\href{http://xxx.lanl.gov/abs/astro-ph/0206089}{{\tt astro-ph/0206089}}].

\bibitem{AspIV}
M.~{Asplund}, N.~{Grevesse}, A.~J. {Sauval}, C.~{Allende Prieto}, and
  D.~{Kiselman}, {\it {Line formation in solar granulation. IV. [O I], O I and
  OH lines and the photospheric O abundance}},  {\em \aap} {\bf 417} (2004)
  751--768, [\href{http://xxx.lanl.gov/abs/astro-ph/0312290}{{\tt
  astro-ph/0312290}}].

\bibitem{AspVI}
M.~{Asplund}, N.~{Grevesse}, A.~J. {Sauval}, C.~{Allende Prieto}, and
  R.~{Blomme}, {\it {Line formation in solar granulation. VI. [C I], C I, CH
  and C$_{2}$ lines and the photospheric C abundance}},  {\em \aap} {\bf 431}
  (2005) 693--705, [\href{http://xxx.lanl.gov/abs/astro-ph/0410681}{{\tt
  astro-ph/0410681}}].

\bibitem{AGS05}
M.~{Asplund}, N.~{Grevesse}, and A.~J. {Sauval}, {\it {The Solar Chemical
  Composition}},  in {\em Cosmic Abundances as Records of Stellar Evolution and
  Nucleosynthesis} (T.~G. {Barnes III} and F.~N. {Bash}, eds.), Astron. Soc.
  Pac., San Francisco, {\em ASP Conf.\ Ser.} {\bf 336} (2005) 25.

\bibitem{ScottVII}
P.~{Scott}, M.~{Asplund}, N.~{Grevesse}, and A.~J. {Sauval}, {\it {Line
  formation in solar granulation. VII. CO lines and the solar C and O isotopic
  abundances}},  {\em \aap} {\bf 456} (2006) 675--688,
  [\href{http://xxx.lanl.gov/abs/astro-ph/0605116}{{\tt astro-ph/0605116}}].

\bibitem{Melendez08}
J.~{Mel{\'e}ndez} and M.~{Asplund}, {\it {Another forbidden solar oxygen
  abundance: the [O I] 5577 {\AA} line}},  {\em \aap} {\bf 490} (2008)
  817--821, [\href{http://xxx.lanl.gov/abs/0808.2796}{{\tt arXiv:0808.2796}}].

\bibitem{Scott09Ni}
P.~{Scott}, M.~{Asplund}, N.~{Grevesse}, and A.~J. {Sauval}, {\it {On the Solar
  Nickel and Oxygen Abundances}},  {\em \apjl} {\bf 691} (2009) L119--L122,
  [\href{http://xxx.lanl.gov/abs/0811.0815}{{\tt arXiv:0811.0815}}].

\bibitem{AGSS}
M.~{Asplund}, N.~{Grevesse}, A.~J. {Sauval}, and P.~{Scott}, {\it {The chemical
  composition of the Sun}},  {\em \araa} {\bf 47} (2009) 481--522,
  [\href{http://xxx.lanl.gov/abs/0909.0948}{{\tt arXiv:0909.0948}}].

\bibitem{Bahcall:2004yr}
J.~N. Bahcall, S.~Basu, M.~Pinsonneault, and A.~M. Serenelli, {\it
  {Helioseismological implications of recent solar abundance determinations}},
  {\em \apj} {\bf 618} (2005) 1049--1056,
  [\href{http://xxx.lanl.gov/abs/astro-ph/0407060}{{\tt astro-ph/0407060}}].

\bibitem{Basu:2004zg}
S.~Basu and H.~Antia, {\it {Constraining solar abundances using
  helioseismology}},  {\em \apj} {\bf 606} (2004) L85,
  [\href{http://xxx.lanl.gov/abs/astro-ph/0403485}{{\tt astro-ph/0403485}}].

\bibitem{Bahcall06}
J.~N. {Bahcall}, A.~M. {Serenelli}, and S.~{Basu}, {\it {10,000 Standard Solar
  Models: A Monte Carlo Simulation}},  {\em \apjs} {\bf 165} (2006) 400--431,
  [\href{http://xxx.lanl.gov/abs/astro-ph/0511337}{{\tt astro-ph/0511337}}].

\bibitem{Yang07}
W.~M. {Yang} and S.~L. {Bi}, {\it {Solar Models with Revised Abundances and
  Opacities}},  {\em \apjl} {\bf 658} (2007) L67--L70,
  [\href{http://xxx.lanl.gov/abs/0805.3644}{{\tt arXiv:0805.3644}}].

\bibitem{Basu08}
S.~{Basu} and H.~M. {Antia}, {\it {Helioseismology and solar abundances}},
  {\em \physrep} {\bf 457} (2008) 217--283,
  [\href{http://xxx.lanl.gov/abs/0711.4590}{{\tt arXiv:0711.4590}}].

\bibitem{Serenelli:2009yc}
A.~Serenelli, S.~Basu, J.~W. Ferguson, and M.~Asplund, {\it {New Solar
  Composition: The Problem With Solar Models Revisited}},  {\em \apj} {\bf 705}
  (2009) L123--L127, [\href{http://xxx.lanl.gov/abs/0909.2668}{{\tt
  arXiv:0909.2668}}].

\bibitem{Bahcall05}
J.~N. {Bahcall}, A.~M. {Serenelli}, and S.~{Basu}, {\it {New Solar Opacities,
  Abundances, Helioseismology, and Neutrino Fluxes}},  {\em \apjl} {\bf 621}
  (2005) L85--L88, [\href{http://xxx.lanl.gov/abs/astro-ph/0412440}{{\tt
  astro-ph/0412440}}].

\bibitem{Badnell05}
N.~R. {Badnell}, M.~A. {Bautista}, {\em et.~al.}, {\it {Updated opacities from
  the Opacity Project}},  {\em \mnras} {\bf 360} (2005) 458--464,
  [\href{http://xxx.lanl.gov/abs/astro-ph/0410744}{{\tt astro-ph/0410744}}].

\bibitem{Arnett05}
D.~{Arnett}, C.~{Meakin}, and P.~A. {Young}, {\it {The Lambert Problem}},  in
  {\em Cosmic Abundances as Records of Stellar Evolution and Nucleosynthesis}
  (T.~G. {Barnes}, III and F.~N. {Bash}, eds.) {\bf 336} (2005) 235.

\bibitem{Charbonnel05}
C.~{Charbonnel} and S.~{Talon}, {\it {Influence of Gravity Waves on the
  Internal Rotation and Li Abundance of Solar-Type Stars}},  {\em \science}
  {\bf 309} (2005) 2189--2191,
  [\href{http://xxx.lanl.gov/abs/astro-ph/0511265}{{\tt astro-ph/0511265}}].

\bibitem{Guzik05}
J.~A. {Guzik}, L.~S. {Watson}, and A.~N. {Cox}, {\it {Can Enhanced Diffusion
  Improve Helioseismic Agreement for Solar Models with Revised Abundances?}},
  {\em \apj} {\bf 627} (2005) 1049--1056,
  [\href{http://xxx.lanl.gov/abs/astro-ph/0502364}{{\tt astro-ph/0502364}}].

\bibitem{Castro07}
M.~{Castro}, S.~{Vauclair}, and O.~{Richard}, {\it {Low abundances of heavy
  elements in the solar outer layers: comparisons of solar models with
  helioseismic inversions}},  {\em \aap} {\bf 463} (2007) 755--758,
  [\href{http://xxx.lanl.gov/abs/astro-ph/0611619}{{\tt astro-ph/0611619}}].

\bibitem{Christensen09}
J.~{Christensen-Dalsgaard}, M.~P. {di Mauro}, G.~{Houdek}, and F.~{Pijpers},
  {\it {On the opacity change required to compensate for the revised solar
  composition}},  {\em \aap} {\bf 494} (2009) 205--208,
  [\href{http://xxx.lanl.gov/abs/0811.1001}{{\tt arXiv:0811.1001}}].

\bibitem{Guzik10}
J.~A. {Guzik} and K.~{Mussack}, {\it {Exploring Mass Loss, Low-Z Accretion, and
  Convective Overshoot in Solar Models to Mitigate the Solar Abundance
  Problem}},  {\em \apj} {\bf 713} (2010) 1108--1119,
  [\href{http://xxx.lanl.gov/abs/1001.0648}{{\tt arXiv:1001.0648}}].

\bibitem{Serenelli11}
A.~M. {Serenelli}, W.~C. {Haxton}, and C.~{Pe{\~n}a-Garay}, {\it {Solar Models
  with Accretion. I. Application to the Solar Abundance Problem}},  {\em \apj}
  {\bf 743} (2011) 24, [\href{http://xxx.lanl.gov/abs/1104.1639}{{\tt
  arXiv:1104.1639}}].

\bibitem{Vincent12}
A.~C. {Vincent}, P.~{Scott}, and R.~{Trampedach}, {\it {Light bosons in the
  photosphere and the solar abundance problem}},  {\em \mnras} {\bf 432} (2013)
  3332--3339, [\href{http://xxx.lanl.gov/abs/1206.4315}{{\tt
  arXiv:1206.4315}}].

\bibitem{Haxton13}
W.~C. {Haxton}, R.~G. {Hamish Robertson}, and A.~M. {Serenelli}, {\it {Solar
  Neutrinos: Status and Prospects}},  {\em \araa} {\bf 51} (2013) 21--61,
  [\href{http://xxx.lanl.gov/abs/1208.5723}{{\tt arXiv:1208.5723}}].

\bibitem{Lopes13b}
I.~{Lopes}, {\it {Probing the Sun's inner core using solar neutrinos: A new
  diagnostic method}},  {\em \prd} {\bf 88} (2013) 045006,
  [\href{http://xxx.lanl.gov/abs/1308.3346}{{\tt arXiv:1308.3346}}].

\bibitem{Smirnov13}
A.~Y. {Smirnov}, {\it {Neutrino 2012: Outlook - theory}},  {\em Nuclear Physics
  B Proceedings Supplements} {\bf 235} (2013) 431--440,
  [\href{http://xxx.lanl.gov/abs/1210.4061}{{\tt arXiv:1210.4061}}].

\bibitem{Nauenberg87}
M.~{Nauenberg}, {\it {Energy transport and evaporation of weakly interacting
  particles in the Sun}},  {\em \prd} {\bf 36} (1987) 1080--1087.

\bibitem{GouldRaffelt90b}
A.~{Gould} and G.~{Raffelt}, {\it {Cosmion energy transfer in stars - The
  Knudsen limit}},  {\em \apj} {\bf 352} (1990) 669--680.

\bibitem{Fitzpatrick13}
A.~L. {Fitzpatrick}, W.~{Haxton}, E.~{Katz}, N.~{Lubbers}, and Y.~{Xu}, {\it
  {The effective field theory of dark matter direct detection}},  {\em \jcap}
  {\bf 2} (2013) 4, [\href{http://xxx.lanl.gov/abs/1203.3542}{{\tt
  arXiv:1203.3542}}].

\bibitem{Kumar13}
J.~{Kumar} and D.~{Marfatia}, {\it {Matrix element analyses of dark matter
  scattering and annihilation}},  {\em \prd} {\bf 88} (2013) 014035,
  [\href{http://xxx.lanl.gov/abs/1305.1611}{{\tt arXiv:1305.1611}}].

\bibitem{Pospelov00}
M.~{Pospelov} and T.~{ter Veldhuis}, {\it {Direct and indirect limits on the
  electro-magnetic form factors of WIMPs}},  {\em \plb} {\bf 480} (2000)
  181--186, [\href{http://xxx.lanl.gov/abs/hep-ph/0003010}{{\tt
  hep-ph/0003010}}].

\bibitem{Sigurdson04}
K.~{Sigurdson}, M.~{Doran}, A.~{Kurylov}, R.~R. {Caldwell}, and
  M.~{Kamionkowski}, {\it {Dark-matter electric and magnetic dipole moments}},
  {\em \prd} {\bf 70} (2004) 083501,
  [\href{http://xxx.lanl.gov/abs/astro-ph/0406355}{{\tt astro-ph/0406355}}].

\bibitem{DelNobile:2013cva}
E.~{Del Nobile}, G.~{Gelmini}, P.~{Gondolo}, and J.-H. {Huh}, {\it {Generalized
  halo independent comparison of direct dark matter detection data}},  {\em
  \jcap} {\bf 10} (2013) 48, [\href{http://xxx.lanl.gov/abs/1306.5273}{{\tt
  arXiv:1306.5273}}].

\bibitem{Sommerfeld}
A.~{Sommerfeld}, {\it {{\"U}ber die Beugung und Bremsung der Elektronen}},
  {\em \annp} {\bf 403} (1931) 257--330.

\bibitem{Hisano05}
J.~{Hisano}, S.~{Matsumoto}, M.~M. {Nojiri}, and O.~{Saito}, {\it
  {Nonperturbative effect on dark matter annihilation and gamma ray signature
  from the galactic center}},  {\em \prd} {\bf 71} (2005) 063528,
  [\href{http://xxx.lanl.gov/abs/hep-ph/0412403}{{\tt hep-ph/0412403}}].

\bibitem{AHDM}
N.~{Arkani-Hamed}, D.~P. {Finkbeiner}, T.~R. {Slatyer}, and N.~{Weiner}, {\it
  {A theory of dark matter}},  {\em \prd} {\bf 79} (2009) 015014,
  [\href{http://xxx.lanl.gov/abs/0810.0713}{{\tt arXiv:0810.0713}}].

\bibitem{CUBPACK}
R.~Cools and A.~Haegemans, {\it Algorithm 824: Cubpack: A package for automatic
  cubature; framework description},  {\em ACM Trans. Math. Softw.} {\bf 29}
  (2003) 287--296.

\bibitem{Krstic}
P.~S. Krsti\ifmmode~\acute{c}\else \'{c}\fi{} and D.~R. Schultz, {\it
  Consistent definitions for, and relationships among, cross sections for
  elastic scattering of hydrogen ions, atoms, and molecules},  {\em \pra} {\bf
  60} (1999) 2118--2130.

\bibitem{Tulin:2013teo}
S.~{Tulin}, H.-B. {Yu}, and K.~M. {Zurek}, {\it {Beyond collisionless dark
  matter: Particle physics dynamics for dark matter halo structure}},  {\em
  \prd} {\bf 87} (2013) 115007, [\href{http://xxx.lanl.gov/abs/1302.3898}{{\tt
  arXiv:1302.3898}}].

\bibitem{Vincent14}
A.~C. Vincent, A.~Serenelli, and P.~Scott {\em in prep}.

\bibitem{Gould87a}
A.~{Gould}, {\it {Weakly interacting massive particle distribution in and
  evaporation from the sun}},  {\em \apj} {\bf 321} (1987) 560--570.

\bibitem{Lopes14}
I.~{Lopes}, P.~{Panci}, and J.~{Silk}, {\it {Helioseismology with long range
  dark matter-baryon interactions}},
  \href{http://xxx.lanl.gov/abs/1402.0682}{{\tt arXiv:1402.0682}}.

\end{thebibliography}\endgroup

\end{document}